\documentclass{aa}
\usepackage{graphicx}
\usepackage{amsmath}
\usepackage[varg]{txfonts}

\usepackage{soul}
\newlength{\linwx}
\usepackage{xcolor}
\usepackage{subcaption}
\usepackage[bookmarks=false]{hyperref}
\usepackage{enumitem}
\usepackage{xfrac}


\hypersetup{colorlinks=true,linkcolor=magenta,urlcolor=purple,citecolor = teal}

\newcommand{\be}{\begin{equation}}
\newcommand{\ee}{\end{equation}}
\newcommand{\infint}{\int_0^{\infty}}

\newcommand{\Sg}{\Sigma_g}
\newcommand{\Sgmid}{\Sigma_{g,0}}

\begin{document}

\title{Influence of grain growth on the thermal structure of protoplanetary discs}
\author{Sofia Savvidou \inst{1} 
\and Bertram Bitsch \inst{1}
\and Michiel Lambrechts\inst{2}
}
\offprints{S. Savvidou,\\ \email{savvidou@mpia.de}}
\institute{
Max-Planck-Institut f\"ur Astronomie, K\"onigstuhl 17, 69117 Heidelberg, Germany
\and
Lund Observatory, Department of Astronomy and Theoretical Physics, Lund University, 22100 Lund, Sweden
}
\date{Received date / Accepted date }
\abstract{
The thermal structure of a protoplanetary disc is regulated by the opacity that dust grains provide. However, previous works have often considered simplified prescriptions for the dust opacity in hydrodynamical disc simulations, for example by considering only a single particle size. In the present work we perform 2D hydrodynamical simulations of protoplanetary discs where the opacity is self-consistently calculated for the dust population, taking into account the particle size, composition and abundance.  We first compare simulations utilizing single grain sizes to two different multi-grain size distributions at different levels of turbulence strengths, parameterized through the $\alpha$-viscosity, and different gas surface densities. 
Assuming a single dust size leads to inaccurate calculations of the thermal structure of discs, because the grain size dominating the opacity increases with orbital radius.
Overall the two grain size distributions, one limited by fragmentation only and the other determined from a more complete fragmentation-coagulation equilibrium, give comparable results for the thermal structure. We find that both grain size distributions give less steep opacity gradients that result in less steep aspect ratio gradients, in comparison to discs with only micrometer sized dust. Moreover, in the discs with a grain size distribution, the innermost (< 5 AU) outward migration region is removed and planets embedded is such discs experience lower migration rates. We also investigate the dependency of the water iceline position on the alpha-viscosity  ($\alpha$), the initial gas surface density ($\Sgmid$) at 1 AU and the dust-to-gas ratio ($f_{DG}$) and find $r_{ice} \propto \alpha^{0.61}\Sgmid^{0.8} f_{DG}^{0.37}$ independently of the distribution used in the disc. The inclusion of the feedback loop between grain growth, opacities and disc thermodynamics allows for more self-consistent simulations of accretion discs and planet formation. 
}
\authorrunning{Savvidou et al.}\maketitle

\section{Introduction}
\label{sec:Introduction}

Protoplanetary discs surround young stars for the first few million years after their formation and they are the birthplaces of planetary systems.  The position of the iceline within the discs influences the formation and growth of planets. 
Planetesimal formation has been found to be enhanced or even initiated there because of water vapor that is diffused outwards from the hot, inner disc and recondenses after the iceline \citep{2013A&A...552A.137R}.  This recondensation increases the abundances of icy pebbles, which have better sticking properties compared to dry aggregates \citep{1997Icar..129..539S,2009ApJ...702.1490W,2015ApJ...798...34G}, causing a pile-up near the iceline and triggering the streaming instability  \citep{2014A&A...572A..72G,2017A&A...602A..21S,2017A&A...608A..92D}.

An increase in the dust surface density after the iceline can also aid in the growth of gas giant planet cores \citep{1988Icar...75..146S}. The location and the evolution of the iceline location can be defining for the innermost boundary of gas giant formation and along with other parameters, such as the disc's mass, it can also determine what kind of planets will be created \citep{2008ApJ...673..502K} and their masses \citep{2015Icar..258..418M,2016Icar..267..368M}. In addition to that, the location of the iceline transition affects the composition of exoplanetary atmospheres \citep{2014ApJ...791L...9M,2016MNRAS.461.3274C,2017MNRAS.469.4102M} .

The location of the iceline is determined by the local temperature in the disc \citep{1981PThPS..70...35H,2000ApJ...528..995S,2004M&PS...39.1859P}. The thermal structure of the discs is thus decisive for planetesimal and planet formation. It is, though, greatly affected by the dust content of the protoplanetary disc and the opacity that the dust grains provide.  This complex interplay is caused by the influence between gas and dust. The relative velocity for each pair of grains is determined by the aerodynamic properties of the grains, namely the Stokes number, and the local properties of the gas, such as the temperature or the volume density \citep{2007A&A...466..413O}. The variety in the relative velocities results in different collisional outcomes between grain sizes, such as coagulation or fragmentation \citep{2008A&A...480..859B,2011A&A...534A..73Z, 2011A&A...525A..11B,2012A&A...539A.148B}. As a result, the dust content of the protoplanetary disc is described by a distribution of grain sizes, with number densities that are not necessarily equally distributed between all existing sizes.  

  Each grain size population has a different opacity, therefore having a distribution instead of a single grain size means that the disc's total opacity will be affected and as a consequence it will affect the resulting structure of the disc.  As stated in \citet{2016SSRv..205...41B}, the opacity, defines the observational characteristics of a protoplanetary disc by influencing the dust thermal continuum emission and the excitation conditions for the gas lines. Additionally, since opacity regulates the amount of light that can be absorbed by the disc, it determines its thermal structure. These reasons make opacity a very important factor of the structure and evolution of a protoplanetary disc. 

\begin{figure}
\centering
\includegraphics[scale=.5]{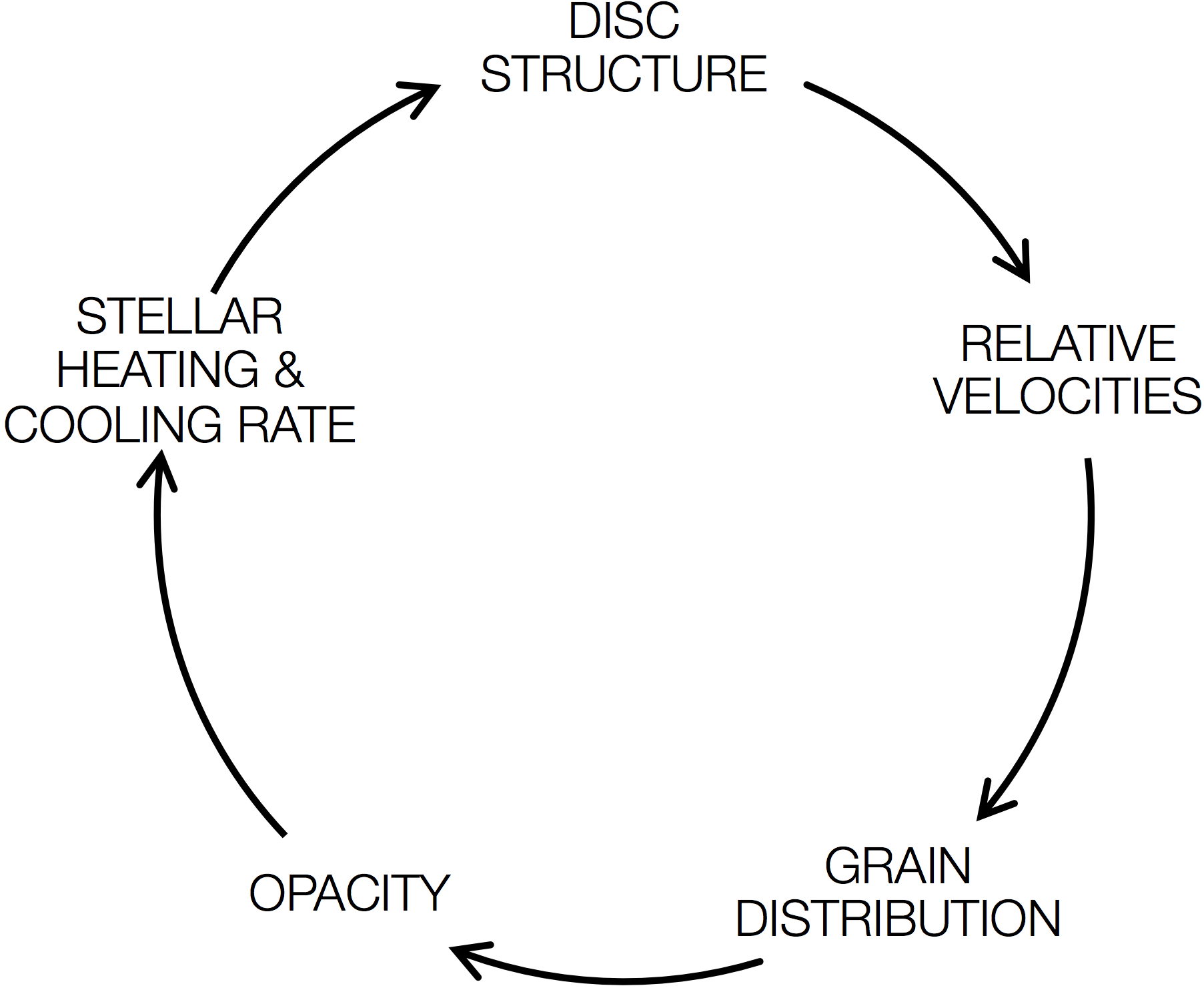}
\caption{Graphical illustration of the feedback loop. The thermal and density structure of a
protoplanetary disc is determined by this loop: temperature  and gas surface density affect the relative velocities for grains of different sizes. Through the relative velocities we find the outcomes of collisions between grains, therefore a grain size distribution is created. The grains are then vertically distributed according to their sizes and the turbulence strength. The spatial distribution of the grain sizes determines the opacity of the disc, which then affects its cooling rate and the stellar heating. This way the temperature and density of the disc change and subsequently its whole structure. }
\label{Fig:Loop}
\end{figure}

The interplay between opacity and the thermal structure creates a feedback loop that we include in hydrodynamical simulations of equilibrium discs (Fig. \ref{Fig:Loop}).  
Even though the goal of theoretical models is to simulate protoplanetary discs as realistically as possible, typically they only include specific parts of the feedback loop, contrary to this work. In the following paragraphs we will introduce what work has been done in parts of the feedback loop.

\emph{Relative velocities and grain size distribution.} Early on, \citet{1969epcf.book.....S} worked on dust growth within the context of planet formation and on the time evolution equation for grain size distributions (often called Smoluchowski equation, \citealt{1916ZPhy...17..557S}). A lot of work was also done on dust dynamics and how they would affect collisional outcomes and, as a consequence, coagulation and fragmentation of dust particles \citep{1980Icar...44..172W,1984Icar...60..553W,1981Icar...45..517N,2008A&A...480..859B}. A grain size distribution has been widely assumed to follow a power-law derived from the equilibrium between coagulation and fragmentation, inspired by the work of \citet{1969JGR....74.2531D} on the number density distribution of objects in the asteroid belt. The number density, thus, can be approximated as $n(s) \propto s^\xi$, where s is the grain size and $\xi$ a constant. Several attempts were made in order to define this constant, mainly through analytical calculations combined with observational data for the interstellar medium grains \citep[e.g.][MRN power-law]{1977ApJ...217..425M}, but also through experimental studies \citep[e.g.][]{1990Icar...83..156D}. 
It was shown by \citet{1996Icar..123..450T} that the $\xi$ constant is independent of the specific parameters of the collisional outcome model, as long as it is self-similar, which in this case means that the outcome of impacts between dust grains depends on the masses of two colliding particles only through their ratio.

However, such a description of a grain size distribution with only one power law is a simplification, since it only takes into account the coagulation/fragmentation equilibrium. More recently, the work on grain size distributions has been aided by laboratory experiments of dust collisions \citep[review by][]{2008ARA&A..46...21B,2010A&A...513A..56G}. Such experiments determine what the collisional outcomes are between particles of equal or different size, for different relative velocities. They also help in creating models to simulate such collisions accurately and they can be used in the effort of understanding which processes are relevant within the context of planetesimal formation in protoplanetary discs \citep[e.g.][]{2010A&A...513A..57Z}. If additional effects are also taken into account, such as cratering or different regimes due to size-dependent relative velocities, then the size distribution is described by broken power laws \citep{2011A&A...525A..11B}. 

The studies that were discussed above focused on the local distribution of grains in a protoplanetary disc patch due to fragmentation and growth by coagulation, and typically assume that the gas disc does not evolve in time and the dust has no effect on the gas.

\emph{Opacity.} As a first step, some work has been done on opacity alone within the context of protoplanetary discs \citep{1993Icar..106...20M,2006ApJ...636.1114D,2014ApJS..210...21C}. The goal of those works is to create a simple opacity model that can describe as realistically as possible the dust opacity and can be then used in disc simulations \citep{2013A&A...549A.124B} or help in the interpretation of disc observations \citep{2018ApJ...869L..45B}. Alongside the theoretical models, several observations of the dust emission have been performed in order to connect opacity with the particle sizes present in the protoplanetary discs \citep{2007prpl.conf..767N,2005ApJ...631.1134A,2007ApJ...671.1800A,2006A&A...446..211R,2009A&A...495..869L,2010A&A...521A..66R,2011ApJ...739L...8R,2012MNRAS.425.3137U}. 

\emph{Disc structure and grain size distribution.} Several works in the recent years aimed to couple the dust and gas components of protoplanetary discs in simulations and in most of the cases such models include a grain size distribution. However, the models that will be discussed here simulate the gas component of a protoplanetary disc and how the dust component is affected by the gas, but the solids do not influence the gas. Even without the back-reactions of dust on gas, modeling grain size distributions can be computationally challenging, given the long list of effects and parameters to be taken into account, especially using N-body like techniques to treat dust particles. As a consequence, some of the first attempts on this kind of models were made using the Monte-Carlo method \citep{2007A&A...461..215O,2008ApJ...684.1291O} and the goal was to examine how the internal structure of dust affects the collisional evolution of the particles and the disc structure. The Monte-Carlo method has been also used in \citet{2010A&A...513A..57Z,2011A&A...534A..73Z}, while in their work the experimental collisional outcomes from \citet{2010A&A...513A..56G} were implemented and the effect of the porosity and settling of the dust grains on the collisional outcomes was tested. \citet{2008A&A...480..859B} and \citet{2010A&A...513A..79B} numerically solve the Smoluchowski equation for the coagulation/fragmentation equilibrium in vertically isothermal steady-state gas discs, while \citet{2012ApJ...752..106O}studied the effects of the dust grain porosity on the dust evolution in a similar disc setup. In the works discussed above the feedback of the dust on the gas disc structure and especially its thermal part, is not taken into account.

\emph{Disc structure, opacity and cooling rate.} The category of models that was described above neglected the effects of opacity, even though the dust opacity regulates the cooling rate of the disc, which affects the disc structure. In recent years, some studies tried to fill this gap by including the effect of dust opacity in disc simulations. \citet{2011ApJ...738..141O} performed 1+1D \footnote{In the  1+1D approach, the vertical structure of each annulus is solved independently and then all of the annuli are used to construct the radial and vertical structure of the disc.} simulation
 focusing on the effect that water-ice opacity has on the location of the iceline. In the aforementioned study the wavelength-dependent opacities of water-ice and silicates are directly used when calculating the radiative transfer. In \citet{2013A&A...549A.124B,2014A&A...564A.135B,2015A&A...575A..28B} the \citet{1994ApJ...427..987B} opacity profile is followed (in \citet{2013A&A...549A.124B} constant opacity discs were also modeled) and 2D simulations (radial and vertical direction, assuming axisymmetry) are performed using the NIRVANA and the FARGOCA code adding radial heat diffusion and stellar irradiation. The effect of the water-ice to silicates ratio on the resulting thermal disc structures has also been studied recently \citep{2016A&A...590A.101B} using the FARGOCA code and the opacity module from the RADMC-3D code to calculate the mean opacities (as in the present work), but the opacity differences for the water-to-silicate fractions considered are then translated into differences in the \citet{1994ApJ...427..987B} opacity model.
The \citet{1994ApJ...427..987B} opacity model gives approximate values for the frequency averaged opacities within specific temperature regimes (e.g. ice grains, evaporation of ice grains, metal grains, etc.) assuming micrometer sized particles. 
The fixed opacity profile then gives the cooling rate and the stellar heating, therefore it defines the disc structure.

Even though including the opacity feedback in disc simulations is an important improvement, the aforementioned studies did not include the effect of grain growth and fragmentation, and thus only employed opacities derived for single grain sizes. In addition to this, all of these studies assumed a uniform dust-to-gas ratio in the vertical direction of the disc which is in contrast to our approach in this work (see section \ref{subsec:vertical}).

\emph{Disc structure, grain size distribution and opacity.} \citet{1997A&A...325..569S} coupled the dust and gas evolution in 1D simulations, while they also took into consideration the grain opacity. For the mean opacity calculations they followed the approach of \citet{1996A&A...311..291H}, which is similar to the \citet{1994ApJ...427..987B} opacity model approach. Moreover, the size distribution follows the \citet{1977ApJ...217..425M} power-law. It was found that since grains determine the opacity, their evolution will subsequently change the opacity and therefore affect the structure and evolution of a protoplanetary disc. Prior to this study, \citet{1988A&A...195..183M} and \citet{1989Icar...80..189M} included the dust component evolution in accretion discs and used the results to perform grain opacity calculations.  In \citet{2001ApJ...551..461S} the coagulation/fragmentation equilibrium is included in order to investigate how the dust emission is affected by the grain size distribution and its corresponding opacity. In this work the size distribution follows the \citet{1977ApJ...217..425M} power-law and opacity was calculated using Mie theory. However, in the studies discussed above the back-reaction of the opacity onto the disc structure was not taken into account. 

In the previous paragraphs some examples were given of the work that has been done in the context of grain growth within protoplanetary discs. Nevertheless, previous models were based on several simplifications, most important of which was that they neglected parts of the feedback loop (Fig. \ref{Fig:Loop}) that defines protoplanetary disc structures \citep[e.g.][]{2011A&A...525A..11B} or used simplified assumptions for the opacity \citep[e.g.][]{2013A&A...549A.124B}. The few attempts that have been made to include the dust feedback on the gas of the disc, were 1D simulations or assumed an isothermal vertical structure for the gas, in contrast to the 2D hydrodynamical models that will be presented here. Secondly, the opacity is either not included in the actual simulations or the opacities were included only for single fixed grain sizes. 

The motivation for this project is to approach a more realistic model for disc structures and their evolution and more specifically to simulate the whole feedback loop including a detailed opacity module. We will consider how grain dynamics and more specifically how grain size distributions affect the opacity and as a consequence the thermal structure of the disc in order to simulate the whole feedback loop. As far as the grain size distribution is concerned, two models were used for the simulations of this project. A simple power-law model following \citet{1977ApJ...217..425M}, hereafter MRN distribution and also a more complex model following \citet{2011A&A...525A..11B}, hereafter BOD distribution. Moreover, an opacity module was included in the 2D hydrodynamical disc simulations in order to more accurately calculate the opacity of the dust grain distribution and account for the back-reactions of dust to gas. In this opacity module, the Rosseland and Planck mean opacities as a function of temperature are used and they are calculated via Mie theory. The simulations were run until the disc reached thermal equilibrium. Such simulations offer us the opportunity to discuss the implications of the resulting disc structures to planet formation and could also serve as the basis to compare with observations (e.g. ALMA images) in future work. 

The structure of this paper is as follows: In Sect. \ref{sec:Methods} we describe the energy equations used in the hydrodynamical simulations, the new opacity module and the two grain size distributions that we included in the code. Also, we mention the input parameters of the simulations and how the disc is set up. Then, in Sect. \ref{sec:Grain size distributions} we compare the discs utilizing the two different grain size distributions for one set of simulation with a nominal turbulence parameter and initial gas surface density. In Sect. \ref{sec:alpha-visc_Sigma0_fDG}, we discuss how the simulations change when we vary the turbulence parameter and how it changes when the surface density is different. The implications of the results are discussed in Sect. \ref{sec:Implications} and a summary follows in Sect. \ref{sec:Summary}.

\section{Methods}
\label{sec:Methods}

In this section we discuss the different methods used in our work. We review the hydrodynamical equations in section \ref{subsec:Hydro sims}, the opacities and how they are calculated in section \ref{subsec:Opa-Temp}. In section \ref{subsec:GSD_methods} we discuss the grain growth mechanism and in section \ref{subsec:vertical} we discuss the effects of vertical grain settling. We then finally describe our simulation setup in section \ref{subsec:Sims_setup}.

\subsection{Hydrodynamical simulations}
\label{subsec:Hydro sims}

Calculations with mean opacities derived from single grain sizes were first introduced into the FARGOCA code by \citet{2014MNRAS.440..683L} and \citet{2014A&A...564A.135B}, who performed 2D and 3D radiation hydrodynamical simulations of discs and planet-disc interactions. The FARGOCA code solves the continuity and the Navier-Stokes equations and uses the flux-limited diffusion approach to radiative transfer.
More specifically, the  time evolution of the energy profile of the protoplanetary disc is determined by
\be \label{eq:E} \frac{\partial E_R}{\partial t} + \nabla \cdot {\bf F} = \rho\kappa_P [B(T) - cE_R] \ee
\be \label{eq:e} \frac{\partial \epsilon}{\partial t} + \nabla \cdot ({\bf u}\cdot\nabla)\epsilon = -P\nabla\cdot {\bf u} -\rho \kappa_P[B(T)-cE_R]+ Q^+ + S~. \ee
The radiative energy density $E_R$ is thus  independent from the thermal energy density $\epsilon$. In the expressions above the blackbody radiation energy is $B(T) = 4\sigma T^4$, where $\sigma$ is the Stefan-Boltzmann constant, $\rho$ is the gas density, $\kappa_P$ the Planck mean opacity (further specified in Sec. \ref{subsec:Opa-Temp}), u the velocity, P is the thermal pressure, $Q^+$ is the viscous dissipation or heating function and $S$ is the stellar heating component \citep{1981ApJ...248..321L,2010ApJ...710.1395D,2011A&A...529A..35C}. 

In our simulations we use the flux-limited diffusion (FLD) for the radiation flux ${\bf F}$ as described in \citet{1981ApJ...248..321L}
\be {\bf F} = -\frac{\lambda c}{\rho\kappa_R }\nabla E_R~. \ee 
In the flux-limited diffusion equation, $c$ is the speed of light, $\alpha_R$ is the radiation constant, $\kappa_R$ is the Rosseland mean opacity and $\lambda$ the flux-limiter of \citet{1989A&A...208...98K}. More details on the energy equations can be found in \cite{Bitsch:2013cd}. The opacities that were introduced in the above equations will be discussed in the following section.

The stellar heating density received by a grid cell of width $\Delta$r is defined as \citep{2010ApJ...710.1395D}:
\be S = F_{\star}e^{-\tau}\frac{1-e^{-\rho\kappa_{\star}\Delta r}}{\Delta r}~,\ee
with $F_{\star} = R_{\star}^2\sigma T_{\star}^4/r^2$ being the stellar flux, $R_{\star}$ the stellar radius, $T_{\star}$ the stellar surface temperature, $\tau$ the radially integrated optical depth (up to each grid cell) and $\kappa_{\star}$ the stellar opacity (further specified in Sec. \ref{subsec:Opa-Temp}).

\subsection{Opacity-Temperature module}
\label{subsec:Opa-Temp}

\begin{figure*}
\centering
\begin{subfigure}{0.33\textwidth}
\includegraphics[width=1.1\textwidth]{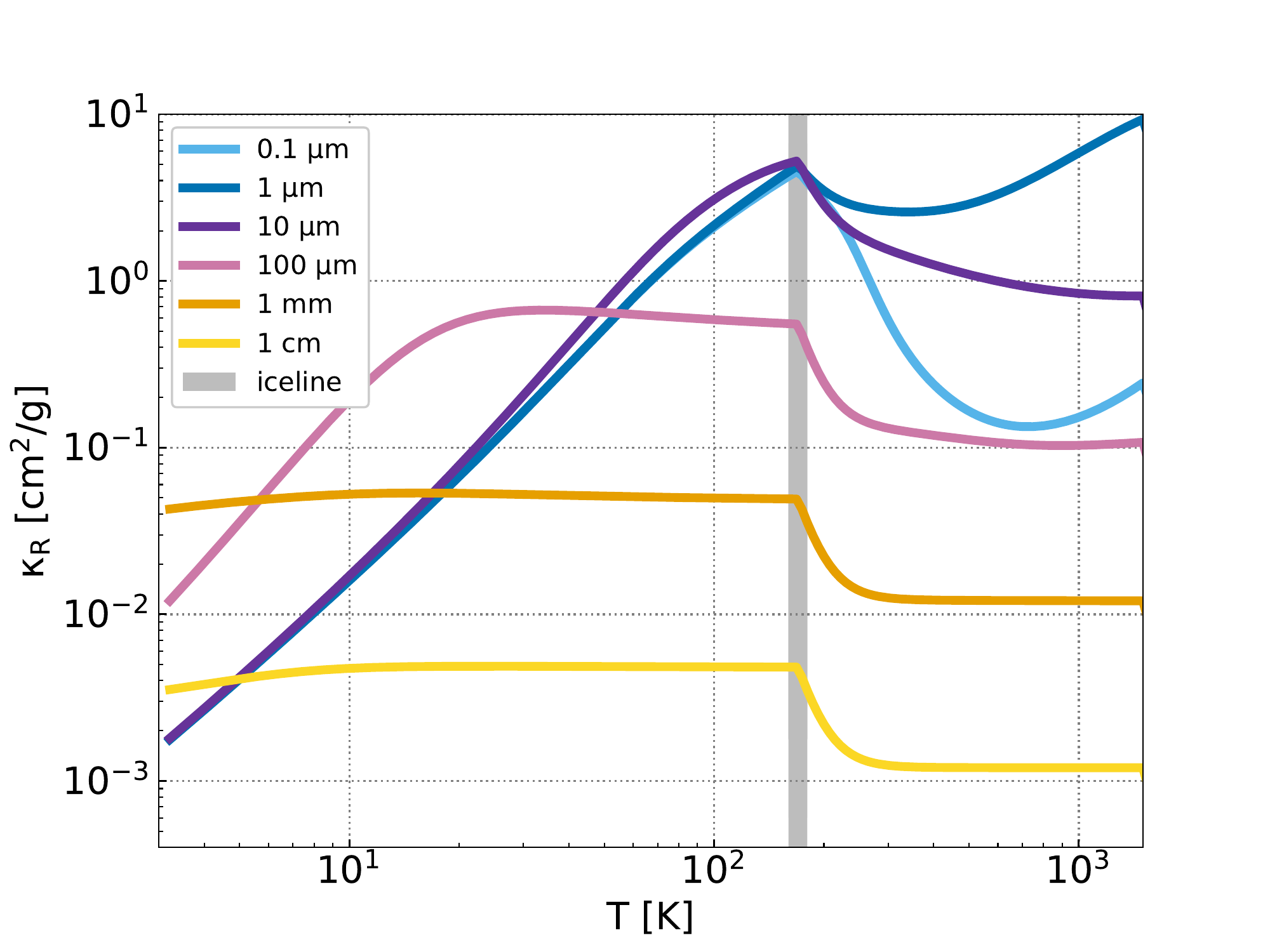}
\end{subfigure}
\begin{subfigure}{0.33\textwidth}
\includegraphics[width=1.1\textwidth]{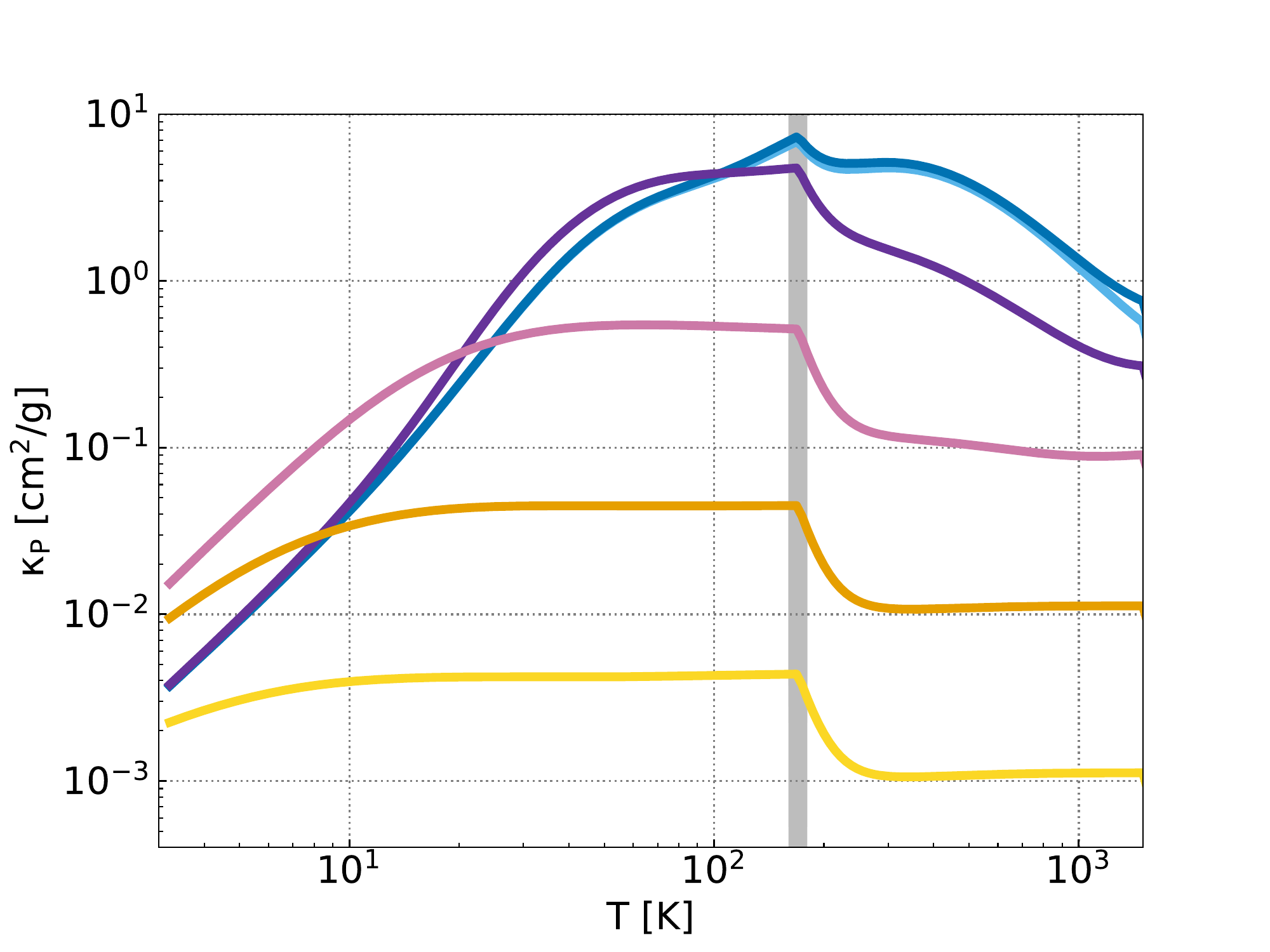}
\end{subfigure}
\begin{subfigure}{0.33\textwidth}
\includegraphics[width=1.1\textwidth]{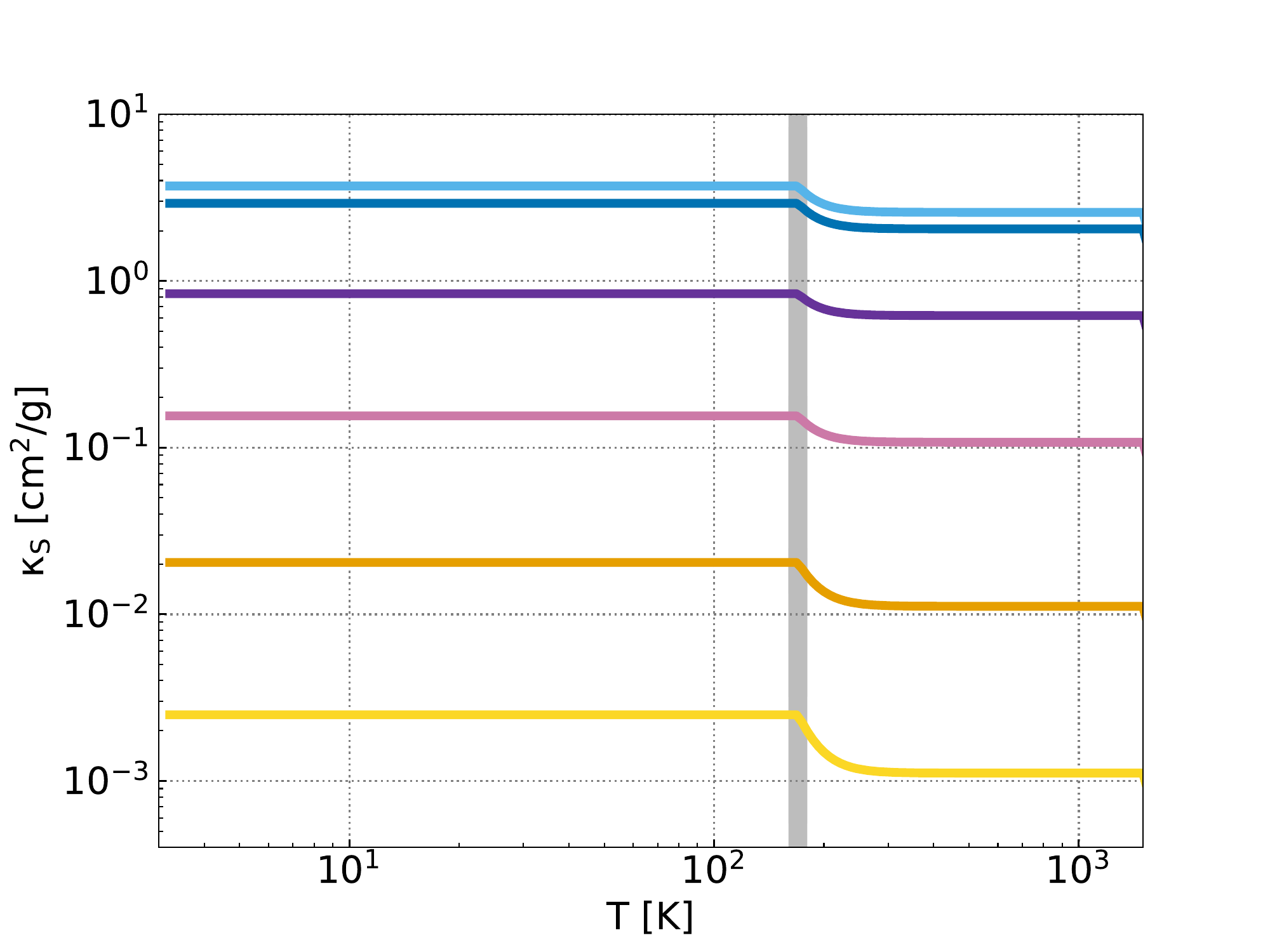}
\end{subfigure}
\caption{Rosseland, Planck and stellar mean opacities (from left to right) as a function of temperature for grains of sizes 0.1, 1, 10, 100 $\mu$m, 1 mm and 1 cm. They are independent of the gas density because they are dominated by the dust component.
These values were calculated using RADMC-3D for a mixture of 50\% silicates, 50\% ice and disc dust-to-gas ratio of 1\%. The gray vertical line shows the location of the water iceline transition (170 K $\pm$ 10 K), causing a transition in opacity due to the evaporation/condensation of dust grains.}
\label{Fig:Opacities}
\end{figure*}

In the energy equations (Eqs. \ref{eq:E} and \ref{eq:e}) of the hydrodynamical simulation, we use the mean opacities that are averaged over all wavelengths. 
If we use the Planck black body radiation energy density distribution $B_{\lambda}(\lambda, T)$ as a weighting function we can define the Planck mean opacity as
\be \label{eq:kappaP}\kappa _P = \frac{\infint \kappa_{\lambda, ns}(T,\rho) B_{\lambda}(\lambda, T)d\lambda}{\infint B_{\lambda}(\lambda ,T) d\lambda}~. \ee
Since the mean free path of thermal radiation in the disc is small compared to the disc's scale height, the radiation field can be considered isotropic, blackbody emission. 

The Rosseland mean opacity uses the temperature derivative of the Planck distribution as a weighting function and is defined as 
\be \kappa_R^{-1} = \frac{\infint \kappa_{\lambda,s}^{-1}(T,\rho) (\partial B_{\lambda} (\lambda , T)/\partial T)d\lambda }{\infint (\partial B_{\lambda} (\lambda , T)/\partial T)d\lambda }~. \ee
It should be noted that scattering processes are neglected (subscript ns) when calculating the wavelength dependent opacities $\kappa_{\lambda}$ for the Planck mean, but are included in the Rosseland mean opacity (subscript s). 

We also consider the stellar radiation and define the stellar or optical opacity as
\be \label{eq:kappaS} \kappa_{\star} = \frac{\infint \kappa_{\lambda,ns}(T,\rho) B_{\lambda}(\lambda,T_{\star})d\lambda}{\infint  B_{\lambda}(\lambda,T_{\star})d\lambda}~. \ee
The stellar opacity is then the Planck mean opacity taking into consideration the stellar radiation temperature instead of the local disc temperature. 

We calculate the mean Rosseland, Planck and stellar opacities as a function of temperature using the RADMC-3D \footnote{\url{http://www.ita.uni-heidelberg.de/~dullemond/software/radmc-3d/}} code. Note that dust opacities are independent of the gas density, as opposed to gas opacities. The latter are not considered in this work as opacities in the disc are dominated by the dust component and the high temperature needed for dust evaporation are not reached within our simulations. The code utilizes Mie-scattering theory and the optical constants for water-ice \citep{2008JGRD..11314220W} and silicates \citep{1994A&A...292..641J,1995A&A...300..503D} in order to calculate the wavelength-dependent opacities, which are then averaged over all wavelengths. The main input parameters are the size of the grains and the dust-to-gas ratio of the disc. We can also choose the dust grain species, silicates, water ice and carbon or the fraction between those in the dust mixture. In this work we include a mixture of 50\% water-ice and 50\% silicates and the dust-to-gas ratio for the calculation of the opacities is 1\%.  Finally, the Rosseland, Planck and stellar mean opacities \citep[see][]{Bitsch:2013cd} are calculated. 

In Fig. \ref{Fig:Opacities} it is illustrated how each mean opacity scales with temperature for six different grain sizes, from 0.1 $\mu$m to 1  cm. The wavelength dependent opacities and subsequently the mean opacities depend on the size parameter $x=\frac{2\pi s}{\lambda}$, but also on the refractive index of the given grain species, which is also itself dependent on wavelength \citep{2008Icar..194..368M}.
By Wien's law the wavelength is inversely proportional to the temperature.
Using the size parameter we find that the regime changes at approximately x = 1 and more specifically at x $\ll$ 1 we have the Rayleigh scattering, whereas at x $\gg$ 1 we have the geometric optics regime \citep{Bohren1998}.  Consequently, if the size of the particle is a lot smaller than the wavelength of the incident radiation, absorption dominates over scattering and the wavelength dependent opacities become independent of grain size. In the case of the larger grain sizes, or when x $\gg$ 1, the opacities become independent of wavelength (and consequently temperature), but depend on the grain size. Most of the regions though lie somewhere in between, which means that calculating the opacity depends on both the grain size with its individual refractive index and the given wavelength or temperature.

The Rosseland mean opacities (Fig. \ref{Fig:Opacities}) for the largest particles of the set (100 $\mu$m, 1 mm and 1 cm) are almost flat, except for the transition around the iceline at 170 K $\pm$ 10 K. At this temperature, ice sublimates and the opacity is then only determined by silicates. For those large particle sizes, the size parameter is greater than 1, therefore we are in the geometric optics regime and the Rosseland mean opacity is independent of temperature. However, we note that for a grain size of 100$\mu$m and temperature below 20K the opacity depends on the temperature. In this region the regime has changed and the opacity is determined by Rayleigh scattering. Equally, the size parameter is well below 1. The same trend can be seen for the smaller particles, namely 0.1,1 and 10 $\mu$m	before the iceline. The opacity of the 10 $\mu$m grain sizes goes into the geometric optics regime after the iceline and tends to become independent of temperature. In the region after the iceline for the smallest particles (0.1 and 1 $\mu$m) the size parameter is closer to 1, so the opacities are also influenced by the refractive indices.

The Planck opacities have a weaker dependency on temperature compared to the Rosseland mean opacities. The stellar opacities depend only on the stellar temperature and grain sizes, but not on the disc temperature, except for the transition at the water ice line, when water rich particles evaporate. Both the Planck and stellar opacities are calculated using only the absorption coefficient, which does not have a strong dependency on wavelength and consequently temperature, as opposed to the extinction coefficient. The Planck mean opacities are calculated taking into account the temperature of the disc, while the stellar opacities, use the temperature of the star, which is constant. 

Using RADMC-3D a number of files is created with the mean opacity values as a function of temperature. These files are then used in the hydrodynamical code (FARGOCA). The opacity calculations from these files are interpolated and in this way we get in the code the appropriate opacity values given the temperature of the grid cell. The reason why the interpolation is done instead of directly calculating opacity using Mie theory is because the computational time would be very long.
We include the direct opacity-temperature calculations for at least 10 grain sizes, from 0.1 $\mu$m to 1 mm or 2 cm. We then create size bins and as a simplification, each grain size within a bin shares the same opacity-temperature calculations (corresponding to the logarithmic mean size of that bin). We note here that these bins are different and for computational reasons wider than the bins used for the calculations of the vertically integrated dust surface densities (see Sec. \ref{subsec:GSD_methods}).

As a comparison we will also use the frequency averaged \citet{1994ApJ...427..987B} opacity law. The opacity in this case, depends on the local temperature and density. There are several transitions in this opacity regime caused by the processes which dominate each temperature region, such as the evaporation of ices interior to the iceline, which is also present in the prescription we are using for the discs with the grain size distributions. However, the greatest difference between the two opacity regimes is that the \citet{1994ApJ...427..987B} opacity law is based on micrometer-sized dust and does not take the opacity provided by all of the dust sizes present in the disc into account. The \citet{1994ApJ...427..987B} law also considers the gas opacities, but these are relevant for high temperatures that will not be reached in the simulations presented here.

\subsection{Grain size distributions}
\label{subsec:GSD_methods}

The collision between two dust grains can result in various outcomes. The possible outcomes are coagulation, fragmentation, cratering and bouncing \cite[review by][]{2008ARA&A..46...21B}. The outcome of a collision is determined by the relative velocities of the colliding bodies and their mass ratio \citep{1977MNRAS.180...57W,2008A&A...480..859B}. 

The relative velocities between grains are determined by the mass of the particles, but they are also greatly affected by the local temperature and the gas scale height. Dust dynamics involve not only collisions between grains, but also with the molecules of the protoplanetary disc's gas. These collisions with the gas cause a lag to the dust particles that leads to relative velocities between themselves. 

\begin{figure}
\centering
\includegraphics[scale=.45]{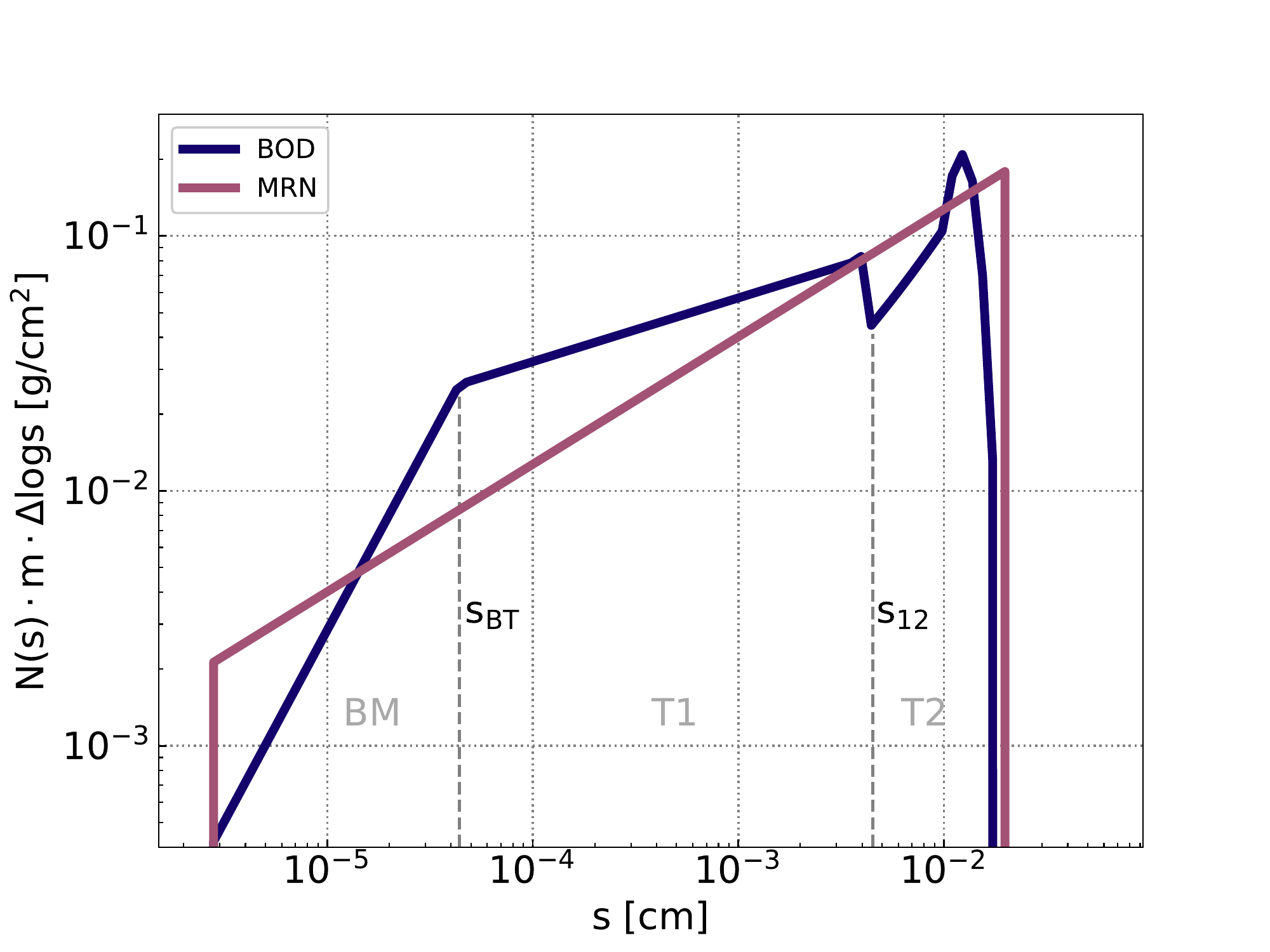}
\caption[Grain size distributions]{Vertically integrated dust surface density distribution per logarithmic bin of grain size as a function of grain size for the two distributions used here, after \citet{2011A&A...525A..11B} (BOD) and \citet{1977ApJ...217..425M} (MRN), at 10 AU, for the simulation with $\alpha = 5\times10^{-3}$ (see Fig. \ref{Fig:Dust} for the distributions over all orbital distances). The dust to gas ratio is 1\%  and the gas surface density is $1000~g/cm^2$ at 1 AU in both cases. For both of the distributions we additionally used $u_f = 1~m/s$ and $\rho_s = 1.6~g/cm^3$. In the BOD, we can distinguish three regions. Small particles follow Brownian motion (BM), then as they grow they follow the turbulent gas motions (T1) and finally they are affected by the turbulence, but get decoupled from the gas as their stopping times are much larger than the turn-over time of the eddies (T2). The barrier $s_{BT}$ is the grain size limit for Brownian motion, while the barrier $s_{12}$ separates the two turbulent regimes. The bump near the end of the distribution is caused by cratering, since large particles only lose part of their mass this way, while small particles can only coagulate and form larger grains. This causes the distribution to be top-heavy. The MRN distribution is a single power-law (Eq.\ref{eq:MRN}). Both grain size distributions have a maximum value determined by the same fragmentation limit (Eq.\ref{eq:s_max}), but they are not the same for the BOD and the MRN distribution because of the self-consistently calculated temperature.}
\label{Fig:Distributions}
\end{figure}

We compare in this work two different grain size distribution models. These models provide the vertically integrated surface density of dust as a function of the grain size. The first and simple model (hereafter MRN) is inspired by the groundwork on dust distributions \citep{1969JGR....74.2531D,1977ApJ...217..425M,1996Icar..123..450T}. At a given distance to the star, the equilibrium between fragmentation and coagulation results in a steady-state size distribution, where the number density of the particles can be written as 
\be n(m)dm \propto m^{-\xi}dm
\ee
or
\be n(s)ds \propto s^{2-3\xi}ds~,
\ee
with m the particle mass, s the particle size and $\xi$ a constant. 

The mass of a specific size, within a size bin $[s_i-ds',s_i+ds'']$ is 
\be
M_{s_i} = \int_{s_i-ds'}^{s_i+ds''} m\cdot n(s) ds \propto\left[\frac{s^{6-3\xi}}{6-3\xi}\right]_{s_i-ds'}^{s_i+ds''}~,
\ee
assuming 5-3$\xi\neq$ -1. The grain sizes for this project are distributed over a logarithmic grid, so $s_i-ds'$ is $\sqrt{s_i\cdot s_{i-1}}$ and  $s_i+ds''$ is $\sqrt{s_i\cdot s_{i+1}}$. The vertically integrated surface density of each grain size bin is then 
\be \Sigma_{d,s_i}  \propto f_{DG}\Sigma_g \left[\left(\sqrt{s_i\cdot s_{i+1}}\right)^{6-3\xi}-\left(\sqrt{s_i\cdot s_{i-1}}\right)^{6-3\xi}\right]~,
\ee
where $f_{DG}$ is the dust-to-gas ratio and $\Sigma_g$ is the gas surface density.

We use a grain size grid, such as $s_{i+1} = c\cdot s_i$ and the assumption that $\xi$=11/6 \citep{1969JGR....74.2531D,1994Icar..107..117W}, so then the expression for the unnormalized vertically integrated surface density for each grain size bin can be simplified to
\be \Sigma_{d,s_i} \propto s_i^{1/2}f_{DG}\Sigma_g~.\ee
The contributions from each grain size are then summed up. In order to get the normalized surface density values we divide each contribution by the aforementioned sum
\be \label{eq:MRN} \tilde{\Sigma}_{d,s_i} =\frac{s_i^{1/2}f_{DG}\Sigma_g}{\sum_i s_i^{1/2}}.\ee

The second and more complex model \citep[][hereafter named BOD]{2011A&A...525A..11B} takes into account fragmentation, coagulation and also cratering, where only part of the mass of the target body is excavated after the collision with a small impactor. The input parameters for this model are the dust and gas surface densities ($\Sigma_{d,0}$ and $\Sigma_{g,0}$), the local disc temperature ($T$), the alpha turbulence parameter ($\alpha$), the volume density of the particles ($\rho_s$) and finally the fragmentation velocity ($u_f$), which is the critical velocity above which all collisions lead to either fragmentation or cratering. The logarithmic grid for the sizes of both distributions is defined as $ s_{i+1} = 1.12s_i$, while the smallest grain size is 0.025 $\mu$m. As mentioned in Sec. \ref{subsec:Opa-Temp} this grain size grid is finer than the size grid we use to determine the opacities. 

Considering that different particle sizes lead to different collision outcomes, this recipe takes into account the relative velocities that particles of different sizes will develop in order to create different regimes for each size. These regimes are created according to size boundaries, within which different power-laws apply for the fit to the size distribution. It should be mentioned that these size boundaries are defined by the corresponding relative velocities of the dust grains. The smallest particles of the distribution follow Brownian motion, which means their motions are affected by collisions with the gas molecules, there is no preferred direction and they do not have angular momentum. The next regime regards larger particles that start to get affected by turbulent mixing. It was also found \citep{2007A&A...466..413O} that when particles have stopping times approximately equal or larger compared to the turn-over time of the smallest eddy of the gas, they start to decouple from the gas, so they follow a different regime. Finally, the distribution has an upper end or a fragmentation barrier above which particles can no longer grow and only fragmentation occurs. 

Between two size boundaries, the distribution is described by a power-law $n(m)\cdot m \cdot s = s_i ^{\delta_i}$ of different powers $\delta_i$, depending on the regime (Brownian motion or turbulent mixing). Within each one of the regimes, the power-law indices are different if the grains are affected by settling, given their sizes and the disc parameters (see Sec. \ref{subsec:vertical} for a discussion on the vertical distribution of grains). The powers for each regime are found in Table \ref{tab:Exponents} and using these we can create a first fit $f(s_i)$. It is necessary then to include a bump caused by cratering and the cut-off effects of the distribution that cause an increase in the fit for large enough particles. This boundary effect is caused by the fact that large particles near the upper boundary of the grain sizes grid do not have larger particles to collide with, but the mass transfer from one size bin to the other needs to be constant to keep a steady-state grain size distribution. Therefore the number density is increased. Similarly, erosion by small impactors slows down the growth of large particles and an increase in the number density is needed to keep the flux constant. More details for this recipe can be found in \citet{2011A&A...525A..11B}. 

Finally the fit is normalized according to the total dust surface density at the given location (Fig. \ref{Fig:Distributions}, also see Sect. 5.2 in \citet{2011A&A...525A..11B}) as in the first model. This fit represents the vertically integrated dust surface densities per logarithmic bin of grain size, N(s)$\cdot$m$\cdot \Delta\log$s, where
\be N(s) = \int_0^{z_{max}} n(s)~dz
\ee
is the vertically integrated number density.

\renewcommand{\arraystretch}{2}
\begin{table}[h]
\centering
\begin{tabular}{l|c|c}
\hline \hline 
							& \multicolumn{2}{c}{$\delta_i$} \\
Regime				& $s_i \leq s_{sett}$ 	& $s_i \geq s_{sett}$ \\
\hline
Brownian motion regime & $\frac{3}{2}$				&  $ \frac{5}{4}$	\\
\hline
Turbulent regime I 			& $ \frac{1}{4}	$			&  $0$	\\
\hline
Turbulent regime II 			&$ \frac{1}{2}	$			&  $ \frac{1}{4}$	\\
\hline \hline 
\end{tabular}
\caption{Power-law exponents for each regime in the grain size distribution \citep{2011A&A...525A..11B}. The distribution in each regime is $n(m) \cdot m \cdot s \propto s_i^{\delta_i}$. }
\label{tab:Exponents}
\end{table}

The maximum grain size or in other words the fragmentation barrier is defined in the simulations with both of the grain size distributions as 
\be \label{eq:s_max} s_{max} \simeq \frac{2\Sg}{\pi \alpha \rho_s}\frac{u_f^2}{c_s^2}~, \ee 
with $\rho_s=1.6~g/cm^2$ the density of each particle, $u_f= 1~m/s$ the fragmentation threshold velocity, $k_B$ the Boltzmann constant, $m_p$ the proton mass and $\mu$ = 2.3 the mean molecular weight in proton masses. The sound speed is given by
\be \label{eq:soundspeed} c_s = \sqrt{\frac{k_BT}{\mu m_p}}~. \ee
The threshold velocity $u_f \sim 1~m/s$ corresponds to the threshold after which collisions between silicates always lead to fragmentation \citep{2000ApJ...533..454P}. However, it has also been experimentally found that water-ice shows a higher threshold velocity, $u_f \sim 10~m/s$ \citep{2015ApJ...798...34G}. We choose to use only the lower fragmentation threshold in the here presented work, but the composition dependency will be studied in future work.

Because of Eq. \ref{eq:s_max}, which applies to both distributions, and the different regime boundaries in BOD, which depend on the local disc parameters (see Fig. \ref{Fig:Distributions} and Table \ref{tab:Exponents}), there is not a global size distribution, but rather a self-consistent spatial distribution of grain sizes both radially and vertically (see also Sec. \ref{subsec:vertical}).
 
It is noteworthy that even though we consider the coagulation/fragmentation equilibrium and the effects of cratering and settling, we neglect in the following work the drift of grains and the effect of bouncing. However, in the simulations presented here we find that the fragmentation barrier is always smaller than the drift barrier. This means that the particles have already fragmented and replenished the smaller pieces before they would have the chance to experience drift. The small particles are less affected by radial drift \citep{1977MNRAS.180...57W} and since they coagulate, an equilibrium forms that drives the grain size distribution. The fragmentation barrier decreases with increasing $\alpha$-viscosity parameter, which is expected, since an increased $\alpha$ leads to increased turbulent relative velocities. The maximum possible grain size also decreases when the fragmentation threshold velocity decreases. Drift is an important effect acting on dust grains in protoplanetary discs, but it is a reasonable simplification to neglect it for the chosen parameters of our simulations. In future work where, for example, the fragmentation threshold velocities are increased or the composition dependency is included, drift is an effect that needs to be taken into consideration.

\subsection{Vertical distribution of grains}
\label{subsec:vertical}

The grains of a given size are vertically distributed according to the following
\be \label{eq:vertical_distribution}
\rho_d = \rho_{d,0}\exp\left(-\frac{z^2}{2H_d^2}\right)~,
\ee
with 
\be \label{eq:rho_d,mid} \rho_{d,0} =  \frac{\Sigma_d}{\sqrt{2\pi}H_d } \ee
the dust density at midplane and the dust scale height derived by \citet{1995Icar..114..237D}
 \be \label{eq:Hdust} H_d = H_g\sqrt{\frac{\alpha}{\alpha + St}}~,\ee
 where $H_d$ and $H_g$ is the dust and gas scale height respectively. The Stokes number of the particles in the Epstein regime is 
 \be \text{\it St}=\tau_f\Omega_K =  \sqrt{\frac{\pi}{8}} \frac{\rho_s s}{\rho_g H_g}~, \ee
 with $\tau_f$ the stopping time of the particles, $\Omega_K$ the Keplerian velocity and $\rho_g$ the gas volume density. At midplane, where $\rho_g = \rho_{g,0}$ (resembling Eq. \ref{eq:rho_d,mid}), the Stokes number is
\be \label{eq:Stokes}St =  \frac{\pi}{2} \frac{\rho_s s}{\Sg}~. \ee

The vertically integrated dust surface densities $\Sigma_d$ as a function of orbital distance are determined by the grain growth and fragmentation equilibrium prescriptions that were introduced in Sec. \ref{subsec:GSD_methods}. The BOD grain size distribution has already taken into account the effect of settling (to calculate the distribution itself), depending on the grain sizes and the disc parameters (see Sect. \ref{subsec:GSD_methods} and Table \ref{tab:Exponents}). We then distribute in our model the grains vertically according to their sizes and how much they are expected to be affected, in a fashion consistent with the assumptions made in BOD (Eqs. \ref{eq:vertical_distribution}-\ref{eq:Stokes}). However, it has been shown that small particles can get trapped in lower altitudes by the concentration of larger grains due to settling \citep{2016ApJ...822..111K}. This effect is not taken into account here as it is beyond the scope of this work, but could be an improvement in future work. 
 
We use the volume density of dust within a grid cell to find the opacity through 
\be \label{eq:kappa} \bar{\kappa} = \sum_i \left(\frac{\rho_{d,i}}{\rho_g}100\right)~\kappa_i~,
\ee 
where $\kappa_i$ is the opacity of each grain size i, as shown in Fig. \ref{Fig:Opacities}, and $\rho_{d,i}/\rho_g$ the dust-to-gas volume density ratio for a given grain size i. In the case of single grain sizes summing is not needed. The dust-to-gas term for the volume densities includes the settling effect (Eq.  \ref{eq:vertical_distribution}). In this expression we multiply the volume density dust-to-gas ratios by 100 to account for the fact that in the calculations of $\kappa_i$ (with the module from RADMC-3D) a dust-to-gas ratio of 1\% was assumed. This way we multiply this $\kappa_i$ with the appropriate factor depending on the volume density dust-to-gas ratio.

\begin{figure}
\centering
\includegraphics[width=.95\columnwidth]{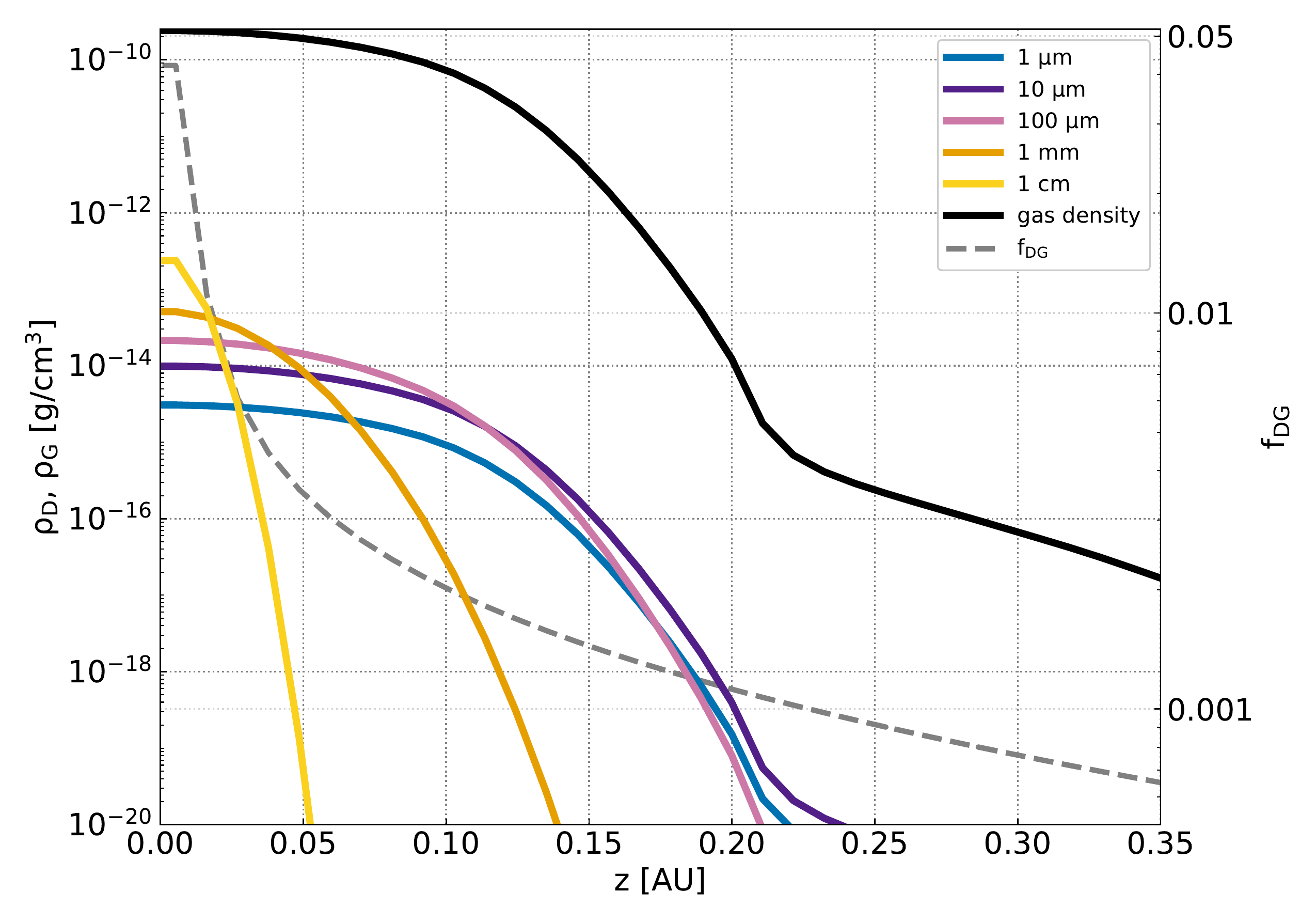}
\caption{Dust density as a function of height for grains of five representative grain sizes within a disc with the BOD, $\alpha=10^{-4}$ and $\Sigma_{g,0} = 1000~g/cm^2$, at 3 AU. The black line shows the gas density of the disc. The dashed gray line shows the dust-to-gas ratio as a function of height for the whole grain size distribution. The dust-to gas ratio of the smallest particles of the sample remain constant, but as grains get larger they get more affected by settling. Consequently, the largest grain (1 cm) is mostly concentrated at the midplane.}
\label{Fig:settling}
\end{figure}

As an example for the effect of settling we show in Fig. \ref{Fig:settling} the dust density as a function of height at 3AU for 5 different grain sizes. In this simulation, the $\alpha$ value is $10^{-4}$, the initial gas surface density at 1 AU is $\Sigma_0 = 1000~g/cm^2$ and the grain size distribution used is the BOD. As a reference, we also plot the gas density to indicate the different volume density dust-to-gas ratios depending on the grain size and the volume density dust-to-gas ratio. 

Considering the $\alpha$ prescription for viscosity of \citet{1973A&A....24..337S} the turbulence must be $\alpha \leq St$ for settling to become important \citep{2009apf..book.....A}.  We see from Eq. \ref{eq:Hdust} that the larger the Stokes number of a particle, the more it will be affected by settling for a given $\alpha$. Additionally, the lower the $\alpha$ value is, the more effective settling will be for even smaller dust particles. For this reason we choose to show an example in Fig. \ref{Fig:settling} of a simulation with $\alpha=10^{-4}$ which is the lowest $\alpha$ value we used in our simulations.

Not only the size, but the location within the disc matters, because the Stokes number for a given grain size depends on the gas density which decreases with the increasing orbital distance. As a consequence, the same particles experience less settling the closer they are to the inner boundary of the disc. 

The smallest particles shown in Figure \ref{Fig:settling}, namely 1 and 10 $\mu$m are not affected by settling, despite the low turbulence strength. Their dust-to-gas ratio remains the same at all heights, so they are well coupled to the gas. Then, the larger the particle, the more effective settling is. The 100 $\mu$m sized dust particles are already affected by settling, but beyond this grain size the difference is even larger. The cm-sized dust, which is nearly the maximum grain size in this simulation, is almost constrained at the midplane. The dust-to-gas ratio (dashed gray line), $\rho_d/\rho_g$, is well below 1\% above z=0.05 AU, but reaches 4\% at midplane.
 
The main difference between various grain sizes is their different opacities as a function of temperature. However, as illustrated in Fig. \ref{Fig:settling}, settling is another important effect, with a distinct efficiency depending on the grain size. Test simulations (presented in Appendix \ref{sec:AppC}) show that a significant settling changes significantly the thermal structure of the disk. Indeed, without it a constant dust-to-gas ratio leads to overestimated opacities above midplane, hence more "puffed-up" inner discs with higher temperatures, which cause a shadowing of the outermost region and prevent it from reaching an equilibrium state. Thus, settling is an important effect which needs to be taken into account in models in order to accurately study the thermal structures of protoplanetary discs.

\subsection{Simulations setup}
\label{subsec:Sims_setup}

 The stellar mass used in the simulations is M$_\star$ = 1M$_\odot$, the temperature is 4370K and the radius is R$_\star$ = 1.5R$_\odot$.  The total dust-to-gas ratio is $f_{DG} = 1\%$. Viscosity in the simulations follows an $\alpha$ prescription \citep{1973A&A....24..337S}, where
 \be \label{eq:viscosity} \nu = \alpha\frac{c_s^2}{\Omega_K}~. \ee
 Recent observations of protoplanetary discs find $\alpha$ values from $10^{-4}$ to $10^{-2}$ or even $10^{-1}$ \citep{2005A&A...442..703H,2009ApJ...700.1502A,2016ApJ...831..122R,2018ApJ...859...21A}, but such large $\alpha$ would cause discs to rapidly expand to great extend \citep{1998ApJ...495..385H} in contrast to observations. We use in this work five sets of simulations with $\alpha=10^{-2}, 5\times10^{-3}, 10^{-3}, 5\times10^{-4}~\text{and}~10^{-4}$ in order to test the effect of turbulence on the thermal structure of the disc. We choose these values in order to include a simulation with $\alpha = 5\times10^{-3}$ so that one can directly compare with the work in \citet{2015A&A...575A..28B} and a simulation with $\alpha = 10^{-3}$ to allow comparison with the discs in \citet{2013A&A...549A.124B}.
 In the simulations with $\alpha=10^{-2}~\text{and}~5\times10^{-3}$ the grid cells are 480$\times$70 (radial-vertical direction) and the disc extends from 2 to 50 AU, while in the simulations with $10^{-3}, 5\times10^{-4}~\text{and}~10^{-4}$ the grid cells are 150$\times$35 and the disc extends from 0.1 to 3.1 AU. 
 
 The gas surface density follows a profile 
 \be \Sigma_{g} = \Sigma_{g,0}\cdot (r/AU)^{-p}~,\ee
  with  $p = 1/2$ and we test two different initial surface densities, $\Sigma_{g,0} = 100~\text{and}~1000~g/cm^2$ for every $\alpha$ value that was mentioned above. We run more combinations of different initial gas surface densities and total dust-to-gas ratios, however these are mainly used in order to produce a fitting for the iceline position as a function of the three parameters, $\alpha$, $\Sigma_{g,0}$ and $f_{DG}$ (see Sect. \ref{subsec:Iceline},  Appendix \ref{sec:AppA} and \ref{sec:AppB}). 
   
Since we simulate equilibrium discs, the surface density profile does not evolve significantly during the simulation, because the thermal equilibrium is reached faster than the viscous evolution equilibrium. At the top of the disc we manually set T=3 K, the temperature of the interstellar medium, so that the disc can be cooled by the upper boundary (as described in \citet{Bitsch:2013cd}). The simulations run for some hundreds of orbits (typically 200-1000 orbits) until they reach thermal equilibrium. Nevertheless, some of the simulations might show signs of convection \citep{2013A&A...550A..52B}, which means that they will remain unstable regardless of integration time.
 
At first, we perform simulations with single grain sizes in order to see the difference in the disc structures between them. Dust grains affect the hydrodynamical simulation through the opacity in each grid cell. Every simulation has a different grain size and the opacity values for this specific size are used (see Fig. \ref{Fig:Opacities}). The simulations of single sizes offer the chance to examine the extend to which different grain sizes affect the disc's evolution and equilibrium structure and predict how much grain growth or a grain size distribution will change the outcome. 
 
In the next step we also consider settling and how it affects large grains. For these simulations we only use single grain sizes and the dust surface density is assumed to be $\Sigma_{d} = f_{DG}\Sigma_{g}$ or specifically $\Sigma_{d} = 0.01\Sigma_{g}$ as before. The difference between discs without and with settling is further discussed in Appendix \ref{sec:AppC}. 

\begin{figure*}
\begin{subfigure}{0.5\textwidth}
\centering
\includegraphics[scale=.5]{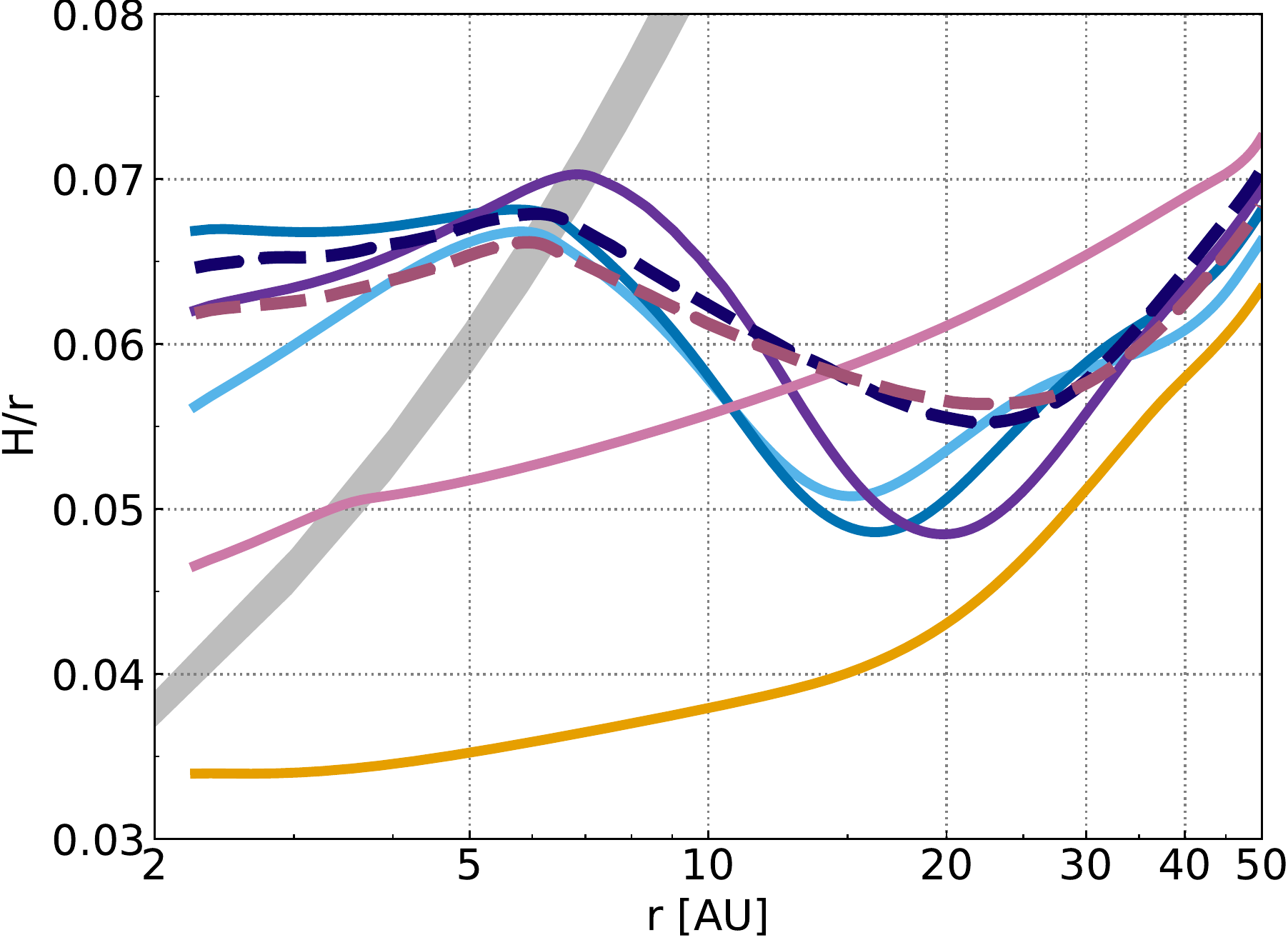}
\caption{}
\label{Fig:ALL_HR_5e3S3O}
\end{subfigure}
\begin{subfigure}{0.5\textwidth}
\centering
\includegraphics[scale=.5]{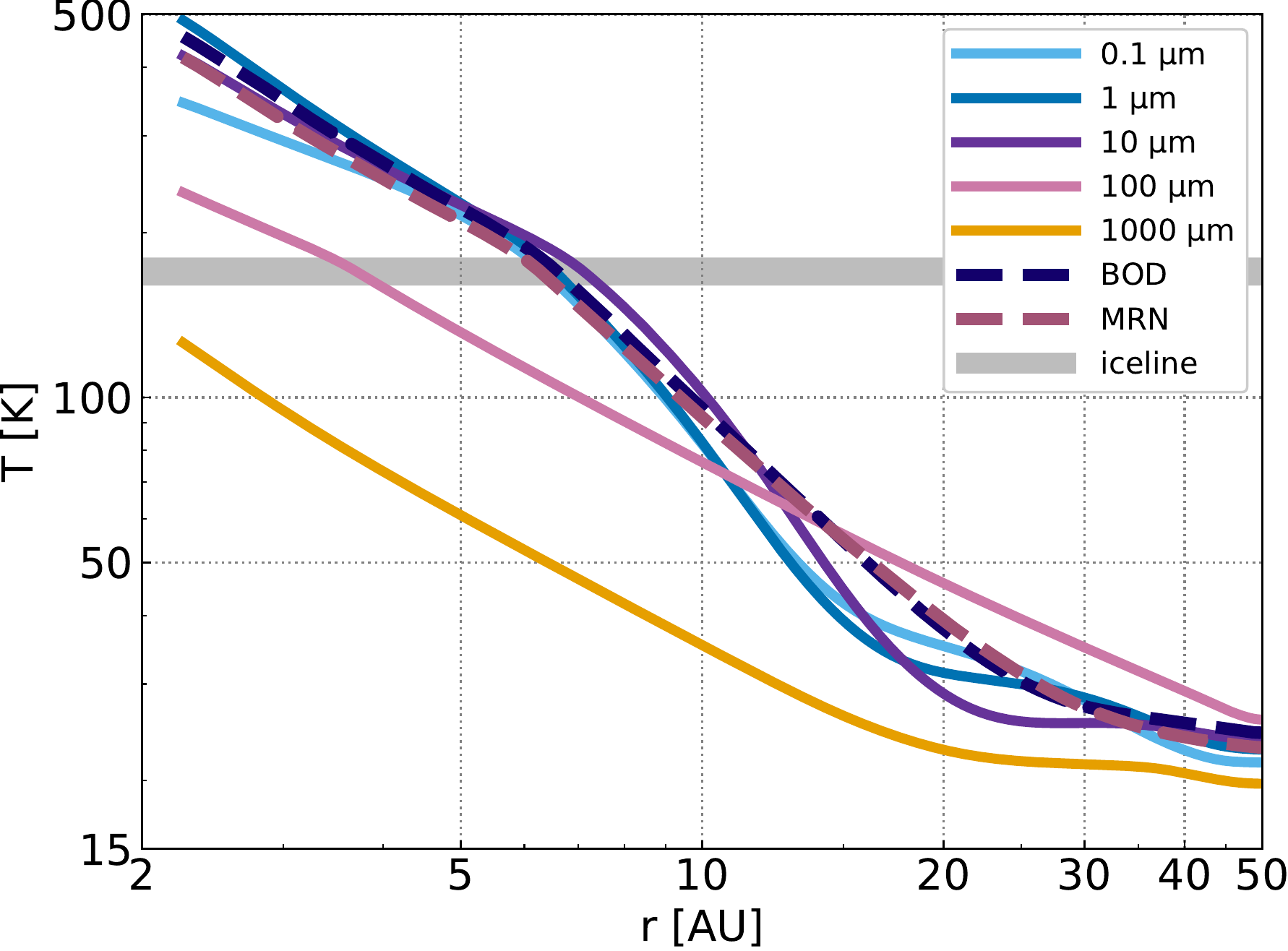}
\caption{}
\label{Fig:ALL_T_5e3S3O}
\end{subfigure}
\caption[]{Aspect ratio (left plot) and midplane temperature (right plot) as a function of orbital distance in AU, for discs with 5 different single grain sizes from 0.1 $\mu$m to 1 mm (see Fig. \ref{Fig:Opacities} for the opacities of those 5 grain sizes).
All of the simulations include viscous heating and stellar irradiation, have $\alpha= 5 \times10^{-3}$ as the turbulence parameter in viscosity, $\Sigma_{g,0} = 1000~g/cm^2$ as the initial gas surface density and the dust-to-gas ratio is $f_{DG} = 1\%$. In this set of simulations we also consider settling, so that we can compare with the simulations that include the grain size distributions. The gray areas in the plot indicate the water iceline transition.  Overplotted with dashed lines are the discs with the MRN distribution in reddish pink and the BOD distribution in dark blue. The simulations with the distributions shows influence from small particles at the inner part of the disc, while the outer parts are mostly affected by larger particles, around 0.1 mm (see Figs. \ref{Fig:Dominant_gs}). The small differences in the aspect ratio and temperature of the discs with the distributions stem from the different slope of the vertically integrated dust surface densities of the two distributions (see Figs. \ref{Fig:Distributions} and \ref{Fig:Dust}).}
\label{Fig:ALL_5e3S3O}
\end{figure*}
 
We, then, include the two grain size distribution models that were discussed in Sect. \ref{subsec:GSD_methods}. The distributions are self-consistently calculated in the code using as input parameters in each time-step and grid cell, the gas surface density (for both distributions) and the temperature (only for the BOD). 
The upper size boundary for the MRN-power law model can be either fixed or follow the same fragmentation barrier formula as the second, more complex model \citep{2011A&A...525A..11B}, but in this work we use the latter. 
 
\section{Grain size distributions}
\label{sec:Grain size distributions}

\subsection{Comparison between the two grain size distributions}
\label{subsec:GSDs}

In this section we compare the simulations utilizing the two different grain size distributions (MRN and BOD). At first, we focus only on the case of $\alpha=5\times10^{-3}$, $\Sigma_{g,0} =1000~g/cm^2$  and $f_{DG} =1\%$. 

In Fig. \ref{Fig:ALL_5e3S3O} we can see the aspect ratio as a function of orbital distance from the star for simulations of different grain sizes along with the two grain size distributions. The gray areas correspond to the iceline transition (T = 170 $\pm$ 10 K). In this area the change in opacity is responsible for the bumps in the aspect ratios. We first focus on the simulation with 0.1 $\mu$m, which roughly corresponds to an unevolved dust population as found in the interstellar medium. The simulation with particles of 0.1 $\mu$m results in an increasing aspect ratio as a function of orbital distance up to 6 AU, where it reaches a maximum and then decreases up to approximately 15 AU. Using 1 $\mu$m-sized particles we see a similar disc structure. The aspect ratio is almost constant for the first few AU and has a small increase around 6 AU. Then it converges with the simulation of 0.1 $\mu$m up to the minimum around 15 AU. If the grain size increases to 10 $\mu$m, then the aspect ratio also increases with distance, features a bump closer to 7.5 AU and decreases with the same slope as the previous two simulations. The larger particles have distinct profiles. Specifically, the aspect ratio of the simulation with particles of 0.1 mm is a monotonically increasing function of orbital distance with a small bump at 3.5 AU. The same can be seen for the simulation with the largest particles, namely 1 mm, but in this case a bump is not visible at any parts of the aspect ratio profile, since the iceline transition does not exist within the boundaries of the simulation. 

The gradients in the aspect ratios for the inner region of the discs are determined by the opacity of the disc. We can compare the opacity gradients in Fig. \ref{Fig:Opacities} with the aspect ratio gradients keeping in mind that the temperature decreases as we move further away from the star. Depending on the temperature at each orbital distance of the disc, we see that the opacity gradients are responsible for the dips and bumps in the aspect ratio profiles of the discs.

\begin{figure*}
\begin{subfigure}{0.5\textwidth}
\centering
\includegraphics[scale=.5]{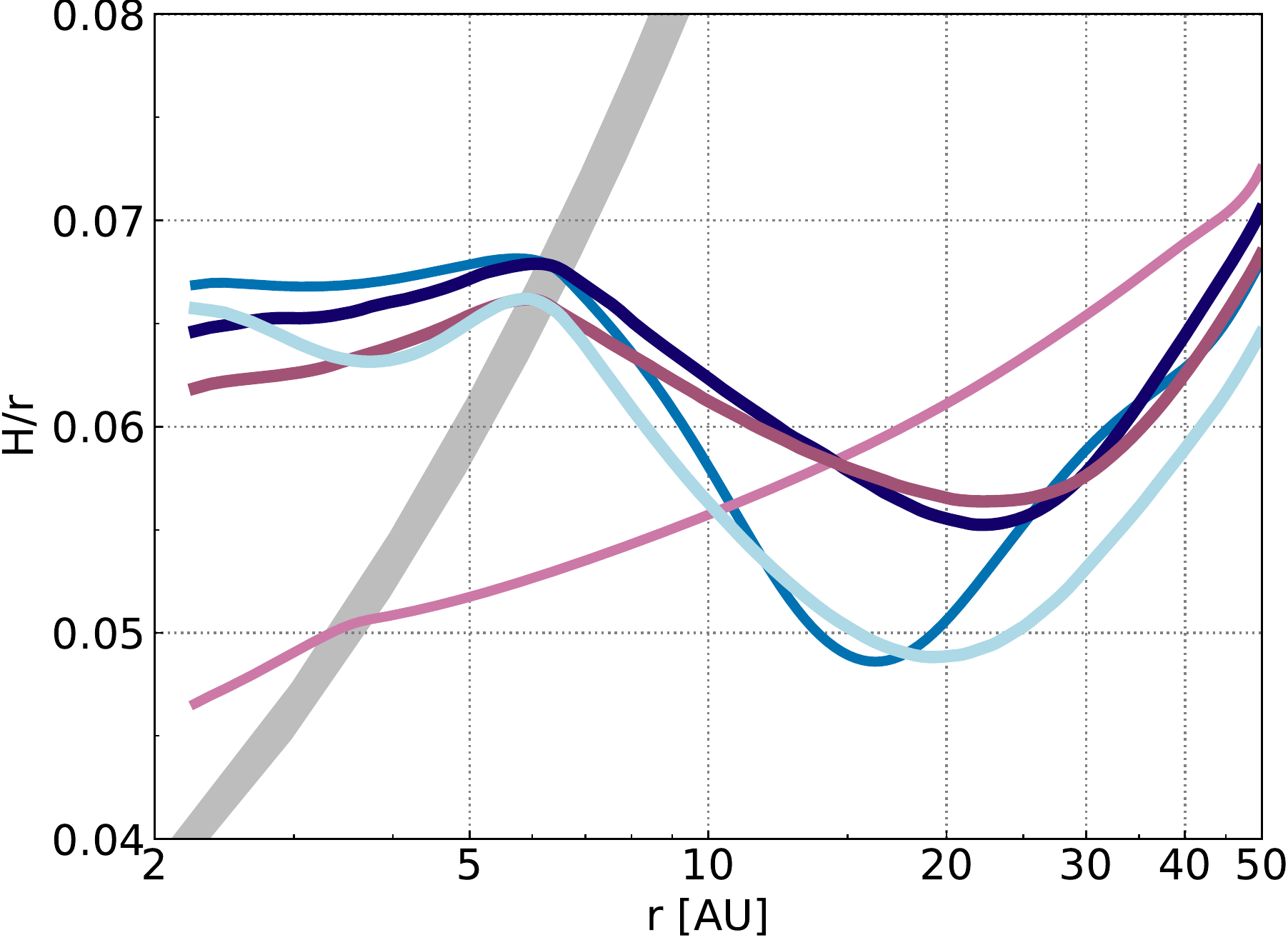}
\end{subfigure}
\begin{subfigure}{0.5\textwidth}
\centering
\includegraphics[scale=.5]{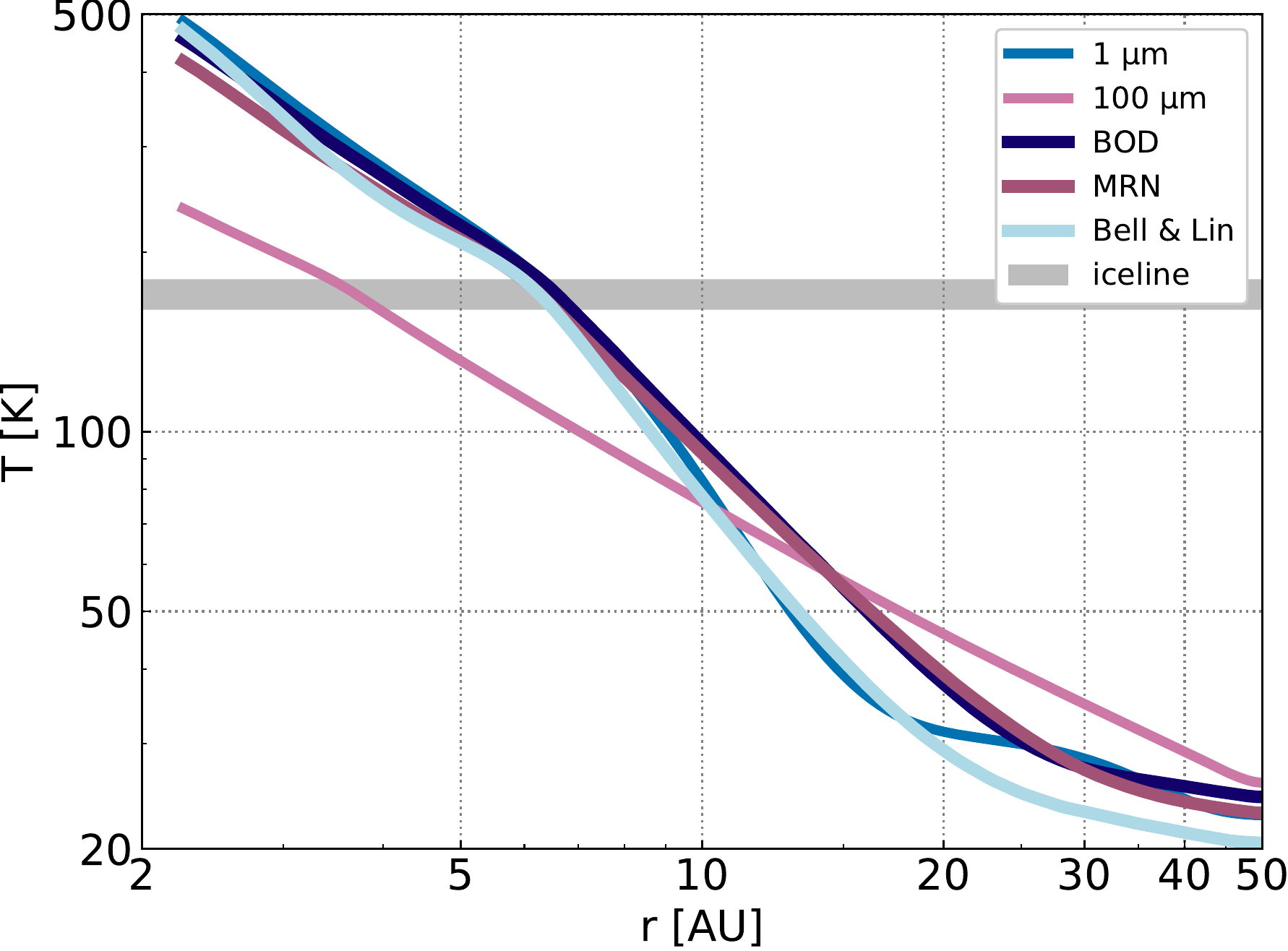}
\end{subfigure}
\caption[]{Aspect ratio (left) and midplane temperature (right) as a function of orbital distance for the discs with the BOD, the MRN distribution and a disc that utilizes the \cite{1994ApJ...427..987B} opacities. All of the simulations have $\alpha= 5 \times10^{-3}$, $\Sigma_{g,0} = 1000~g/cm^2$ and $f_{DG} = 1\%$.
We also show the aspect ratio of the simulations with 1 and 100 $\mu$m for reference. The \cite{1994ApJ...427..987B} opacities are based on micrometer sized particles resulting in comparable aspect ratios. The main differences with the discs including the full grain size distributions are the steeper radial temperature (and thus aspect ratio) gradient for the disc with the \cite{1994ApJ...427..987B} opacities and the reversed slope within 4 AU. These influence the migration speed and direction of planets embedded in those protoplanetary discs (see Figs. \ref{Fig:Migration maps_BOD}, \ref{Fig:Migration maps_MRN}, \ref{Fig:Migration maps_BL}). The gray areas correspond to the water iceline transition.}
\label{Fig:GSDsmicroBL_5e3O}
\end{figure*}

\begin{figure*}
\begin{subfigure}{0.5\textwidth}
\includegraphics[width=\columnwidth]{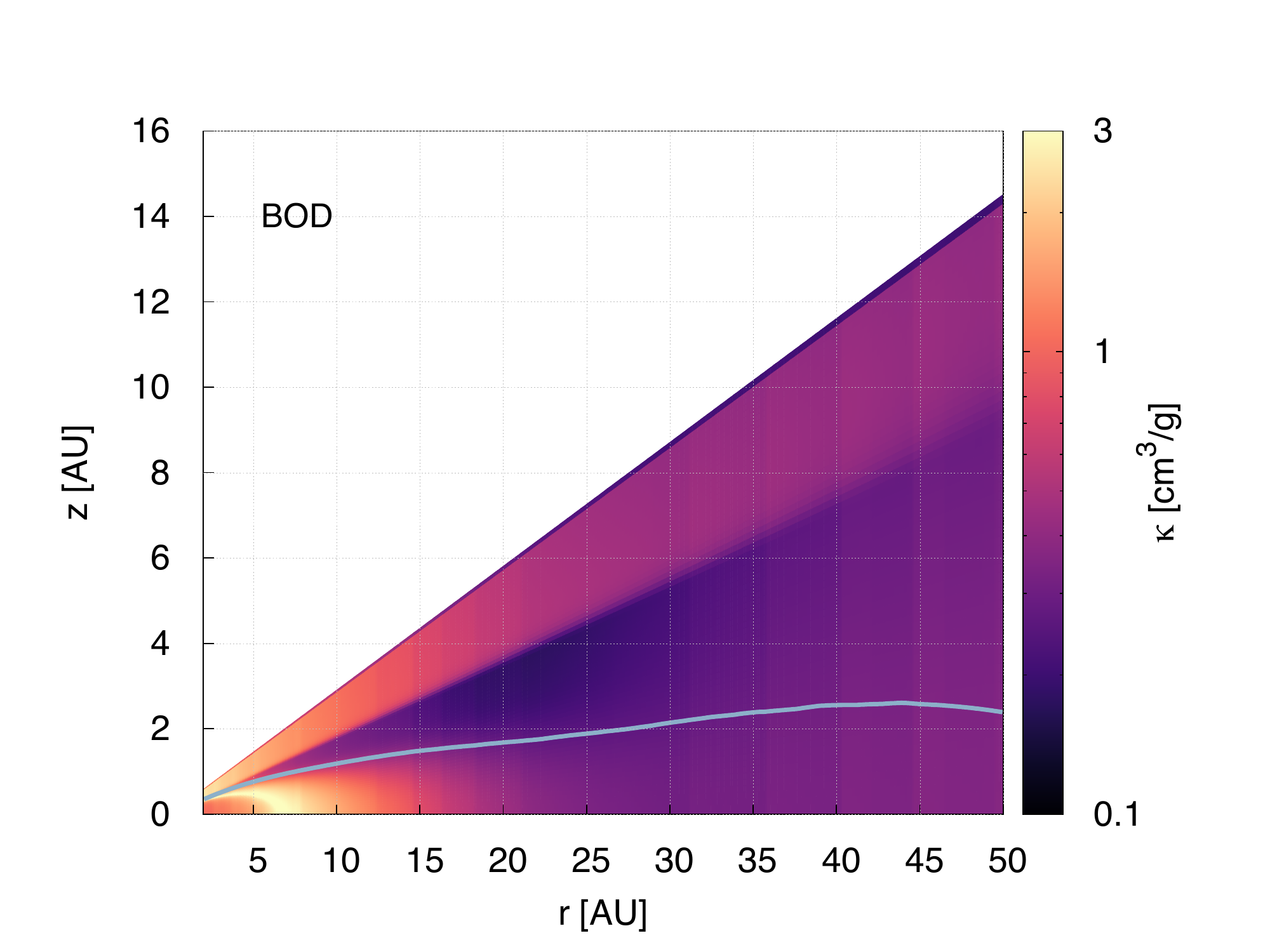}
\end{subfigure}
\begin{subfigure}{0.5\textwidth}
\includegraphics[width=\columnwidth]{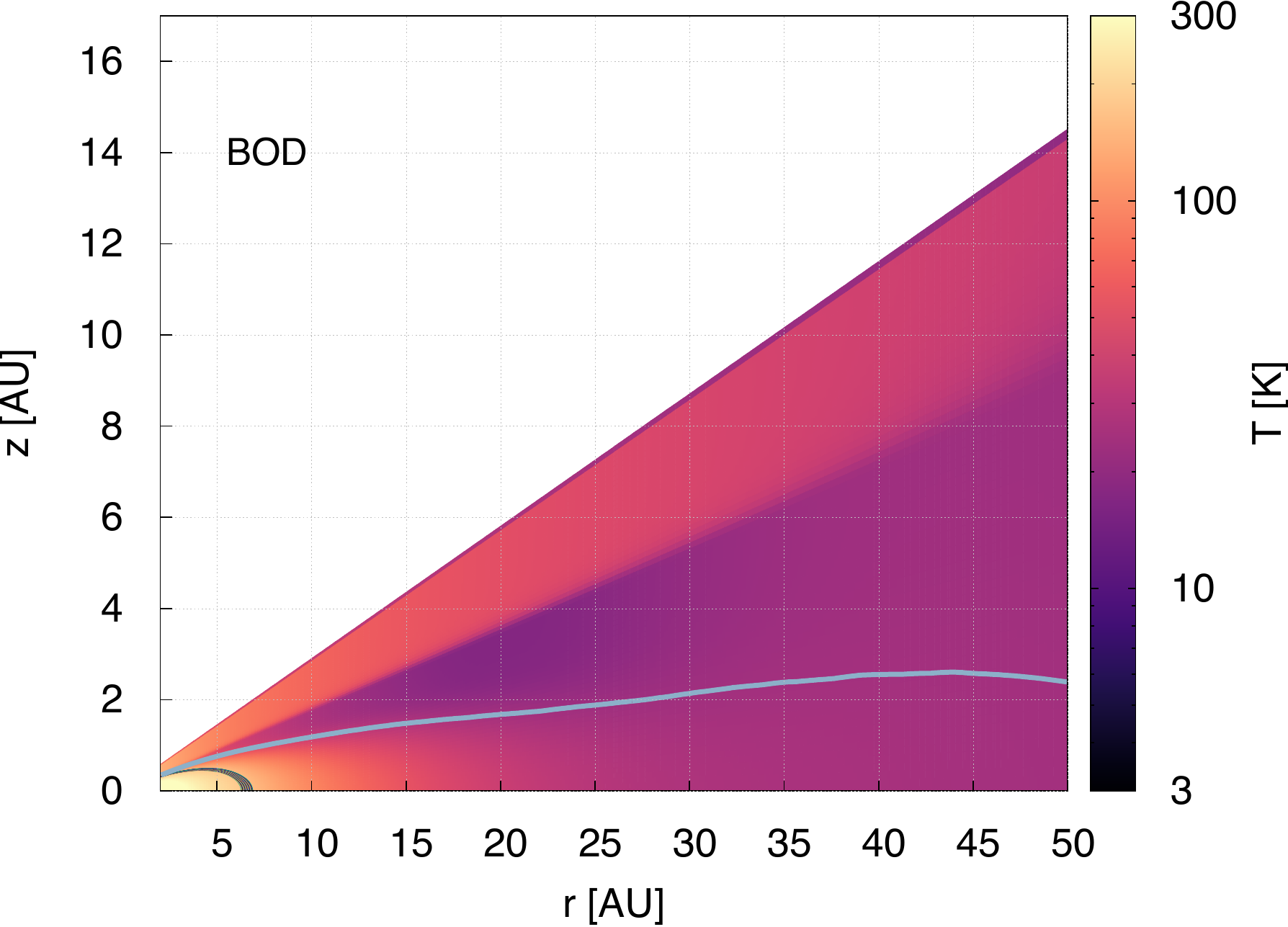}
\end{subfigure}
\begin{subfigure}{0.5\textwidth}
\includegraphics[width=\columnwidth]{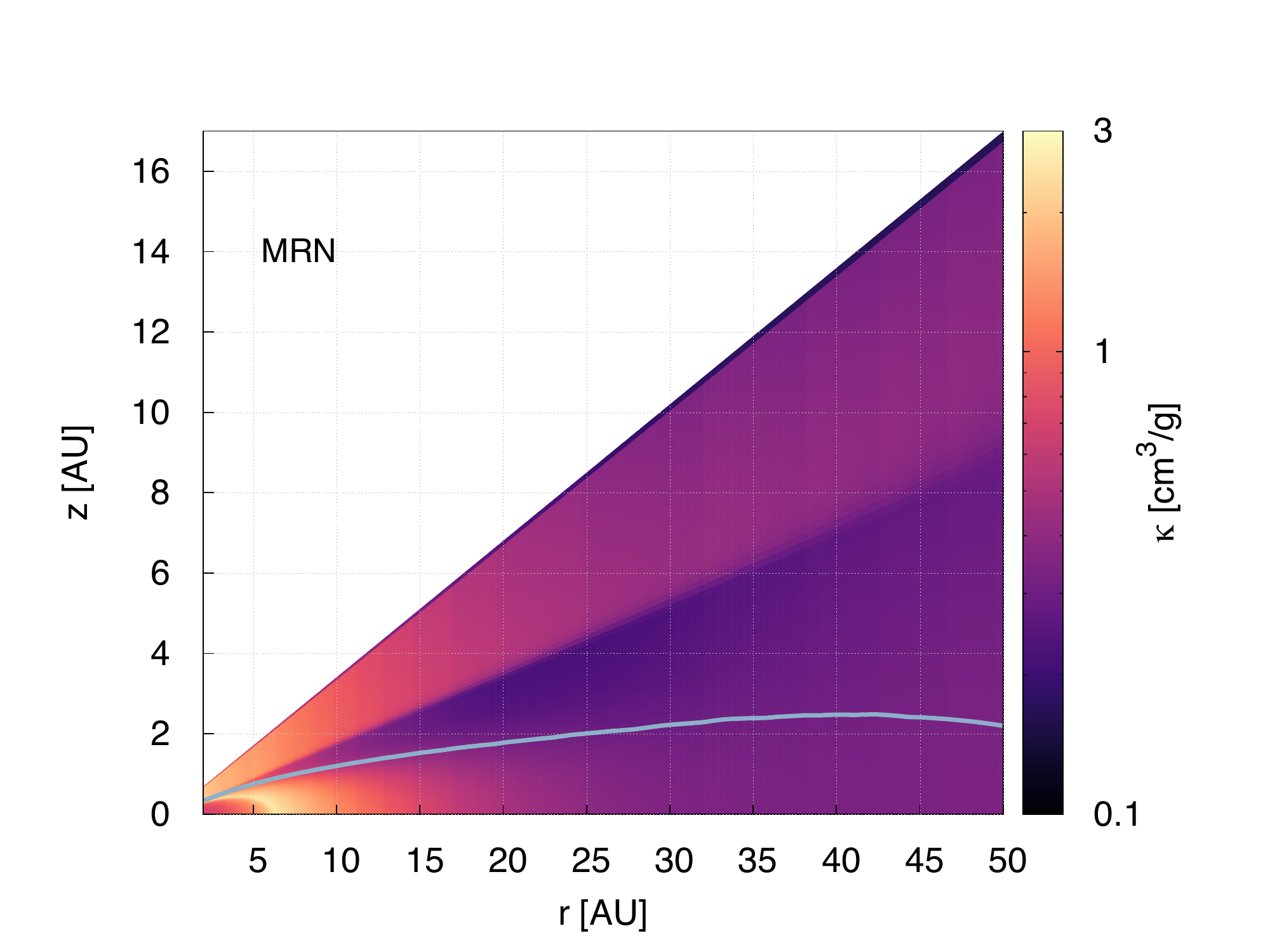}
\end{subfigure}
\begin{subfigure}{0.5\textwidth}
\includegraphics[width=\columnwidth]{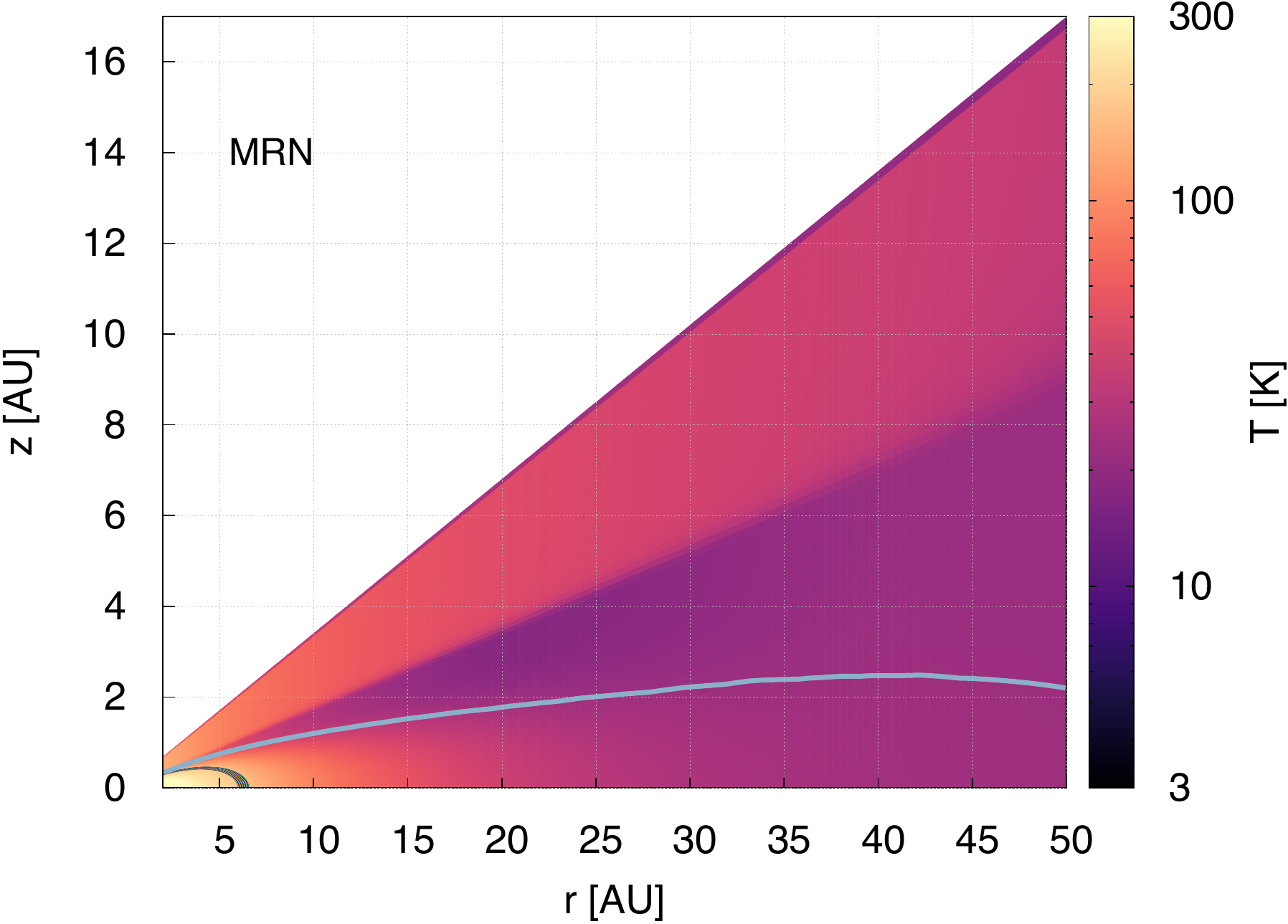}
\end{subfigure}

\caption{Mean Rosseland opacity values for the disc that includes the BOD distribution (top) and the MRN (bottom). These opacities correspond to the discs with the grain size distributions which were presented in Figs. \ref{Fig:ALL_5e3S3O} and \ref{Fig:GSDsmicroBL_5e3O}. The highest opacity values are found at the iceline transition (gray band in the right plots). Moving outwards, the disc gets colder and the opacity of the disc decreases. The light blue line is the location where the vertically integrated optical depth reaches unity ($\tau$=1), so it divides the disc in the optically thick lower region and the optically thin upper region. Above this line the opacities decrease due to the efficient cooling of the disc. The uppermost layers show increased opacities caused by the direct stellar heating, which increases the temperature. The same gradients can be seen in the temperature plots (right).}
\label{Fig:Opacity_nominal}
\end{figure*}

The outer region of the discs is dominated by stellar irradiation, which causes the flaring of the discs. The simulation with the BOD shows influence from the smaller particles at the inner parts of the disc, but moving outwards the aspect ratio gets affected by larger particles, around 100 $\mu$m. A more detailed analysis of the grain sizes that contribute the most to the opacity of the disc will be done in Sect. \ref{subsec:dominant gs}. The simulation with the MRN distribution shows similar aspect ratio gradients. For this case, with high $\alpha$, there is only a minimal difference between the dust surface densities, which leads to similar opacities in total. Both discs are affected by grains of similar sizes and for this reason the aspect ratios are almost the same there. 

In Fig. \ref{Fig:ALL_T_5e3S3O} we show the midplane temperature as a function of orbital distance for the same set of simulations. We can see that the changes in the temperature gradients correspond  to the changes in the aspect ratio gradients. The gray horizontal line is again the iceline transition. The simulation with the millimeter-sized particles does not reach the iceline temperature within the extend of the simulated disc and thus does not feature the aspect ratio bump. 

We can compare in Fig. \ref{Fig:GSDsmicroBL_5e3O} the resulting aspect ratios and midplane temperatures of the simulations with the BOD and MRN distributions with a simulation utilizing the opacities from \cite{1994ApJ...427..987B}. This opacity profile is based on micrometer sized particles, hence it is expected that it resembles the simulation with only micrometer sized particles. It should be pointed out that the \cite{1994ApJ...427..987B} prescription includes the gas opacities as well, but these are relevant for high temperatures that are not reached within the extend of the discs here.
We notice in Fig. \ref{Fig:GSDsmicroBL_5e3O} that the gradient after the iceline of the simulation with the \cite{1994ApJ...427..987B} prescription or only micrometer sized particles is much steeper than the corresponding gradient in the simulations of the full distributions. This is an important difference as the gradients in the aspect ratio affect the migration speed of planets that could be embedded in such a protoplanetary disc (see Sect. \ref{subsec:Planet migration}). 

In conclusion, we find that including either the BOD or the MRN distribution leads to comparable results. The differences between the two grain size distributions tested in this work for different values of the turbulence parameter $\alpha$ and different surface densities will be discussed in Sect. \ref{sec:alpha-visc_Sigma0_fDG}. Prior to that, a more extended discussion on the dust surface densities, dominant grain sizes, opacities and temperatures follows in the next paragraphs.

\subsection{Opacities and temperature}
\label{subsec:opacities}

The opacity and temperature within the disc for the BOD and MRN distribution is plotted in Fig. \ref{Fig:Opacity_nominal} as a function of orbital distance on the x-axis and height on the y-axis. The total opacity of the disc is determined by accounting for the contribution of each grain size according to Eq. \ref{eq:kappa}.

The highest opacity values in the figures correspond to the iceline as it can be also seen in the temperature plot (gray band). Almost every particle size has its highest opacity at the iceline, consequently the total opacity of the disc is the highest at the iceline, as it  can be seen already in Fig. \ref{Fig:Opacities}.  It was already briefly mentioned in Sect. \ref{subsec:GSDs} (detailed discussion in Sect. \ref{subsec:dominant gs}) that the dominant grain sizes at the inner disc are small, therefore we see the same pattern in the opacity of the disc around the iceline as the opacities of the small particles, with the bump around the transition. For this reason we are tracing the iceline at the opacity plot.  

The total opacity is scaled down in the simulation with the MRN distribution compared to the one with the BOD, which is what we also find in the aspect ratio. Since the opacities have the same pattern and since the bumps in the aspect ratio are caused by opacity transitions it is expected to find there the same gradient. 

Viscous dissipation is the dominant source of heating for the inner parts of the disc, while stellar irradiation becomes important at larger orbital distances and more importantly for the upper layers of the disc \citep{2001ApJ...560..957D,Bitsch:2013cd}. Since the upper layers are heated up, the opacities are also higher there. If we move vertically up, the iceline moves inwards as viscous heating becomes weaker. The specific radius and height at which stellar heating begins to affect the structure of the disc is, among other parameters, influenced by the strength of turbulence. The dependence of the disc's thermal structure on the turbulence strength is discussed in Sect. \ref{sec:alpha-visc_Sigma0_fDG} and more opacity plots as a function of orbital distance and height can be found in Appendix \ref{sec:AppB}.

In Fig. \ref{Fig:Opacity_nominal} the $\tau$ = 1 line is overplotted (light blue line). The optical depth $\tau$ is defined as
\be \tau = \int_{z_{max}}^0 \kappa_R \rho_g dz~, \ee 
therefore it increases as the height z is decreasing. The $\tau=1$ line marks the difference between the optically thick and the optically thin medium. When $\tau \geq 1$, then the disc is optically thick or in other words, the mean free path of the photons is much smaller than the length scale over which temperature changes. In optically thin parts of the disc, photons can "freely" travel out of the disc. The $\tau=1$ line thus marks the region of the disc where cooling becomes efficient. A $\tau=1$ line close to midplane corresponds to lower opacities, which results in a cooler disc. Even though the regions above this line are optically thin and cool down very efficiently, the uppermost layers are directly heated by stellar irradiation and we also see an increase in the opacity.

The transition from the optically thin part of the disc to the optically thick (moving from the top layers towards the midplane) is also where the boundary for observations would be if these were integrated over all wavelengths. Mid-infrared wavelength observations of the optically thick disc probe the temperature of the dust "photosphere", the effective surface layer of the disc. Observations are in general carried out at various wavelengths, thus probing different grain sizes and different information for the disc \citep{Andrews_2015}.  The optical depth relevant for such observations might differ for individual grain sizes.

\subsection{Dust surface densities}
\label{subsec:dust surface densities}

The vertically integrated dust surface densities per orbital distance and grain size are presented in Appendix \ref{sec:AppA}.  The maximum grain sizes in both of the distributions depend on the local temperature and gas surface density, which change with time. Additionally, all of the boundary sizes of the BOD distribution depend on the local gas surface density. 

We stress here the loop that is created; the dust surface densities play a major role in determining the opacity of the disc (Eq. \ref{eq:kappa}), which then influences the cooling rate and the stellar heating and thus changes the temperature. The shift in the temperature leads to a new fragmentation barrier (and regime boundaries for the BOD), hence the dust surface densities change and so forth. Given the fact that this loop exists between the dust and the gas, it is important to consider the self-consistent calculations of the dust surface densities in the simulations.

\subsection{Dominant grain sizes}
\label{subsec:dominant gs}

\begin{figure*}
\begin{subfigure}{\textwidth}
\centering
\includegraphics[width=.8\textwidth]{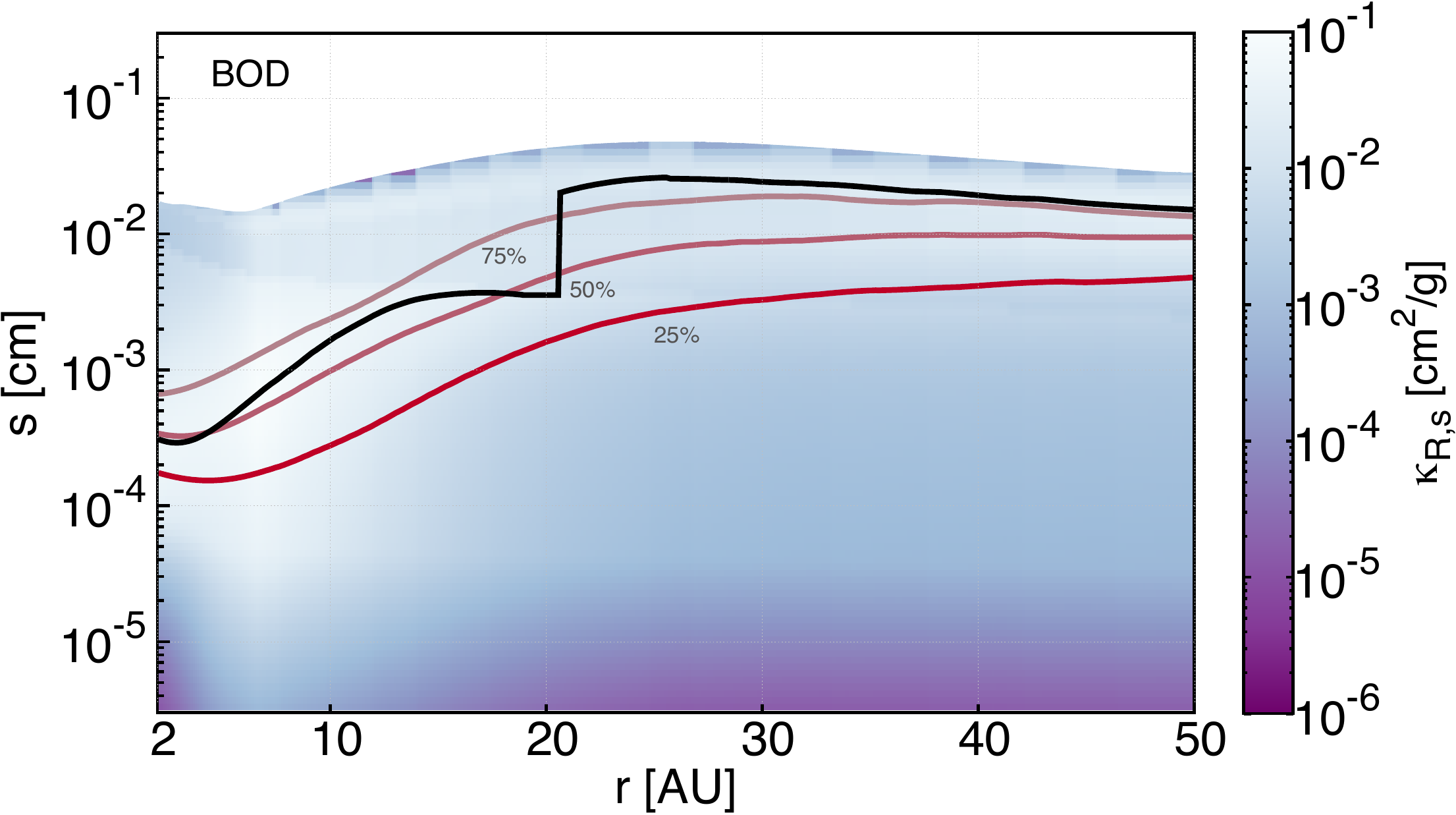}
\end{subfigure}

\begin{subfigure}{\textwidth}
\centering
\includegraphics[width=.8\textwidth]{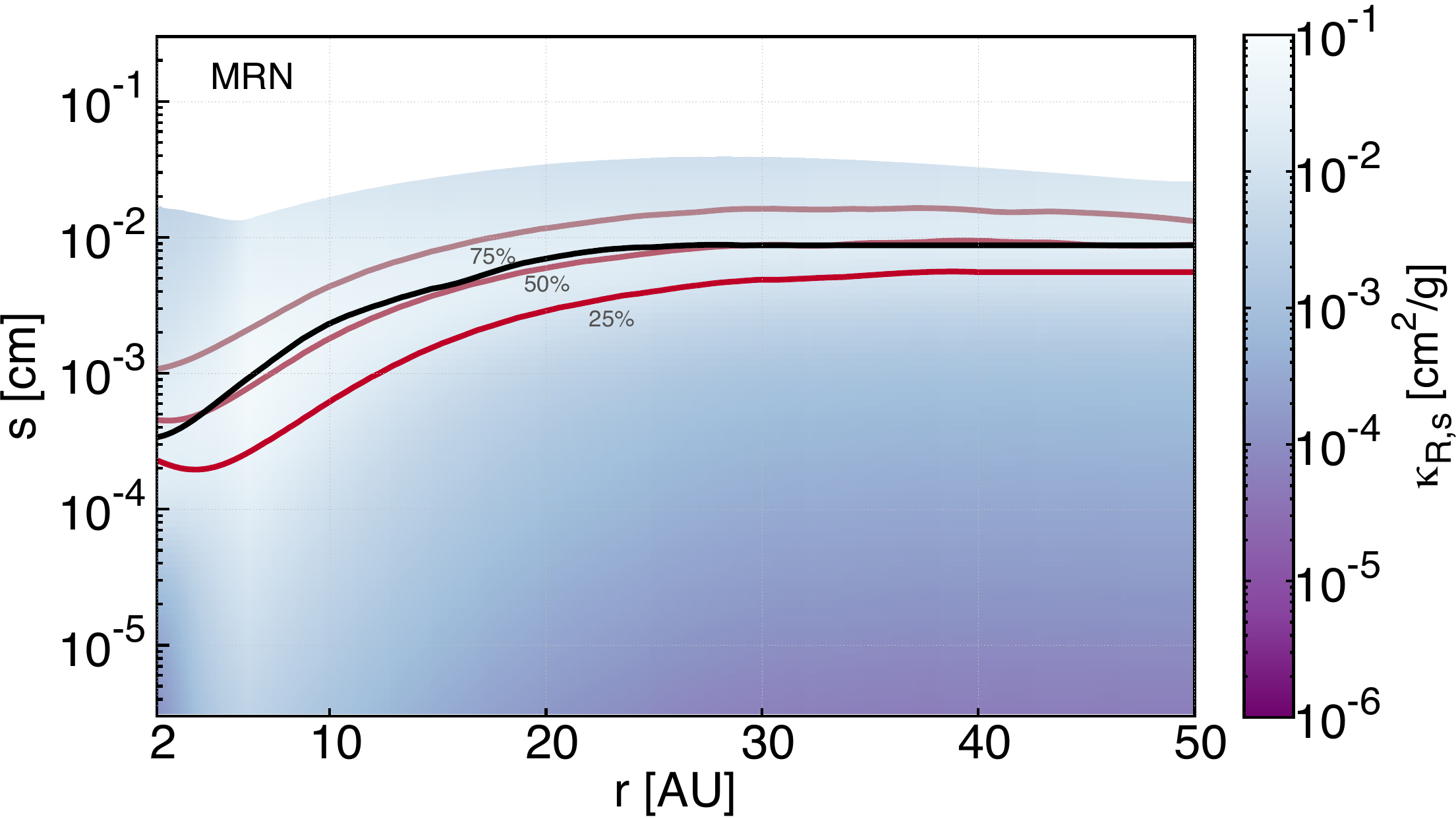}
\end{subfigure}

\caption{Contribution to the midplane mean Rosseland opacity per grain size for the simulation with the BOD on the top and the MRN distribution on the bottom, for the nominal disc parameters $\alpha= 5 \times10^{-3}$, $\Sigma_{g,0} = 1000~g/cm^2$ and $f_{DG} = 1\%$. The black line indicates the grain sizes that contribute the most to the opacity of the disc as a function of orbital distance. The BOD distribution causes a jump in these dominant grain sizes because of the dip in the dust surface density (see Fig. \ref{Fig:Distributions}) in the transition between the two turbulence regimes. The total opacity at each radius is the sum of the contribution from each grain size, or in other words it is the sum of the corresponding column.  For each grain size, the maximum opacities can be found at around 6-6.5 AU, where the iceline is located. The red lines show the percentage of the contribution from the grains below the corresponding line. This clearly illustrates that the dominant grain size does not necessarily determine the whole opacity. At the same time we can see that the smallest grain sizes (up to roughly 50 $\mu$m beyond 20 AU) contribute the least to the total opacity.}
\label{Fig:Dominant_gs}
\end{figure*}

Depending on the local gas disc parameters, the grain size which plays the role of the dominant opacity source will change. We find the individual contribution of each grain size to the total opacity of the disc through Eq. \ref{eq:kappa}. For each set of particles of the same size we calculate its contribution, $\frac{\rho_{d,i}}{\rho_g}100\kappa_i$ and we present it as a function of orbital distance in Fig. \ref{Fig:Dominant_gs} for the nominal case of $\alpha=5\times10^{-3},~ \Sigma_{g,0}=1000~g/cm^2$. In order for a grain size to dominate the opacity it needs a combination of high dust-to-gas volume density ratio (for this specific particle size) and high opacity at the given part of the disc. 

\begin{figure*}
\begin{subfigure}{0.49\textwidth}
\includegraphics[width=\textwidth]{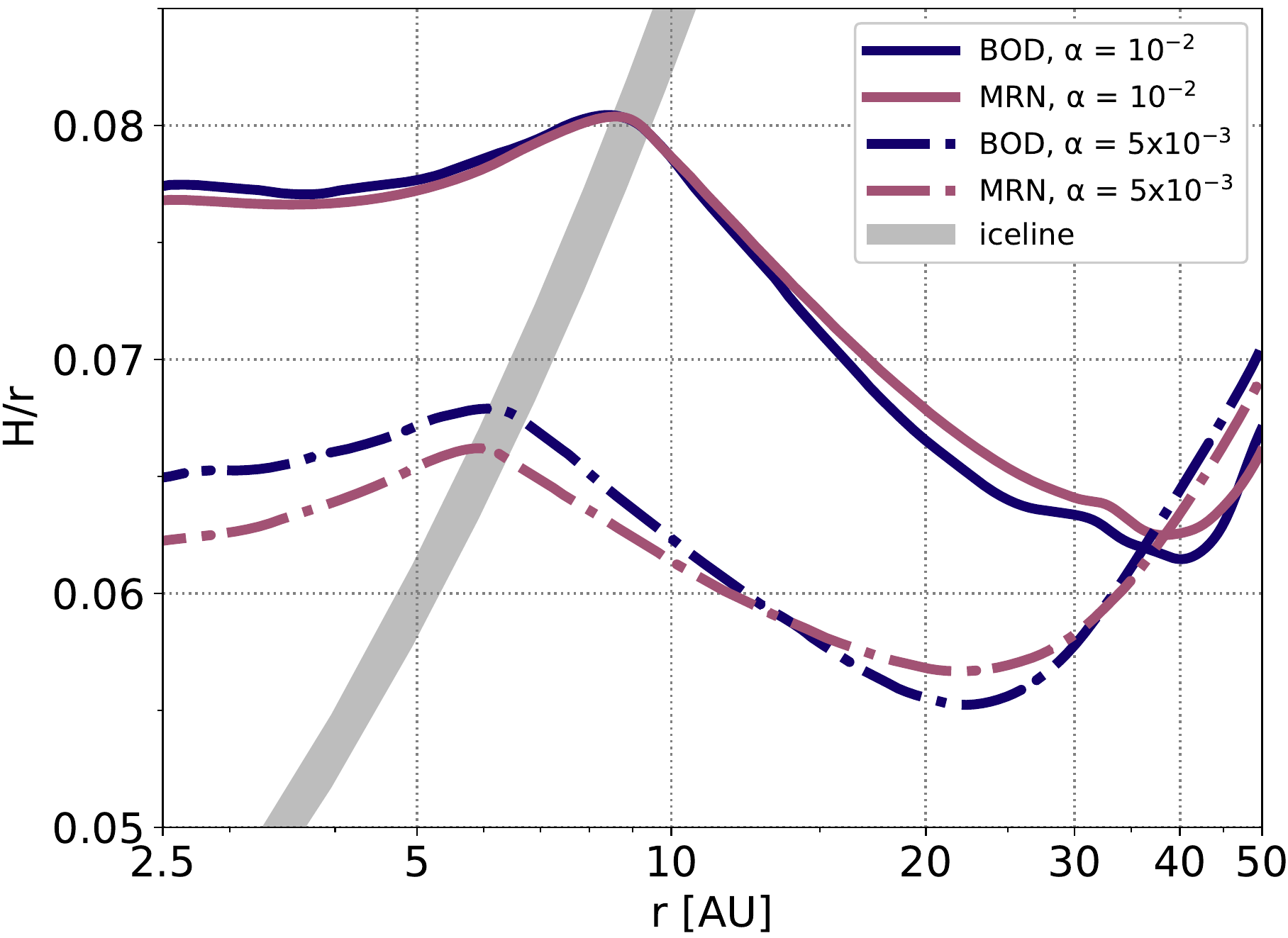}
\end{subfigure}
\begin{subfigure}{0.5\textwidth}
\includegraphics[width=\textwidth]{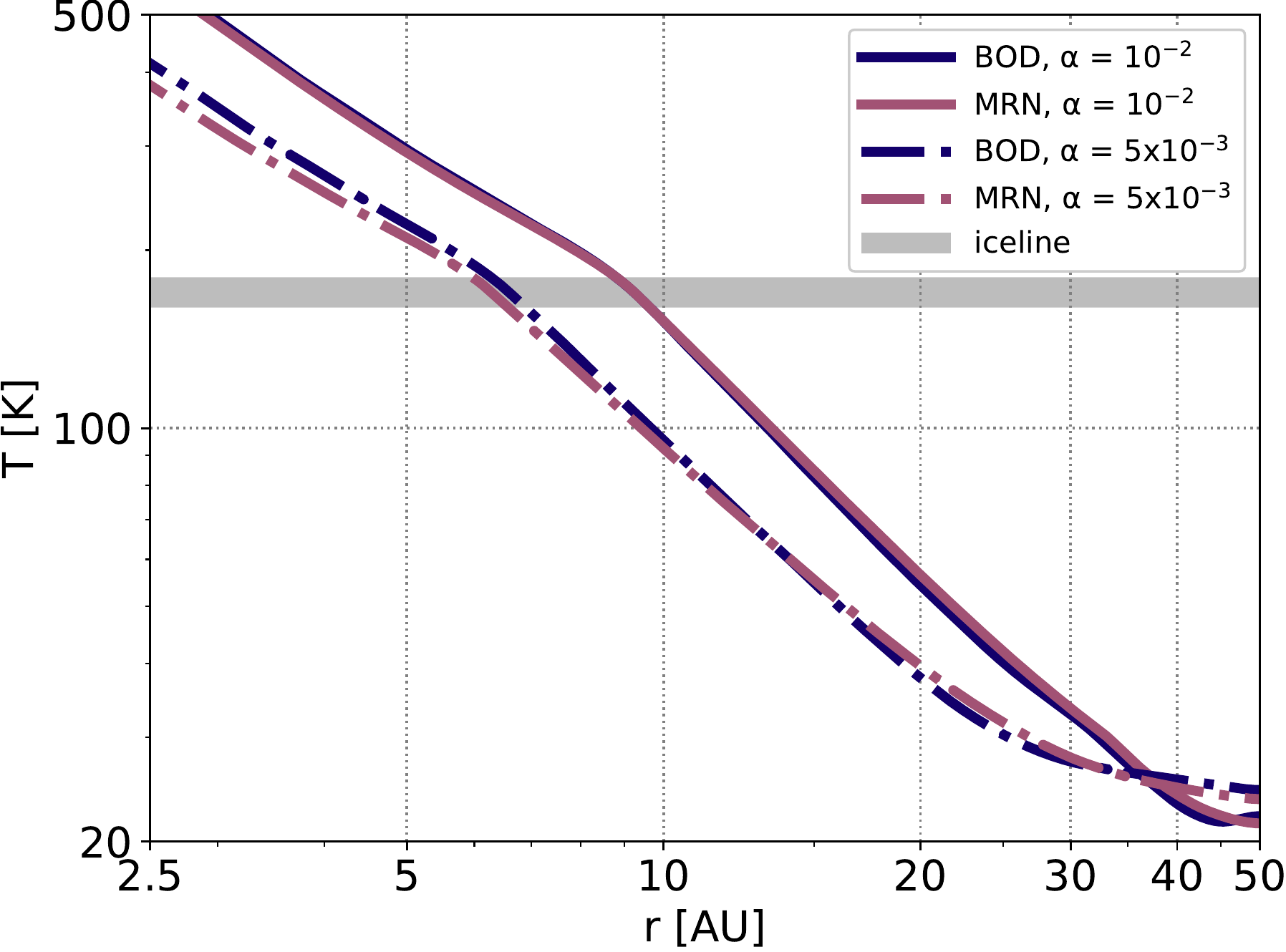}
\end{subfigure}
\begin{subfigure}{0.49\textwidth}
\includegraphics[width=\textwidth]{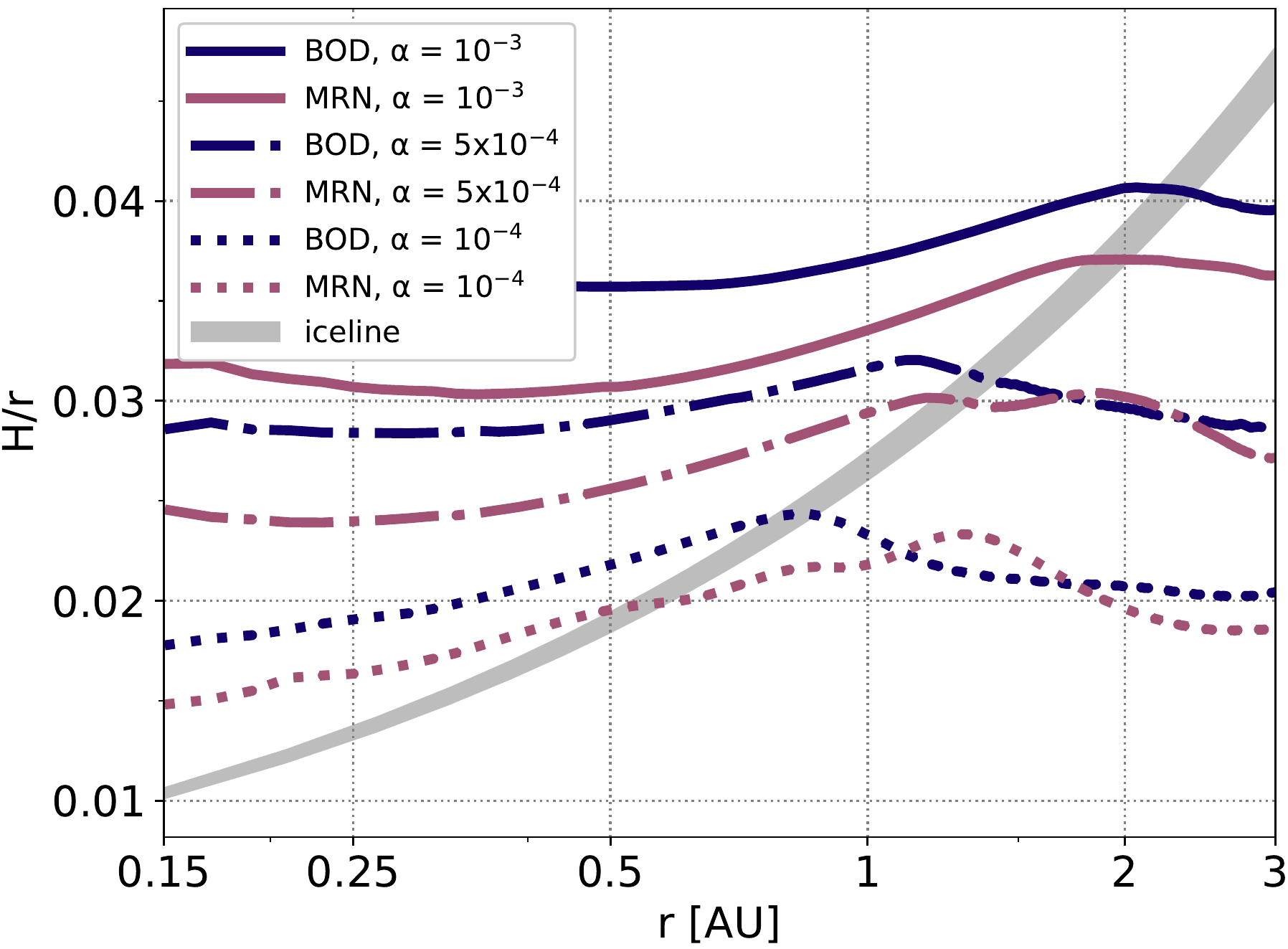}
\end{subfigure}
\begin{subfigure}{0.5\textwidth}
\includegraphics[width=\textwidth]{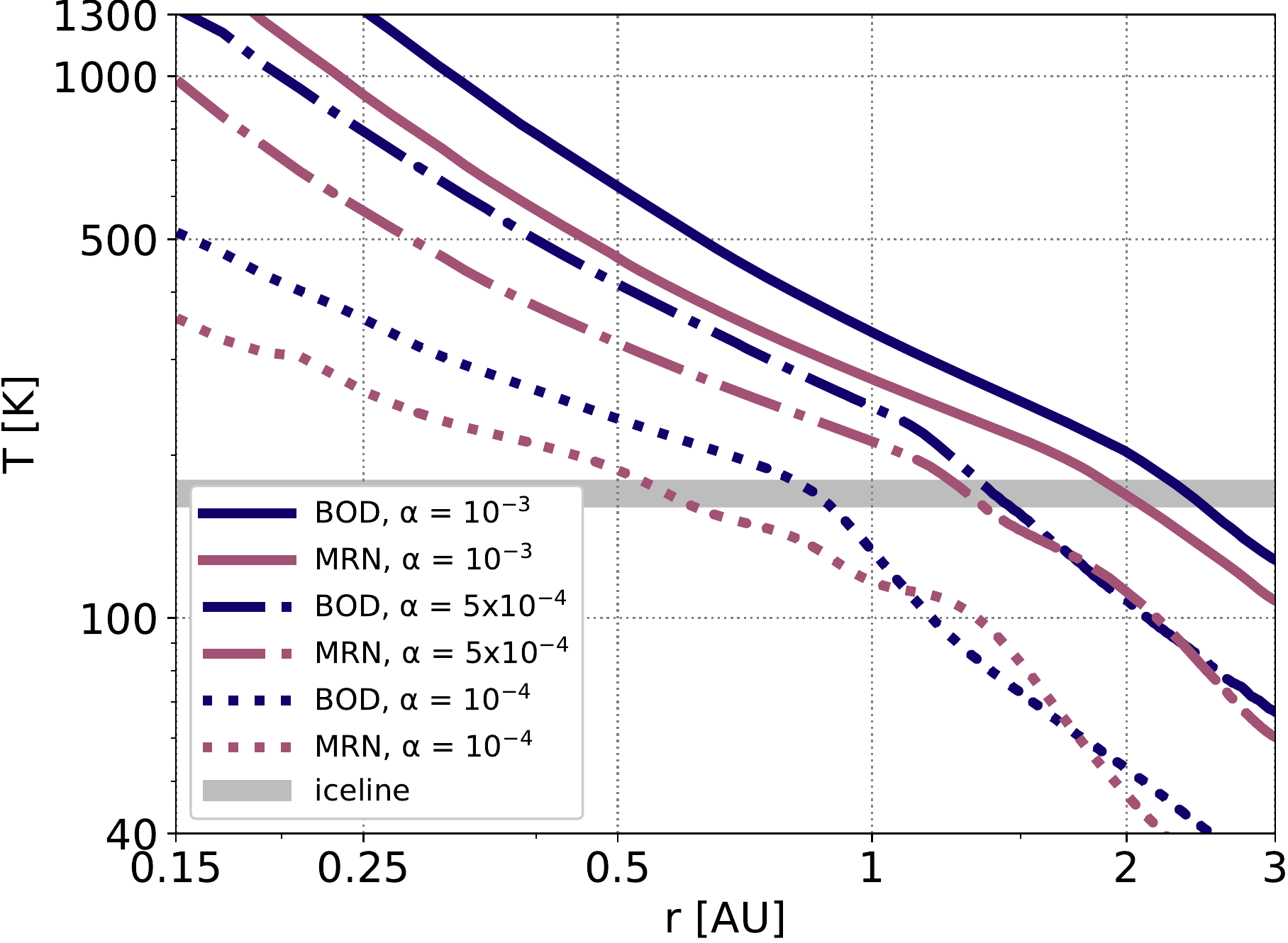}
\end{subfigure}

\caption{Aspect ratio as a function of orbital distance for the discs with high $\alpha$-viscosity values (top left), namely $\alpha = 10^{-2}$, $5\times10^{-3}$ and low $\alpha$-viscosity values (bottom left), namely $\alpha = 10^{-3},~5\times10^{-4}$ and $5\times10^{-4}$, for the two grain size distributions (BOD in dark blue, MRN in reddish pink). The iceline moves inwards as $\alpha$ decreases, due to reduced viscous heating. The wiggles that can be seen in parts of the low $\alpha$ discs are caused by convection. The right plots show the temperature as a function of orbital distance for the discs with high $\alpha$-viscosity values (top), namely $\alpha = 10^{-2}$, $5\times10^{-3}$ and for low $\alpha$-viscosity values (bottom), namely $\alpha = 10^{-3},~5\times10^{-4}$ and $10^{-4}$.}
\label{Fig:GSDs_highandlowalpha}
\end{figure*}

We can see in the same figure, plotted as a black solid line, that the inner disc with either the BOD or the MRN distribution is influenced by very small particles, around 3 $\mu$m. In other words, those small particles have the maximum contribution to the total opacity or in other words their opacity dominates the opacity of the disc. The dominant grain sizes in the disc with the BOD grain size distribution feature a jump at around 20 AU. This jump is caused by the dip in the distribution in the transition between the first turbulence regime and the second (see Fig. \ref{Fig:Distributions}), after which particles are large enough to get completely decoupled from the gas. After this jump the dominant grain size is near 200 $\mu$m. The dominant grain size in the disc with the MRN distribution smoothly increases in the inner regions and then remains constant at around 90 $\mu$m exterior to 20 AU.

At the inner, hot parts of the disc, the grains around the micrometer size have surface densities which are around an order of magnitude lower than the larger grains (see Appendix \ref{sec:AppA}). However, they have the highest opacity by several orders of magnitude (see Fig. \ref{Fig:Opacities}) at these high temperatures. This results in them being the dominant particles at that region. Farther out, the temperature of the disc decreases and as consequence, larger particles carry the highest opacity. The decreased temperature means that the opacities of the larger particles get comparable with those of the smaller particles. At the same time, the difference between the very small and the largest grain sizes slightly increases, aiding the dominance of the largest grains.

In Fig. \ref{Fig:Dominant_gs} the grain sizes below the dark red line give 25\% of the contribution to the total opacity, thus the grain sizes above this line give 75\% contribution. The line above this divides the grain sizes in two groups, each contributing 50\% to the opacity of the disc. In the same way the uppermost line has grain sizes with 75\% of the contribution beneath it and sizes with 25\% contribution above it. These lines show the general trend of the contribution to the opacity, which mainly comes from the small grains in the inner disc and by the large grains in the outer disc.

For a given grain size, the contribution to the opacity shows similar patterns as the dust surface densities. The maximum opacities per grain size are seen at approximately 6.5 AU. This location corresponds to the iceline and it is expected to have the highest opacity contribution by almost all of the grain sizes (Fig. \ref{Fig:Opacities}). In conclusion, what defines the grain size with the maximum contribution to the total opacity is the combination of the dust surface densities and the opacity of each grain size at a specific orbital distance (which is determined by the local temperature). Once more it is evident that the self-consistent calculations within the feedback loop (Fig. \ref{Fig:Loop}) are crucial to the disc structure evolution. 

\vspace{-3mm}
\section{Dependence on viscosity, gas surface density and dust-to-gas ratio} 
\label{sec:alpha-visc_Sigma0_fDG}

Decreasing the $\alpha$-viscosity values is expected to affect the discs in two ways. First of all, the viscous heating decreases, therefore the discs will cool down and their aspect ratio will be lower (see Fig. \ref{Fig:GSDs_highandlowalpha}). Secondly, the lower turbulence means that particles will face less destructive collisions and thus they will be able to grow to larger sizes. The larger grains have in general lower opacities, which means that the discs experience an additional cooling because of the change in the opacities. The general trend that we show in Fig. \ref{Fig:GSDs_highandlowalpha} is that the aspect ratio indeed decreases as the turbulence parameter decreases. The location of the iceline also moves further in. 

The aspect ratios and corresponding midplane temperatures in the low $\alpha$ models (bottom plots in Fig. \ref{Fig:GSDs_highandlowalpha}) show some wiggles due to convection. Convection is caused by the vertical temperature gradient which depends on the opacity gradient in the vertical direction and is present in the optically thick regions of the disc \citep{2013A&A...550A..52B}. 
This also implies that as the grains are more vertically diffused in the higher $\alpha$ case and the vertical temperature gradients are less steep,
the effect should be less strong.  Indeed, convection is also present in the high $\alpha$ simulations, but only at the inner, hotter regions of the disc. All regions that are affected by convection experience some sort of instability, so that reaching a steady state is very hard, if not impossible.

The vertically integrated dust surface densities as a function of orbital distance and grain size for the simulations with the two grain size distributions and the rest of the $\alpha$ values ($10^{-2}, 10^{-3}, 5 \times 10^{-4}, 10^{-4}$) are presented in Fig. \ref{Fig:Dust} in the appendix. In Sect. \ref{sec:Grain size distributions} we presented the vertically integrated dust surface densities for the nominal value of $\alpha = 5 \times 10^{-3}$. 

\begin{figure}
\begin{subfigure}{\textwidth}
\includegraphics[width=.5\textwidth]{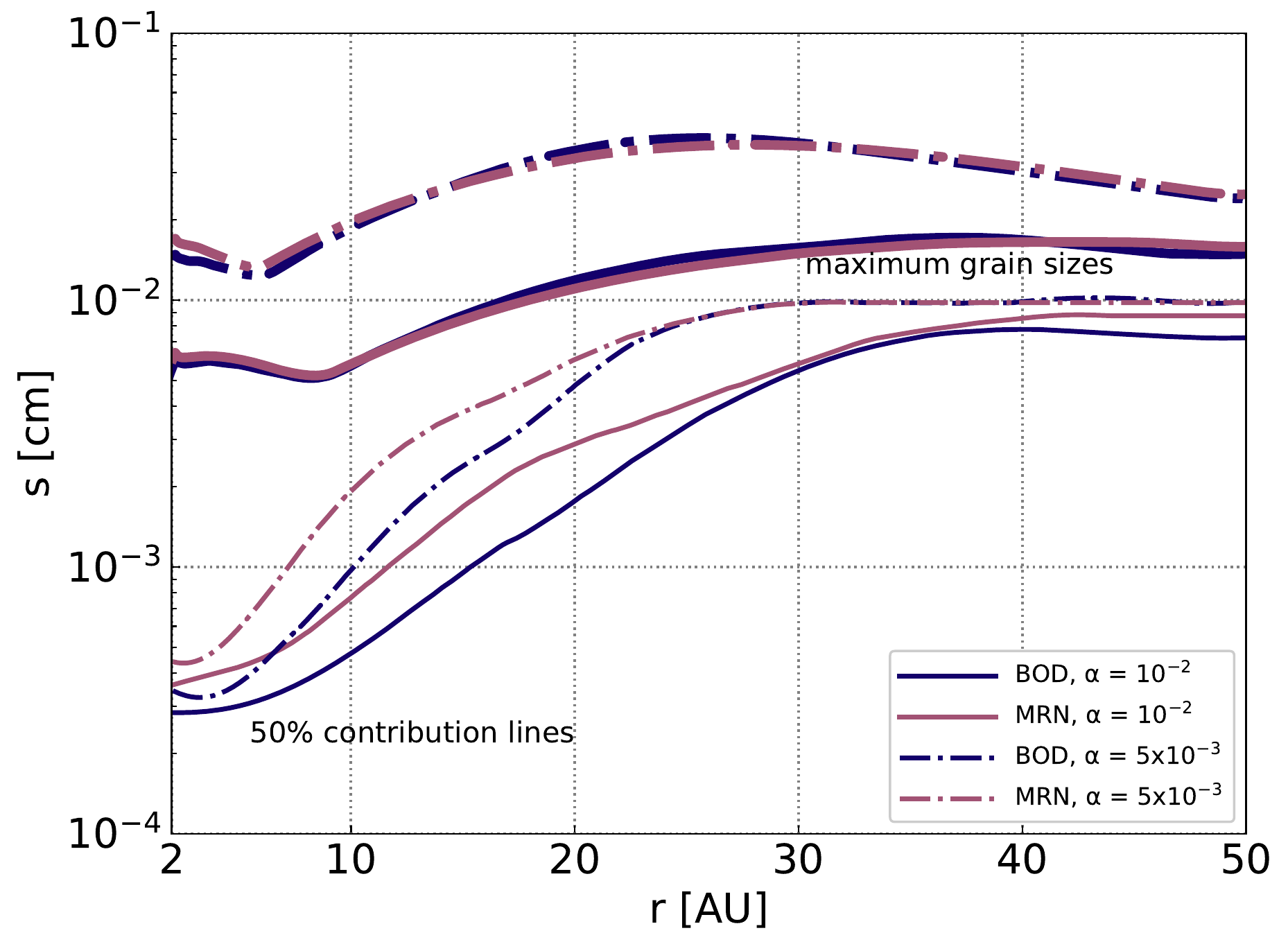}
\end{subfigure}

\begin{subfigure}{\textwidth}
\includegraphics[width=.5\textwidth]{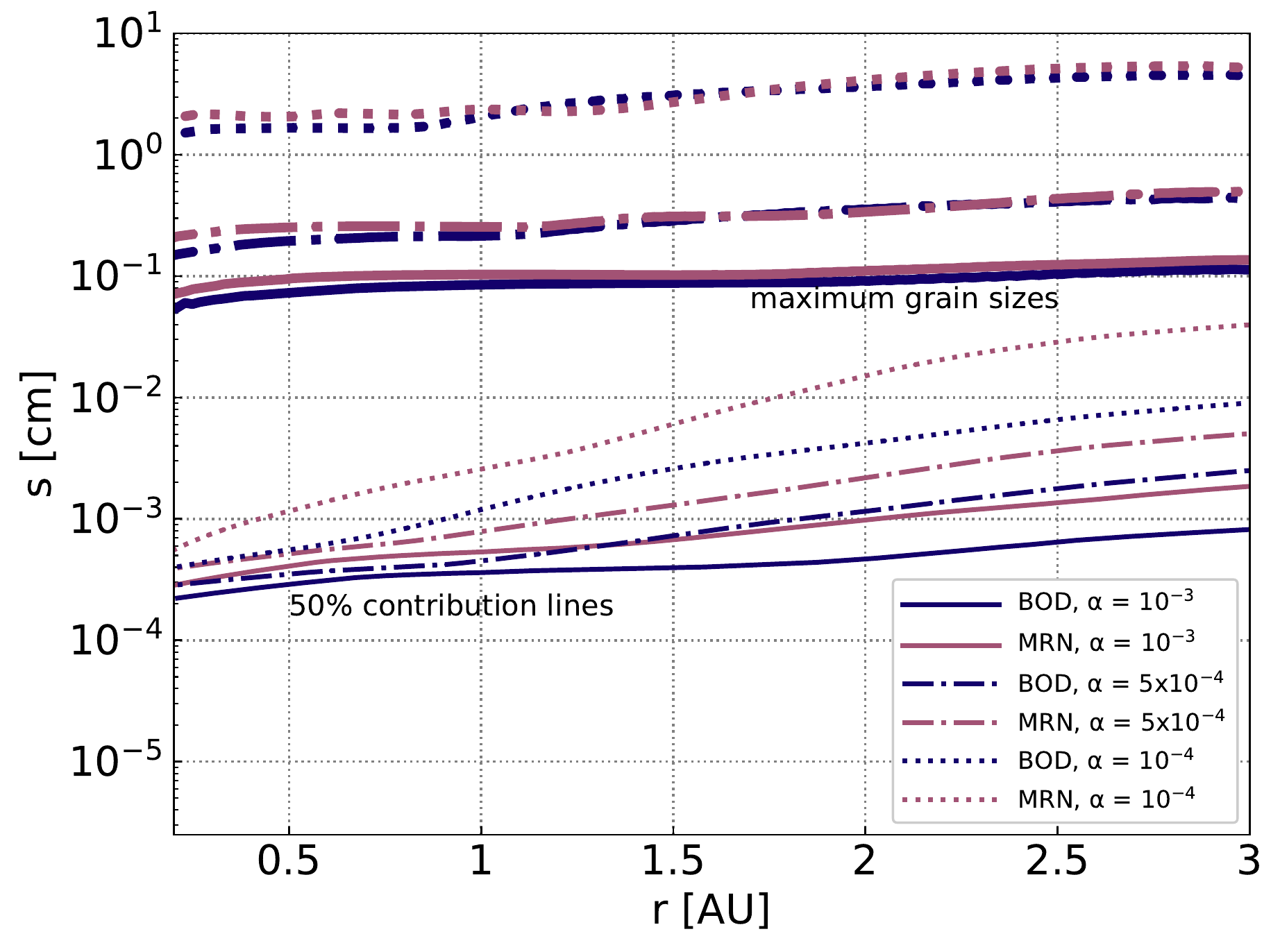}
\end{subfigure}

\caption{50\% contribution lines as a function of orbital distance for the discs with the high $\alpha$ values at the top and the low $\alpha$ values at the bottom. The group of thicker lines at the top correspond to the maximum grain sizes of each disc, so that the 50\% contribution lines below divide the grain sizes into two groups, which contribute equally to the total opacity of the disc (see the red lines in Fig. \ref{Fig:Dominant_gs}). 
The difference between the two grain size distributions is small. As $\alpha$-viscosity decreases, the maximum grain size increases and more influence to the total opacity comes from larger grains.}
\label{Fig:Dominant_gs_comparison}
\end{figure}

We also plotted the 50\% contribution line for the same $\alpha$ value in Fig. \ref{Fig:Dominant_gs}. This line divides the grain sizes into two groups that contribute equally to the total opacity at the given orbital distance. In Fig. \ref{Fig:Dominant_gs_comparison} we present the comparison of the 50\% contribution lines as a function of orbital distance and $\alpha$-viscosity. We plot also, in thicker lines of the same color, the maximum grain size in each disc so that the two groups of grain sizes with equal contribution to the opacity of the disc can be seen. The lower boundary of this plot corresponds to the minimum grain size (constant in all of the simulations). 

We find that the position of this 50\% contribution line is similar almost independently of the grain size distribution utilized in the model. Similarly the position of the 50\% contribution line within the grain size range (vertical axis in Figs. \ref{Fig:Dominant_gs} \& \ref{Fig:Dominant_gs_comparison}) does not change significantly as $\alpha$ decreases (see Fig. \ref{Fig:Dominant_gs_comparison}), but at the same time the maximum grain size increases. The very large particles ($\ge$ 100 $\mu$m) have significantly lower opacities and thus each order of magnitude added in grain size only adds a small contribution to the total opacity of the disc. 

\begin{figure*}
\centering
\includegraphics[width=.8\textwidth]{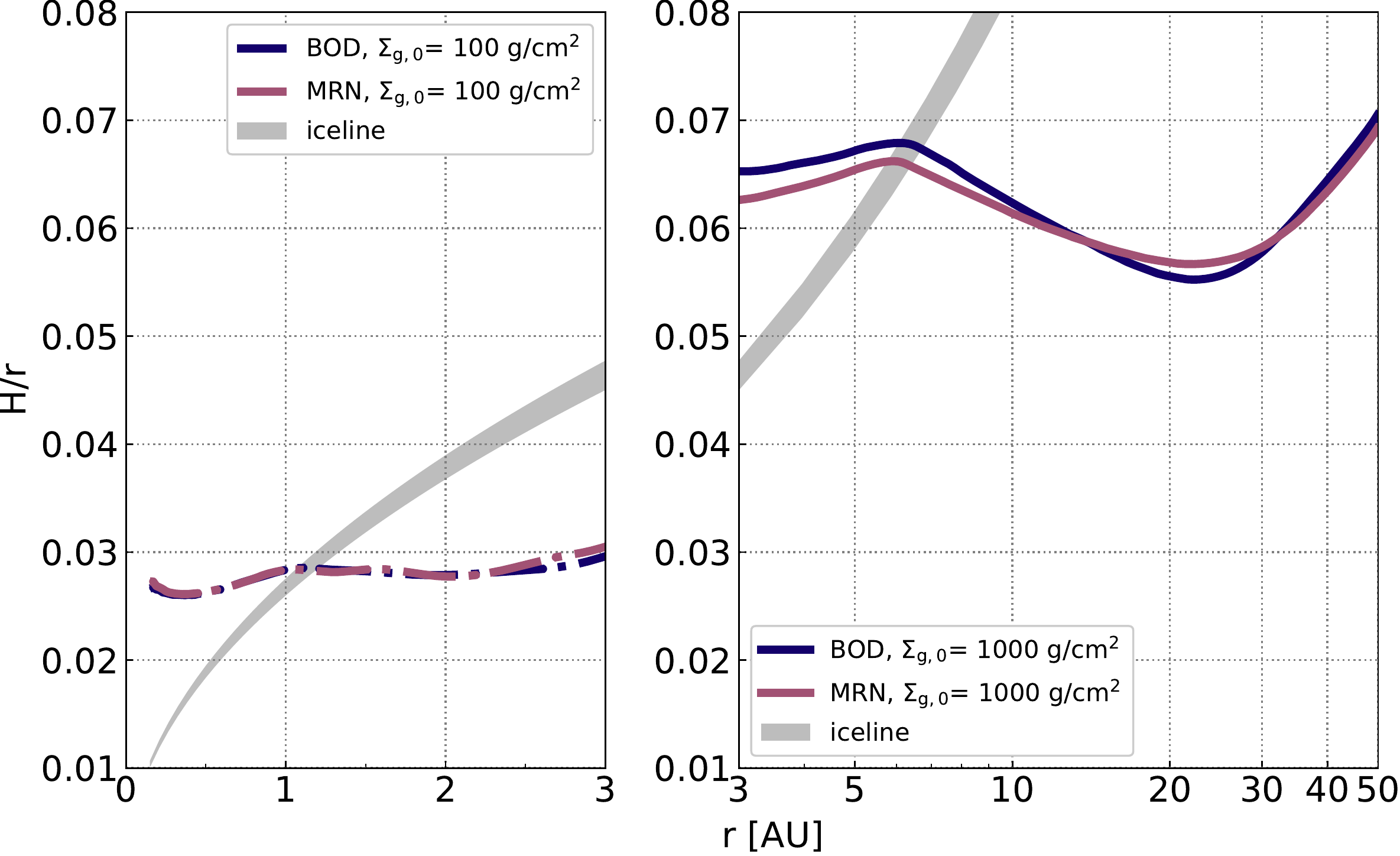}
\caption{Aspect ratio as a function of orbital distance for two different surface densities ($\Sigma_{g,0} = 100~g/cm^2$ and $\Sigma_{g,0} = 1000~g/cm^2$).  The aspect ratio decreases when the gas surface density decreases (while keeping the $\alpha$-viscosity constant at the nominal value of $5\times10^{-3}$) due to reduced viscous heating and increased cooling, which is inversely proportional to the density of the disc. As a result the position of the water ice line moves inwards, close to 1 AU. This inward movement of the water ice line due to a reduced gas surface density follows the trend of previous disc evolution simulations that show an inward movement of the water ice line as the disc evolves and the gas surface density reduces (e.g. \citet{2011ApJ...738..141O,2015A&A...575A..28B,2015A&A...577A..65B}). The low gas surface density results in almost the same aspect ratio profile for the discs with the two distributions.}
\label{Fig:GasSurfDensity}
\end{figure*}

If we decrease the gas surface density then the total dust surface density also scales down. The reduced surface density results in a colder disc because of two effects. On one hand the viscous heating decreases and on the other hand the radiative cooling increases, as it is inversely proportional to the disc's density.
This is what we find in the simulations with the lowest initial gas surface density we used, namely $\Sigma_{g,0} = 100 g/cm^2$ (Fig. \ref{Fig:GasSurfDensity}). The discs with lower surface density are much colder, therefore have significantly lower aspect ratio. In this case the difference in the aspect ratio of the discs with the two distributions almost vanishes completely, therefore the position of the iceline is also practically the same for the two discs. This is explained by the fact that the dust surface densities of the two distributions are comparable, leading to contributions to the opacity from similar grain sizes (see Fig.\ref{Fig:Dust}).  

On the contrary, if we increase the dust-to-gas ratio the total dust surface density is by definition enhanced. This means that the viscous heating is higher and the cooling rate decreased, resulting in hotter discs with higher aspect ratios. In addition to that, the opacity increases because of the enhanced total dust surface densities and as a consequence the optically thick region of the discs extends to higher heights compared to the discs with lower dust-to-gas ratio (see example of a disc with $f_{DG}=3\%$ in Appendix \ref{sec:AppA}).

The opacities as a function of orbital distance and height for the discs with the two distributions and the rest of the $\alpha$ values ($10^{-2}, 10^{-3}, 5 \times 10^{-4}, 10^{-4}$) can be found in Fig. \ref{Fig:Opacity} in the appendix. The turbulence strength affects the viscous heating and the orbital distance and height at which stellar heating takes over. This direct influence to the thermal structure of the disc leads to different opacities at each position in the disc, depending on the $\alpha$-viscosity value. A decrease in the gas surface density, as mentioned, decreases the viscous heating and increases the cooling rate, but the lower temperature decreases the opacity of the disc and cooling is enhanced. A comparison between the opacities in discs with all of the $\alpha$ values can be found in Fig. \ref{Fig:Opacity} in the appendix. We also show in Appendix \ref{sec:AppA} the opacity of the disc with the lowest initial gas surface density $\Sigma_{g,0}=100~g/cm^2$ (with $\alpha=10^{-2}$, $f_{DG}=1\%$) and the opacity of the disc with the highest dust-to-gas ratio, $f_{DG}=3\%$ (with  $\alpha=10^{-2}$ again and $\Sigma_{g,0}=1000~g/cm^2$).

\section{Implications}
\label{sec:Implications}
\subsection{Iceline}
\label{subsec:Iceline}

The location of the iceline can be theoretically calculated by considering the viscous and stellar radiation heating and the partial pressure of water vapor \citep[e.g.][]{2004M&PS...39.1859P,2005ApJ...620..994D,2006Icar..181..178C}. But it is also well predicted by setting a single sublimation temperature \citep[e.g][]{1981PThPS..70...35H,2011Icar..212..416M} as we do here as well. Nevertheless, the location where this sublimation temperature is reached depends on several parameters, as described by the feedback loop (Fig. \ref{Fig:Loop}). 

Planet formation studies indicate that the iceline in protoplanetary disc models should be outside of 1 AU, the Earth’s orbit. Otherwise, mechanisms like blocking the inward flow of pebbles from the outer disc by growing planets need to be invoked to keep the planets in the inner system dry \citep{2016Icar..267..368M}.
If we take into consideration the observations of the composition of the bodies in the asteroid belt, the iceline at the time of formation of the asteroids would be at $\sim$2.7 AU. But if icy planetesimals were formed at such small orbital distances and contributed to the formation of the terrestrial planets, we would observe larger amounts of water on Earth than what we observe today. The composition of asteroids from the inner region of the asteroid belt suggests that at their time of formation they should have been interior to the iceline. 

Evolving disc models indicate that at the time of the formation of terrestrial planets the iceline has already moved towards 1 AU \citep{2015A&A...575A..28B}.
In \citet{2011ApJ...738..141O} it is suggested that in order to reach a better conclusion about the location of the iceline a grain size distribution is required, rather than uniform dust grain sizes. In \cite{2007ApJ...654..606G} the decoupling of dust particles from gas is discussed as a potential influence to the thermal structure of the disc. \citet{2006ApJ...640.1115L} argue that in order to move the iceline outside 3 AU one possible solution would be to increase the opacity used in their model. Here we do not include the opacity of a single grain size, but use an evolving grain size distribution that regulates the opacity of the disc more realistically, because we take into account their individual contributions. 

All of these suggested effects have been therefore taken into account in the here presented work where we study the influence of $\alpha$-viscosity, initial gas surface density and total dust-to-gas ratio on the position of the iceline. In Appendix \ref{sec:AppB} we present the simulations which were used and the procedure that was followed to do the fitting of the iceline position as a function of those three disc parameters. We find that the location of the iceline is independent of the grain size distribution which was utilized in the disc and it follows 

\be 
\label{eq:Iceline_fitting} 
r_{ice}=9.2\cdot\left(\frac{\alpha}{0.01}\right)^{0.61}\cdot\left(\frac{\Sigma_{g,0}}{1000~g/cm^2}\right)^{0.8} 
\cdot\left(\frac{f_{DG}}{0.01}\right)^{0.37}~\text{AU.}
\ee

\begin{figure}
\centering
\includegraphics[width=\columnwidth]{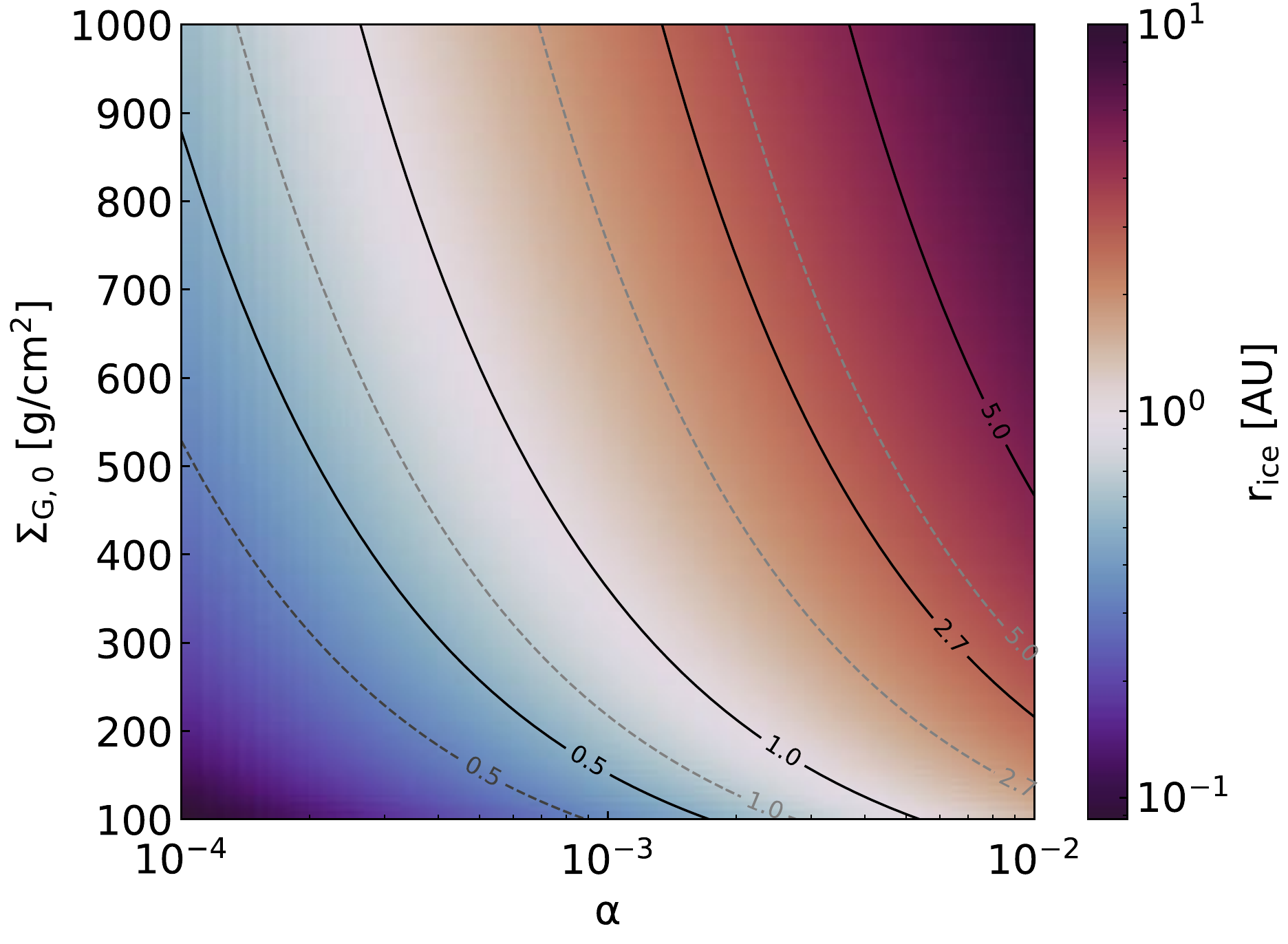}
\caption{Iceline position as a function of $\alpha$ turbulence and initial gas surface density at 1AU with a constant $f_{DG}=1\%$ (Eq. \ref{eq:Iceline_fitting}). The iceline transition is defined as $T=170\pm10~K$. The black lines mark $r_{ice} = 0.5, 1, 2.7\text{ and }5~AU$. Higher viscosity or gas surface density leads to hotter discs, with the iceline located at greater distances from the star. The same applies to higher total dust-to-gas ratio. The gray dashed lines mark $r_{ice} = 0.5, 1, 2.7\text{ and }5~AU$ for a disc with $f_{DG}=3\%$.}
\label{Fig:Iceline}
\end{figure}

In order to investigate the theoretical background of the above power law fit we can start, as in \citet{Bitsch:2013cd}, by considering the heating and cooling balance, $Q^+ = Q^-$, which means
\be 2\sigma T_{eff}^4 = \frac{9}{4}\Sigma_g\nu\Omega_K^2~. \ee
Given that the midplane temperature can be expressed as $T_{mid} = \left(\frac{3\tau_d}{4}\right)^{1/4} T_{eff}$,  the above equation becomes
\be \frac{8\sigma}{3\tau_d}T_{mid}^4 = \frac{9}{4}\Sigma_g\nu \Omega_K^2~.
\ee
We also substitute viscosity with the $\alpha$ prescription (Eq. \ref{eq:viscosity} and Eq. \ref{eq:soundspeed} for the sound speed) and express the vertical optical depth as $\tau_d = \frac{1}{2}\Sigma_g f_{DG} \kappa$, so we get
\be T_{mid}^3 = \left(\frac{27}{64} \frac{k_B}{\sigma\mu m_H}\right) \Sigma_g^2 ~f_{DG}~\kappa~ \alpha~ \Omega_K~. \ee 
The surface density profile follows $\Sigma_g = \Sigma_{g,0} \left(\frac{r}{AU}\right)^{-1/2}$ and $\Omega_K = \sqrt{\frac{GM_*}{r^3}}$, so we obtain
\be T_{mid}^3  \propto \Sigma_{g,0}^2 ~f_{DG}~\kappa~\alpha~r^{-5/2}~.\ee

We can then solve for the position of the iceline $r=r_{ice}$ where $T_{mid}=T_{ice}$
\be \label{eq:Iceline_theoretical} r_{ice} \propto  \Sigma_{g,0}^{4/5} ~ f_{DG}^{2/5} ~\kappa^{2/5}~\alpha^{2/5}~. \ee
We thus find that the power-law indices for the dependencies on $\Sigma_{g,0}$ and $f_{DG}$ are very similar to what we find in our fitting (Eq. \ref{eq:Iceline_fitting}). Comparing Eqs. \ref{eq:Iceline_fitting} and \ref{eq:Iceline_theoretical} suggests that at the iceline $\kappa\propto\alpha^{1/2}$, but is almost independent of $\Sigma_{g,0}$ and $ f_{DG}$. The reason for this dependency has no easy analytical explanation, but it appears to be the outcome of the feedback between the disc structure and the dust evolution. This further illustrates, as we also discuss in Sec. \ref{subsec:dominant gs}, that we cannot rely on single grain size opacities to accurately describe the disc structure. 

The position of the iceline as a function of the $\alpha$-turbulence parameter and the initial gas surface density $\Sigma_{g,0}$ from our fitting formula is presented in Fig. \ref{Fig:Iceline}, for discs with constant $f_{DG}=1\%$. The iceline transition is defined as the location where $T=(170\pm10)~K$. Increasing values of either one of the three parameters, $\alpha$, $\Sigma_{g,0}$, $f_{DG}$, leads to hotter discs so that the iceline moves closer to the star (see Sect. \ref{sec:alpha-visc_Sigma0_fDG}). In the models with any of the grain size distributions, the iceline is located outside 1 AU for $\alpha\geq2.6\times 10^{-4}$ and exterior to 2.7 AU only when $\alpha \geq 1.4 \times 10^{-3}$ for a disc with $\Sigma_{g,0} = 1000~g/cm^2$ and $f_{DG}=1\%$. However, this also depends on the disc's surface density and total dust-to-gas ratio. If the surface density reduces in time, the disc becomes cooler and the ice line moves inwards, even for the high viscosity cases.  For example, for $\Sigma_{g,0} = 100~g/cm^2$ the iceline is located outside 1 AU for $\alpha \geq 5.4 \times 10^{-3}$ and exterior to 2.7 AU only when $\alpha \geq 2.8 \times 10^{-2}$ (again with a constant $f_{DG}=1\%$). In contrast, if the total dust-to-gas ratio is increased to $3\%$, then we find the iceline outside 1 AU even for $\alpha \geq 1.4 \times 10^{-4}$ and outside 2.7 AU for $\alpha \geq 6.9 \times 10^{-4}$.

 We can conclude that utilizing Mie theory for the opacities of the grains and taking into account a distribution of grain sizes helps in keeping the iceline sufficiently far out from the star, especially for high $\alpha$ and $\Sigma_{g,0}$ values. In general, the location of the iceline might  also depend on the composition of the grains, which will be examined in future work. In contrast to  \citet{2011ApJ...738..141O}, who suggest that grain size distributions might help to keep the ice line at larger distances compared to single grain size discs, we actually find the opposite.  Including a distribution in a disc simulation results in a similar position of the iceline to the discs with the smallest grain sizes (0.1 and 1 $\mu$m). At low viscosities the opacity is dominated by larger grains and thus the disc becomes colder. Unrealistic single grain opacities (typically of micrometer size particles) result in discs that are too hot. This implies that potentially other heating source are needed to keep the iceline at large orbital distances, especially if viscous heating is low \citep{2019ApJ...872...98M}. 

\subsection{Planet migration}
\label{subsec:Planet migration}

The protoplanetary disc's structure also affects planet migration. Very roughly said, if the aspect ratio increases with orbital distance then planets migrate inwards (Type-I migration, \citet{2011MNRAS.410..293P}). On the contrary, if the aspect ratio is a decreasing function of orbital distance, planets will migrate outwards \citep{2013A&A...549A.124B, 2014A&A...564A.135B, 2015A&A...575A..28B}, if the viscosity is large enough \citep{2008ApJ...672.1054B}. 

We will focus here on the results with $\alpha = 5\times10^{-3}$, a viscosity large enough to trigger outward migration by the entropy driven corotation torque. For the discs with the grain size distributions, we see an aspect ratio which is a decreasing function of orbital distance beyond the iceline, therefore in those disc regions planets could migrate outwards. Interior to the iceline the aspect ratio is an increasing function of orbital distance which means that planets embedded in this region of the disc would only migrate inwards. The minima in the aspect ratio are locations where planets could get trapped and if (as it is more likely) more than one planet existed, they could get into resonance and remain at those fixed orbital distances until the local parameters of the discs changed sufficiently to force them to migrate again \citep{2012ApJ...750...34H,2013A&A...553L...2C,2014A&A...569A..56C,2017MNRAS.470.1750I,2019arXiv190208772I}.

\begin{figure}
\centering
\includegraphics[width=\columnwidth]{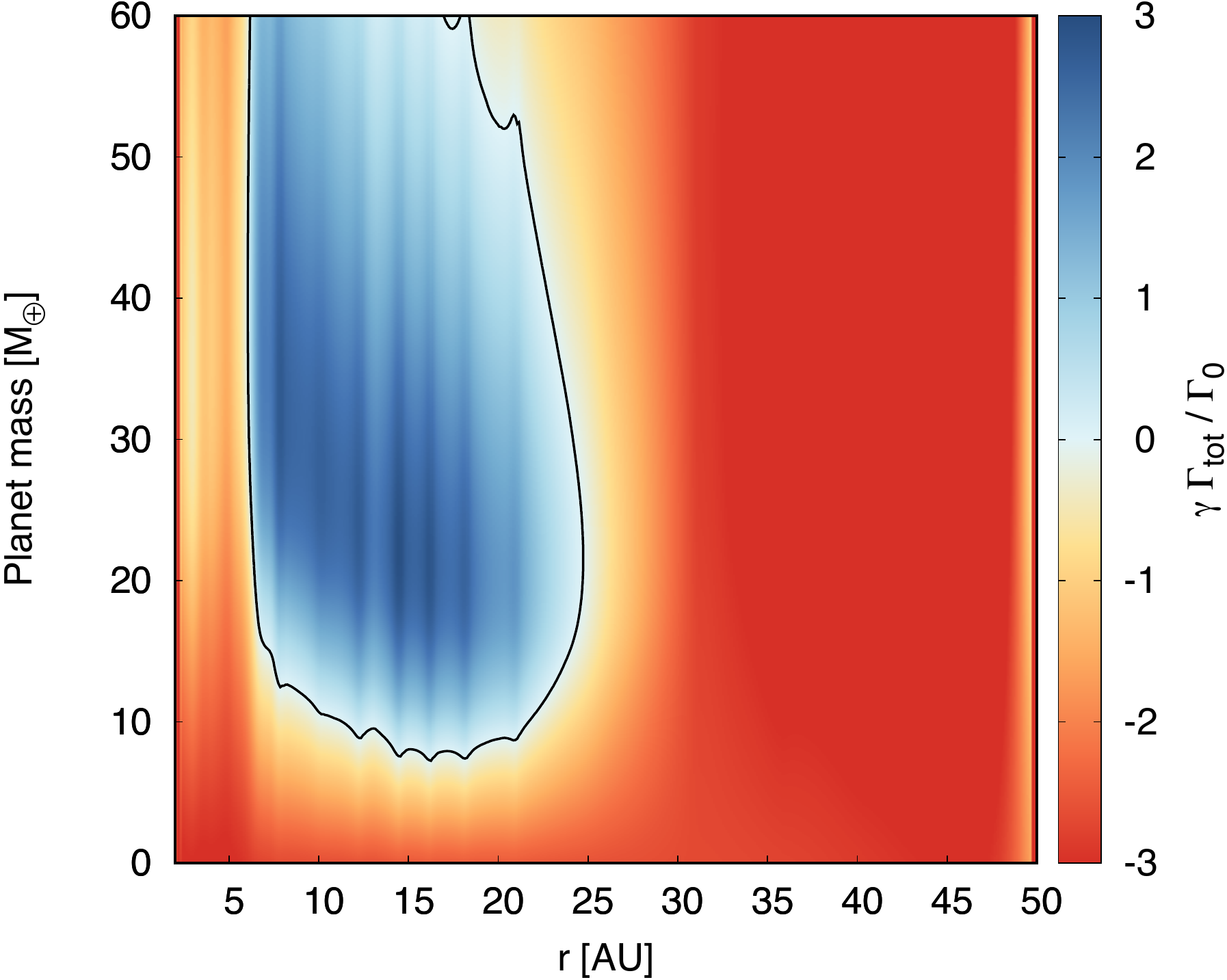}
\caption{Torque acting on planets with different masses for the disc utilizing the BOD distribution for the nominal viscosity of $\alpha = 5\times10^{-3}$. The black line encircles the regions of outward migration and corresponds to the region of the disc where the aspect ratio decreases as a function of the orbital distance. The temperature at the same region shows the steepest gradient.}
\label{Fig:Migration maps_BOD}
\end{figure}

\begin{figure}
\centering
\includegraphics[width=\columnwidth]{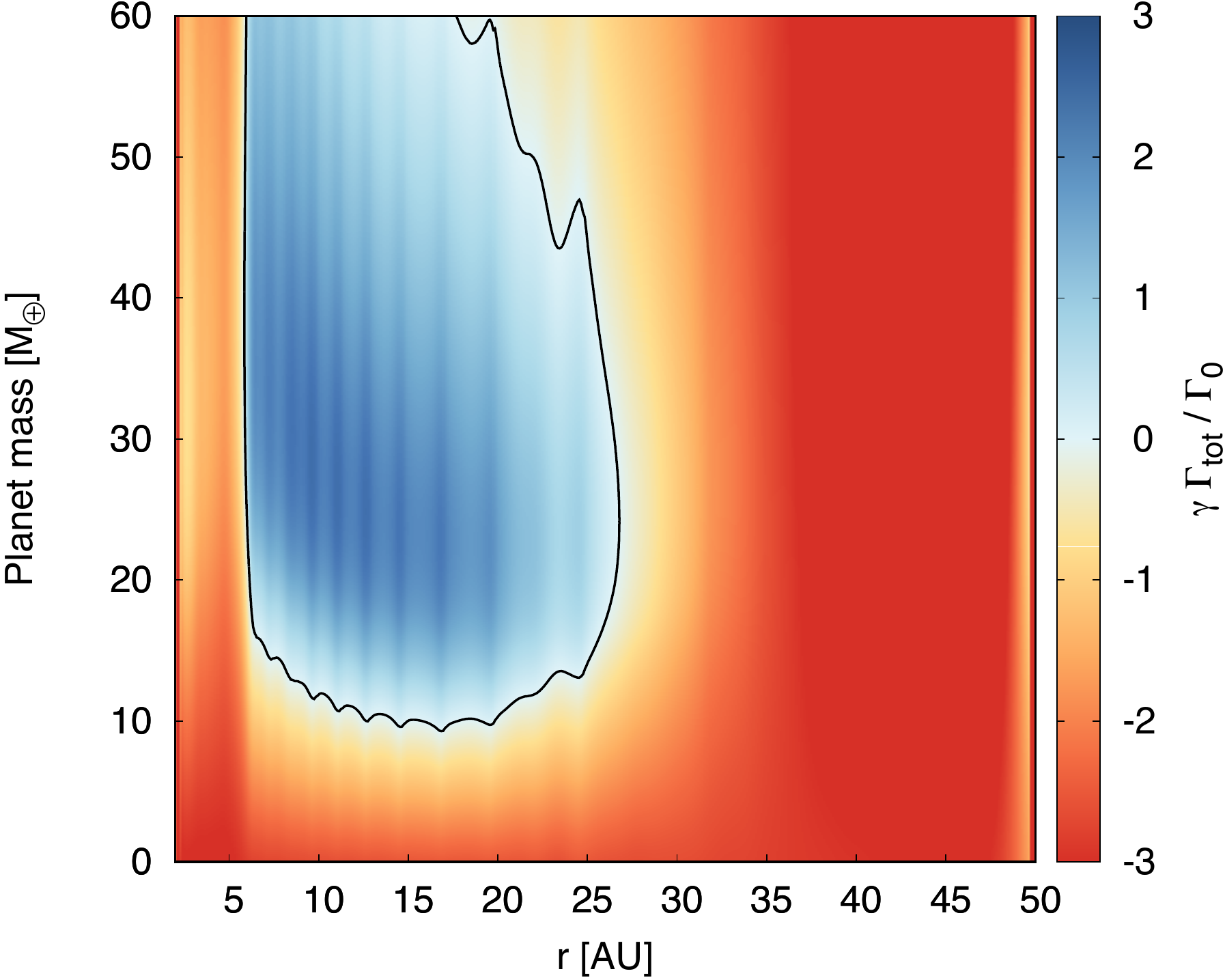}
\caption{Same as Fig. \ref{Fig:Migration maps_BOD} for the disc with the MRN distribution. The difference to the BOD distribution is small regarding the size of the region of outward migration, however, the torque is weaker for the MRN distribution. This could lead to different migration and growth behavior of planets forming in the outer disc.}
\label{Fig:Migration maps_MRN}
\end{figure}

\begin{figure}
\centering
\includegraphics[width=\columnwidth]{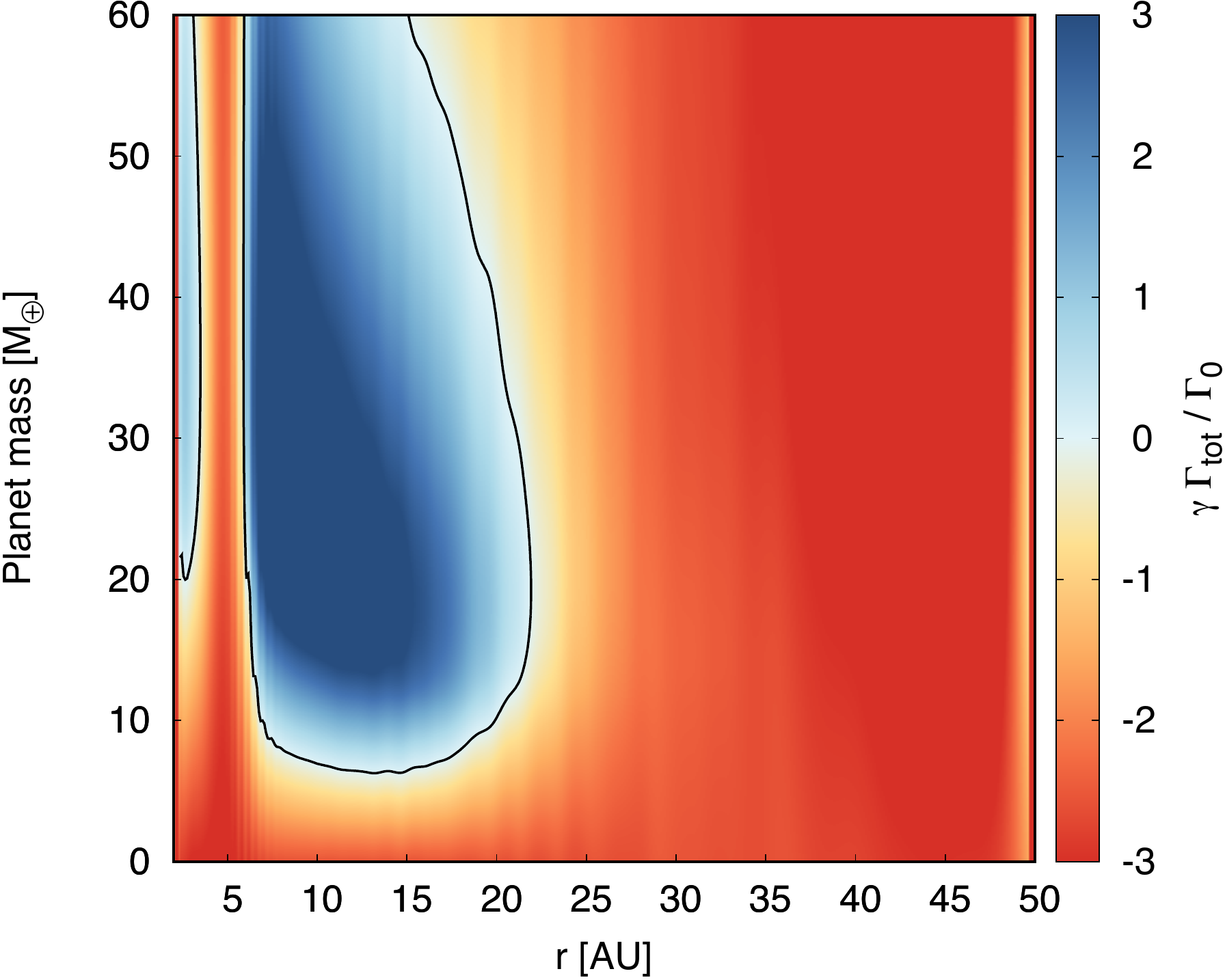}
\caption{Same as Fig. \ref{Fig:Migration maps_BOD} for the disc with the  \citet{1994ApJ...427..987B} opacity profile. In addition to the BOD and MRN grain size distribution the disc simulated with the \citet{1994ApJ...427..987B} opacity law shows an inner region of outward migration, which is caused by another bump in the H/r profile in the inner disc (Fig.5).}

\label{Fig:Migration maps_BL}
\end{figure}

Simulations with single particle sizes larger than 100 $\mu$m show a monotonically increasing aspect ratio with orbital distance (Fig. \ref{Fig:ALL_5e3S3O}), implying that planets in these colder discs would migrate inwards. The speed of the migration scales with the inverse of the square of the aspect ratio. If the aspect ratio decreases, then the migration is faster \citep[for a review see][]{2014prpl.conf..667B}. Therefore, generally speaking, migration would be faster in the single grain size models with large particles and the exact migration speed of planets depends on their exact position in the protoplanetary disc.

To predict more precisely the migration of planets in the discs presented here, we include the migration maps for two discs with the distributions (Figs. \ref{Fig:Migration maps_BOD} and \ref{Fig:Migration maps_MRN}). Migration rates are derived from the torque formula of \citet{2011MNRAS.410..293P}.
For comparison we also show the migration map of the disc with the \citet{1994ApJ...427..987B} opacities (Fig. \ref{Fig:Migration maps_BL}), as these were the opacities used in \citet{2015A&A...575A..28B} in Fig.18. The black solid line in Figs. \ref{Fig:Migration maps_BOD}, \ref{Fig:Migration maps_MRN}, \ref{Fig:Migration maps_BL} encircles the regions of outward migration. Planets with masses less than approximately 10 $M_{\oplus}$ always migrate inwards in the simulations with the distributions, whereas for the simulation using the \cite{1994ApJ...427..987B}, pure micrometer opacities, inward migration is the only possibility for planets less than 6 $M_{\oplus}$. 

The innermost region of outward migration in Fig.\ref{Fig:Migration maps_BL} corresponds to the area where the aspect ratio decreases as the orbital distance increases (Fig. \ref{Fig:GSDsmicroBL_5e3O}). We can also see that this area has a steeper temperature gradient (Fig. \ref{Fig:GSDsmicroBL_5e3O}) both interior and exterior to the water iceline transition. The increased torques and consequently migration speeds at the outer region of outward migration between 5 and 20 AU in the \citet{1994ApJ...427..987B} disc simulations are caused by the steep increase of opacity for temperatures larger than 170 K, which is not the case for the BOD and MRN distribution, as these distributions are not dominated by micrometer grains, in contrast to the \citet{1994ApJ...427..987B} opacity model. This illustrates that the grain sizes which dominate or contribute the most to the opacity of the disc have significant implications, not only for the disc structure itself, but also indirectly for the planets embedded in that disc. In addition, this has important effects for the formation of planetary systems, because the migration rates determine how close planets can migrate towards each other, which sets the stability of the planetary system \citep{2012Icar..221..624M,2006ApJ...652L.133C}.

\subsection{Implications for planet formation and protoplanetary disc simulations}
\label{subsec:Planet formation}

Our here presented simulations are the first step towards more self-consistent protoplanetary disc structure and evolution simulations as well as planet formation simulations. Planet formation in the pebble assisted core accretion scenario rely crucially on the pebble sizes and distributions \citep[e.g.][for review]{2010A&A...520A..43O,2012A&A...544A..32L,2017AREPS..45..359J}, as well as on the disc structure to calculate the planet migration rates as the planets grow \citep{2015A&A...582A.112B}. The here presented model opens the avenue to simulations with self-consistent disc structures and pebble sizes, which can then be accreted onto planets. This can increase the accuracy of future planet formation simulations by pebble accretion.

The here used FARGOCA code also allows for 3D hydrodynamical simulations with embedded planets. A combination with the presented model of thermal structures calculated from full grain size distributions allows a very detailed comparison with observations, which are more advanced that the mostly used 2D isothermal simulations followed by 3D radiative transfer (e.g. \citealt{2018ApJ...869L..47Z}). This could potentially change our interpretation of observed protoplanetary discs featuring substructures potentially caused by planets.

\section{Summary}
\label{sec:Summary}
We perform 2D hydrodynamical simulations including the whole feedback loop shown in Fig. \ref{Fig:Loop}. Specifically, we include and test two full grain size distributions and mean opacities (calculated via Mie theory) and study their influence on the disc structure. The particles have a minimal size of 0.025 $\mu$m and the upper boundary is regulated by the fragmentation barrier (Eq. \ref{eq:s_max}). We test five different $\alpha$-viscosity values ($10^{-2}, 5\times10^{-3}, 10^{-3}, 5\times 10 ^{-4}, 10^{-4}$), five values of initial gas surface density $\Sigma_{g,0}$ (100, 250, 500, 750 and 1000 $g/cm^2$) and five values of dust-to-gas ratio $f_{DG}$ (1\%, 1.5\%, 2\%, 2.5\%, 3\%). We also perform simulations with only single grain sizes and with the \citet{1994ApJ...427..987B} opacity law for comparison and in order to understand to greater extend the influence of grains of different sizes on the thermal structures of protoplanetary discs.

The dust component in protoplanetary discs is believed to follow a size distribution, regulated by a coagulation-fragmentation equilibrium \citep{2008A&A...480..859B,2011A&A...525A..11B}. We utilize and compare two different grain size distributions. The first and simple model (MRN) \citep{1969JGR....74.2531D,1977ApJ...217..425M,1996Icar..123..450T} results from the equilibrium between fragmentation and coagulation, whereas the second and more complex model \citep[][BOD]{2011A&A...525A..11B} takes into account fragmentation, coagulation and also cratering and adjusts the dust surface densities according to the grain sizes and how they compare to the size of the gas molecules and the gas turbulent eddies. 

The dust surface densities are calculated as dictated by the aforementioned grain size distributions and the dust grains are also vertically distributed according to their sizes taking into account the effect of settling. We also have a spatial distribution radially, since the size distribution depends on the local disc parameters and changes self-consistently. In conclusion, a whole loop of growth, fragmentation, and settling of the resulting grains for each vertical slice of the disk is modeled in our simulations and updated at every timestep according to the local disc parameters. 

We show disc structures calculated with the full grain size distributions and single grain sizes in Figs. \ref{Fig:ALL_5e3S3O} and \ref{Fig:ALL_T_5e3S3O}. Additionally we show that the grain sizes which dominate or contribute the most to the opacity of the disc are not the same at all orbital distances of the disc (Figs. \ref{Fig:Dominant_gs} and \ref{Fig:Dominant_gs_comparison}). As a consequence, the opacity prescriptions which assume a single dust size lead to inaccurate calculations of the thermal structures of the discs. 

It is also important to stress that the dust surface densities, or in other words the distribution of mass among the grain sizes, play a major role in determining the disc opacity (Eq. \ref{eq:kappa}), which in turn influences the cooling rate and the stellar heating and changes the temperature and surface density of the gas. This shift in the local disc parameters leads to a new fragmentation barrier (and regime boundaries for the BOD), therefore the dust surface densities change and so on. For this reason it is important to include the self-consistent calculations of the dust surface densities in the simulations.

The two grain size distributions show minimal differences in the dust surface densities (Fig. \ref{Fig:Dust}). The reason for this is that both of the grain size distributions we have used in the discs, feature the same fragmentation barrier. Therefore the grain size range in the discs with either one of the distributions is similar. Any difference between the discs with the BOD distribution and the discs with the MRN distribution comes mainly from the difference in the surface densities as a function of grain size (see Figs. \ref{Fig:Distributions} and \ref{Fig:Dust}), which is usually smaller than an order of magnitude. The dominant grain sizes (Sec. \ref{subsec:dominant gs}) might not be the same, because of the small differences in the dust surface densities per grain size, but the total opacity of the disc is similar independently of the grain size distribution. 

With this accurate prescription we investigate the dependency of the iceline position on the $\alpha$-viscosity, the initial gas surface density and the dust-to-gas ratio, where we see the effect of the feedback loop and find $r_{ice} \propto \alpha^{0.61}\Sigma_{g,0}^{0.8}f_{DG}^{0.37}$  (Eq. \ref{eq:Iceline_fitting}, Fig. \ref{Fig:Iceline}) independently of the grain size distribution utilized in the disc model. Specifically, for high gas surface density ($\Sigma_{g,0}=1000~g/cm^2$) the position of the iceline is exterior to 1 AU for $\alpha\geq2.64\times 10^{-4}$ and exterior to 2.7 AU only when $\alpha \ge 1.35 \times 10^{-3}$. For higher values than the nominal $f_{DG}=1\%$ we find that the iceline moves closer to the star as it is expected by the enhanced dust surface densities and the consequent hotter discs. However, for the nominal $f_{DG}=1\%$, lowering the gas surface density results in colder discs and the iceline is below 2.7 AU, even for the high viscosity models. 

The changes in the aspect ratio gradient as a function of orbital distance affect the regions where outward migration is possible for planets that could be embedded in the disc (Figs. \ref{Fig:Migration maps_BOD}, \ref{Fig:Migration maps_MRN}, \ref{Fig:Migration maps_BL}). Utilizing an $\alpha$-viscosity of $5\times10^{-3}$, $\Sigma_{g,0} = 1000~g/cm^2$ and $f_{DG}=1\%$ we find that the regions where outward migration could be possible in the discs with the two distributions are similar to the one present in a disc with the \citet{1994ApJ...427..987B} opacities, that feature only micrometer sized grains, at around 5-15 AU for planets with masses greater than 10 M$_{\oplus}$. However, the region is more extended for the disc with the BOD distribution (up to 20 AU) and the disc with the \citet{1994ApJ...427..987B} opacities has one more outward migration regions, near the inner boundary (2-3 AU), which is not present in the discs with the grain size distributions.

We can hence conclude that given the complexity and computational expense of the BOD distribution and the fact that it does not take into account radial drift or bouncing of the dust particles it is not necessary to prefer it over a simple MRN-like power-law distribution.  

As the iceline can be the starting point for planetesimal formation \citep{2014A&A...572A..72G,2017A&A...602A..21S,2017A&A...608A..92D} it is important to have as realistic models as possible, therefore include the feedback loop of grain growth and thermodynamics in hydrodynamical models (Fig. \ref{Fig:Loop}). Given also the fact that dust in protoplanetary discs follows a size distribution regulated by a coagulation-fragmentation equilibrium, the opacity prescription of a single grain size is not able to accurately calculate the thermal structures of discs.

The here presented model has some limitations that we wish to further investigate in future work. Both of the distributions tested here neglect radial drift \citep{2008A&A...480..859B,2012A&A...539A.148B} and bouncing \citep{2010A&A...513A..57Z,2018A&A...611A..18L} which can be detrimental for grain growth and in general affect the dust dynamics and subsequently the disc's thermal structure. Also the onset of convection in some regions and for a subset of $\alpha$ and $\Sigma_{g,0}$ values might change the vertical distribution of the grains beyond settling and turbulent stirring by viscosity, as taken into account here. A very important future step is to model accretion discs instead of equilibrium discs and in this way we will be able to also study different evolutionary steps of the protoplanetary disc. The particle composition and abundances are also determinant for dust dynamics and opacities, so it is important to consider a population that is as realistic as possible and use more accurate fragmentation velocities depending on our dust composition. Similarly, we are assuming a dust population consisting of 50\% silicates and 50\% water-ice, but we can relax this assumption and test different fractions (as done for example in \citet{2016A&A...590A.101B}).

It is hence evident that the prescription that we used and presented for this work opens up new avenues for protoplanetary disc simulations and planet formation. The inclusion of the feedback loop of grain growth and disc thermodynamics leads to more self consistent simulations of protoplanetary accretion discs and planet formation simulations in the pebble accretion scenario. Eventually, such models target a more precise comparison of protoplanetary disc observations to simulations  that allow us to move away from simple 2D isothermal models with post-processing of radiation transfer.

\begin{acknowledgements}
B.B. and S.S. thank the European Research Council (ERC Starting Grant 757448-PAMDORA) for their financial support. M.L. thanks the Knut and Alice Wallenberg Foundation (grant 2017.0287, PI A. Johansen). S.S is a Fellow of the International Max Planck Research School for Astronomy and Cosmic Physics at the University of Heidelberg (IMPRS-HD). We would like to thank the anonymous referee for the valuable comments and suggestions that helped us improve the manuscript.

\end{acknowledgements}

\bibliographystyle{aa}
\bibliography{Savvidou19}

\begin{appendix}

\section{Dust surface densities for different $\alpha$-viscosity values}
\label{sec:AppA}

We present here the vertically integrated dust surface densities and the opacities for the simulations using the rest of the $\alpha$ values and the two grain size distributions. The maximum grain sizes of the BOD and MRN discs are approximately the same, because they follow the same fragmentation barrier formula (Eq. \ref{eq:s_max}). The vertically integrated dust surface densities are around one order of magnitude {\bf lower} in the discs with the BOD grain size distribution compared to the discs with the MRN distribution for small particles (Fig. \ref{Fig:Dust}). On the other hand they are always comparable for the largest particles in the discs. This is already evident by the shape of the two grain size distributions (Fig. \ref{Fig:Distributions}). 

As $\alpha$ decreases, the surface densities of the smallest particles in the discs with the BOD distribution are several orders of magnitude lower than these of the largest particles. The gradients are smoother in the discs with the MRN distribution as $\alpha$ decreases. The fragmentation barrier depends on the initial gas surface density so when the latter decreases, the maximum grain size gets smaller (Eq. \ref{eq:s_max}). 

 In Fig. \ref{Fig:Opacity} we show the opacities as a function of orbital distance and height for a selection of the simulated discs for this work. The $\tau$=1 line is located at similar heights in all of the discs with high $\alpha$-viscosity (around 2 AU at the outer edge). The same applies to the low $\alpha$-viscosity discs where the $\tau$=1 line is always around 0.15 AU at the outer edge. As expected the optically thick region is extended towards higher altitudes with higher total dust-to-gas ratio and contained near the midplane for low gas surface densities. Above the $\tau$=1 line opacity always decreases as cooling is more efficient. However, the uppermost layers show increased opacities because of the stellar irradiation that directly heats them up.

\begin{figure*}
\begin{subfigure}{0.5\textwidth}
\includegraphics[width=.8\textwidth]{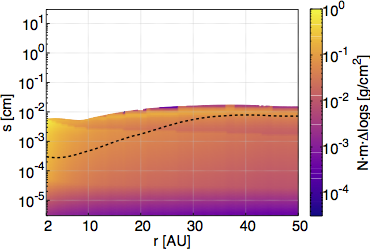}
\caption{BOD, $\alpha=10^{-2}$, $\Sigma_{g,0} = 1000~g/cm^2$, $f_{DG}=1\%$}
\end{subfigure}
\begin{subfigure}{0.5\textwidth}
\includegraphics[width=.8\textwidth]{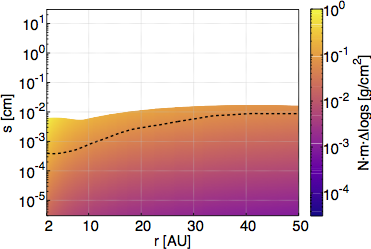}
\caption{MRN, $\alpha=10^{-2}$, $\Sigma_{g,0} = 1000~g/cm^2$, $f_{DG}=1\%$}
\end{subfigure}
\par\bigskip

\begin{subfigure}{0.5\textwidth}
\includegraphics[width=.8\textwidth]{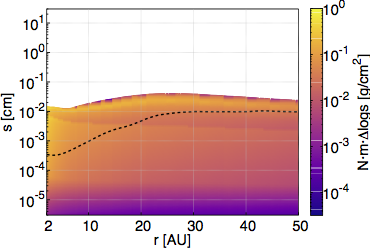}
\caption{BOD, $\alpha=5\times10^{-3}$, $\Sigma_{g,0} = 1000~g/cm^2$, $f_{DG}=1\%$}
\end{subfigure}
\begin{subfigure}{0.5\textwidth}
\includegraphics[width=.8\textwidth]{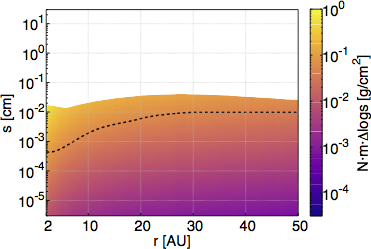}
\caption{MRN, $\alpha=5\times10^{-3}$, $\Sigma_{g,0} = ~1000g/cm^2$, $f_{DG}=1\%$}
\end{subfigure}
\par\bigskip

\begin{subfigure}{0.5\textwidth}
\includegraphics[width=.8\textwidth]{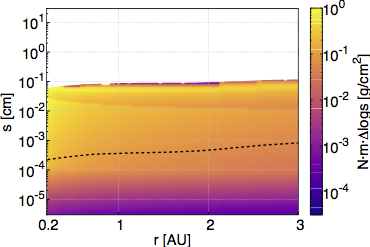}
\caption{BOD, $\alpha=10^{-3}$, $\Sigma_{g,0} = 1000~g/cm^2$, $f_{DG}=1\%$}
\end{subfigure}
\begin{subfigure}{0.5\textwidth}
\includegraphics[width=.8\textwidth]{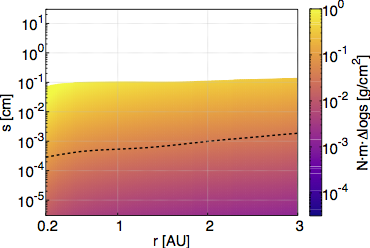}
\caption{MRN, $\alpha=10^{-3}$, $\Sigma_{g,0} = 1000~g/cm^2$, $f_{DG}=1\%$}
\end{subfigure}
\par\bigskip

\begin{subfigure}{0.5\textwidth}
\includegraphics[width=.8\textwidth]{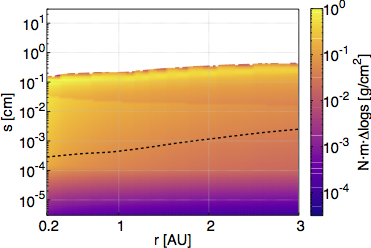}
\caption{BOD, $\alpha=5\times10^{-4}$, $\Sigma_{g,0} = 1000~g/cm^2$, $f_{DG}=1\%$}
\end{subfigure}
\begin{subfigure}{0.5\textwidth}
\includegraphics[width=.8\textwidth]{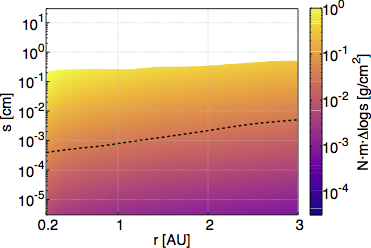}
\caption{MRN, $\alpha=5\times10^{-4}$, $\Sigma_{g,0} = 1000~g/cm^2$, $f_{DG}=1\%$}
\end{subfigure}
\par\bigskip

\end{figure*}

\begin{figure*}\ContinuedFloat
\begin{subfigure}{0.5\textwidth}
\includegraphics[width=.8\textwidth]{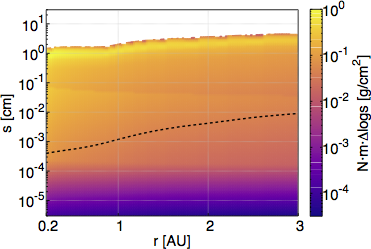}
\caption{BOD, $\alpha=10^{-4}$, $\Sigma_{g,0} = 1000~g/cm^2$, $f_{DG}=1\%$}
\end{subfigure}
\begin{subfigure}{0.5\textwidth}
\includegraphics[width=.8\textwidth]{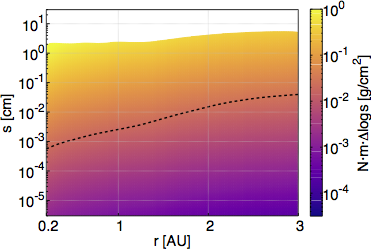}
\caption{MRN, $\alpha=10^{-4}$, $\Sigma_{g,0} = 1000~g/cm^2$, $f_{DG}=1\%$}
\end{subfigure}
\par\bigskip

\begin{subfigure}{0.5\textwidth}
\includegraphics[width=.8\textwidth]{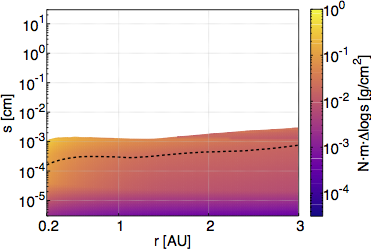}
\caption{BOD, $\alpha=1\times10^{-2}$, $\Sigma_{g,0} = 100~g/cm^2$, $f_{DG}=1\%$}
\end{subfigure}
\begin{subfigure}{0.5\textwidth}
\includegraphics[width=.8\textwidth]{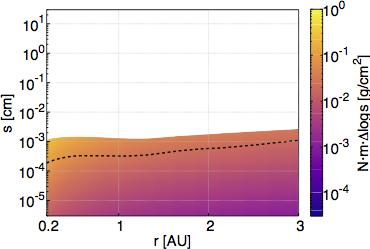}
\caption{MRN, $\alpha=1\times10^{-2}$, $\Sigma_{g,0} = 100~g/cm^2$, $f_{DG}=1\%$}
\end{subfigure}
\par\bigskip

\begin{subfigure}{0.5\textwidth}
\includegraphics[width=.8\textwidth]{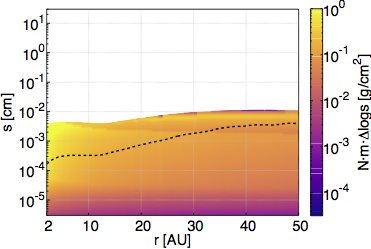}
\caption{BOD, $\alpha=1\times10^{-2}$, $\Sigma_{g,0} = 1000~g/cm^2$, $f_{DG}=3\%$}
\end{subfigure}
\begin{subfigure}{0.5\textwidth}
\includegraphics[width=.8\textwidth]{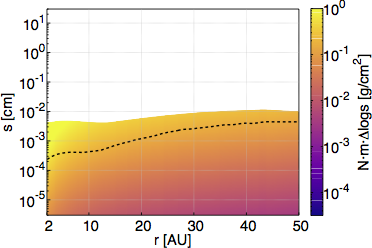}
\caption{MRN, $\alpha=1\times10^{-2}$, $\Sigma_{g,0} = 1000~g/cm^2$, $f_{DG}=3\%$}
\end{subfigure}
\caption{Dust surface densities as a function of orbital distance and grain size for the different $\alpha$ values used here, and for additional simulations with the lowest gas surface densities and with the highest dust-to-gas ratio that we have tried. When the turbulence is reduced, the maximum grain size increases (since less destructive collisions are expected). In addition to that, the reduced $\alpha$-viscosity allows the discs to become cooler, so the opacity of the larger grains ($\sim$mm) becomes comparable or larger than that of the smaller ($\sim\mu$m) grains. For $\alpha=10^{-4}$ we find that the grains grow up to a few centimeters in the discs with either one of the distributions. The dashed lines divide the grain sizes into two groups which contribute equally to the total opacity of the disc (see Figs. \ref{Fig:Dominant_gs} and \ref{Fig:Dominant_gs_comparison}).
}
\label{Fig:Dust}
\end{figure*}

\begin{figure*}

\begin{subfigure}{0.5\textwidth}
\includegraphics[width=.8\textwidth]{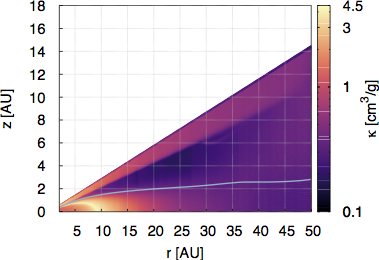}
\caption{BOD, $\alpha=10^{-2}$, $\Sigma_{g,0} = 1000~g/cm^2$, $f_{DG}=1\%$}
\end{subfigure}
\begin{subfigure}{0.5\textwidth}
\includegraphics[width=.8\textwidth]{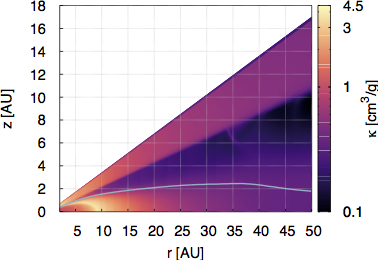}
\caption{MRN, $\alpha=10^{-2}$, $\Sigma_{g,0} = 1000~g/cm^2$, $f_{DG}=1\%$}
\end{subfigure}
\par\bigskip

\begin{subfigure}{0.5\textwidth}
\includegraphics[width=.8\textwidth]{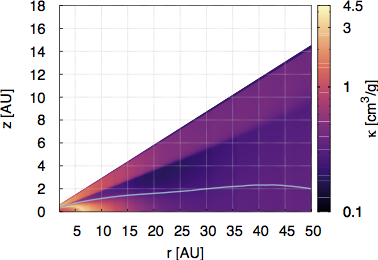}
\caption{BOD, $\alpha=5\times10^{-3}$, $\Sigma_{g,0} = 1000~g/cm^2$, $f_{DG}=1\%$}
\end{subfigure}
\begin{subfigure}{0.5\textwidth}
\includegraphics[width=.8\textwidth]{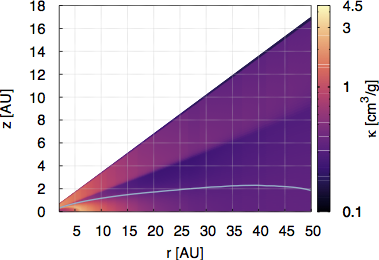}
\caption{MRN, $\alpha=5\times10^{-3}$, $\Sigma_{g,0} = 1000~g/cm^2$, $f_{DG}=1\%$}
\end{subfigure}
\par\bigskip

\begin{subfigure}{0.5\textwidth}
\includegraphics[width=.8\textwidth]{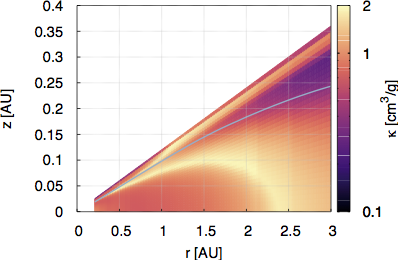}
\caption{BOD, $\alpha=10^{-3}$, $\Sigma_{g,0} = 1000~g/cm^2$, $f_{DG}=1\%$}
\end{subfigure}
\begin{subfigure}{0.5\textwidth}
\includegraphics[width=.8\textwidth]{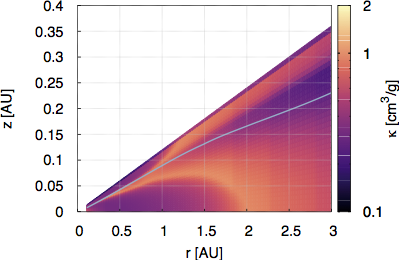}
\caption{MRN, $\alpha=10^{-3}$, $\Sigma_{g,0} = 1000~g/cm^2$, $f_{DG}=1\%$}
\end{subfigure}
\par\bigskip

\begin{subfigure}{0.5\textwidth}
\includegraphics[width=.8\textwidth]{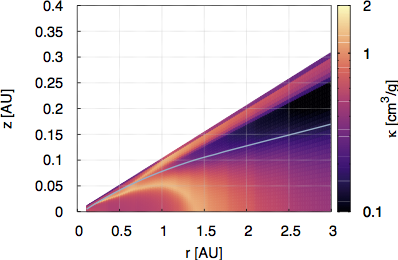}
\caption{BOD, $\alpha=5\times10^{-4}$, $\Sigma_{g,0} = 1000~g/cm^2$, $f_{DG}=1\%$}
\end{subfigure}
\begin{subfigure}{0.5\textwidth}
\includegraphics[width=.8\textwidth]{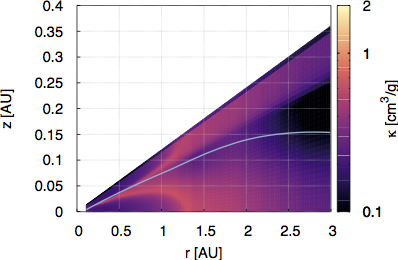}
\caption{MRN, $\alpha=5\times10^{-4}$, $\Sigma_{g,0} = 1000~g/cm^2$, $f_{DG}=1\%$}
\end{subfigure}
\end{figure*}

\begin{figure*}\ContinuedFloat
\begin{subfigure}{0.5\textwidth}
\includegraphics[width=.8\textwidth]{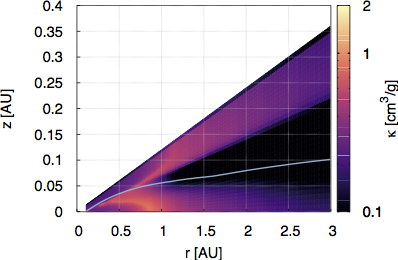}
\caption{BOD, $\alpha=10^{-4}$, $\Sigma_{g,0} = 1000~g/cm^2$, $f_{DG}=1\%$}
\end{subfigure}
\begin{subfigure}{0.5\textwidth}
\includegraphics[width=.8\textwidth]{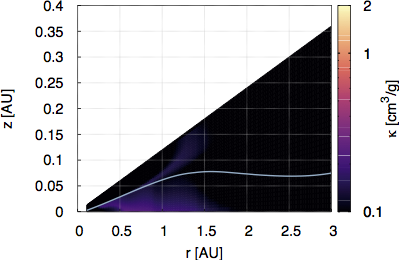}
\caption{MRN, $\alpha=10^{-4}$, $\Sigma_{g,0} = 1000~g/cm^2$, $f_{DG}=1\%$}
\end{subfigure}
\par\bigskip

\begin{subfigure}{0.5\textwidth}
\includegraphics[width=.8\textwidth]{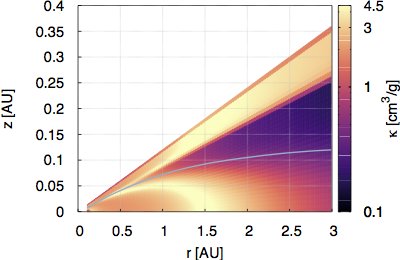}
\caption{BOD, $\alpha=1\times10^{-2}$, $\Sigma_{g,0} = 100~g/cm^2$, $f_{DG}=1\%$}
\end{subfigure}
\begin{subfigure}{0.5\textwidth}
\includegraphics[width=.8\textwidth]{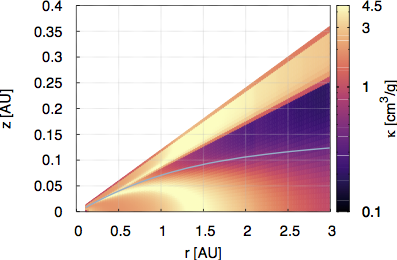}
\caption{MRN, $\alpha=1\times10^{-2}$, $\Sigma_{g,0} = 100~g/cm^2$, $f_{DG}=1\%$}
\end{subfigure}
\par\bigskip

\begin{subfigure}{0.5\textwidth}
\includegraphics[width=.8\textwidth]{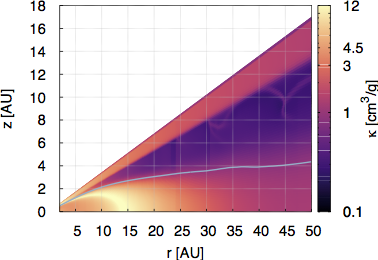}
\caption{BOD, $\alpha=1\times10^{-2}$, $\Sigma_{g,0} = 1000~g/cm^2$, $f_{DG}=3\%$}
\end{subfigure}
\begin{subfigure}{0.5\textwidth}
\includegraphics[width=.8\textwidth]{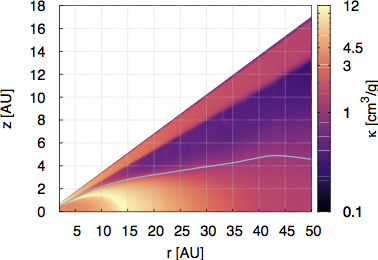}
\caption{MRN, $\alpha=1\times10^{-2}$, $\Sigma_{g,0} = 1000~g/cm^2$, $f_{DG}=3\%$}
\end{subfigure}

\caption{Mean Rosseland opacities as a function of orbital distance and height for the different $\alpha$ values, the lowest gas surface densities and the highest dust-to-gas ratio. Due to the larger grain sizes for the MRN distribution (Fig. \ref{Fig:Dust}), the opacities for the MRN distribution are also generally lower compared to the BOD distribution. The light blue line corresponds to optical depth $\tau=1$ integrated vertically starting from infinity towards midplane, so it divides the optically thin (above) and thick region (below). }
\label{Fig:Opacity}
\end{figure*}

\section{Iceline position as a function of $\alpha$-turbulence, initial gas surface density and dust-to-gas ratio}
\label{sec:AppB}

\begin{figure*}
\centering
\begin{subfigure}{\columnwidth}
\includegraphics[width=1.05\textwidth]{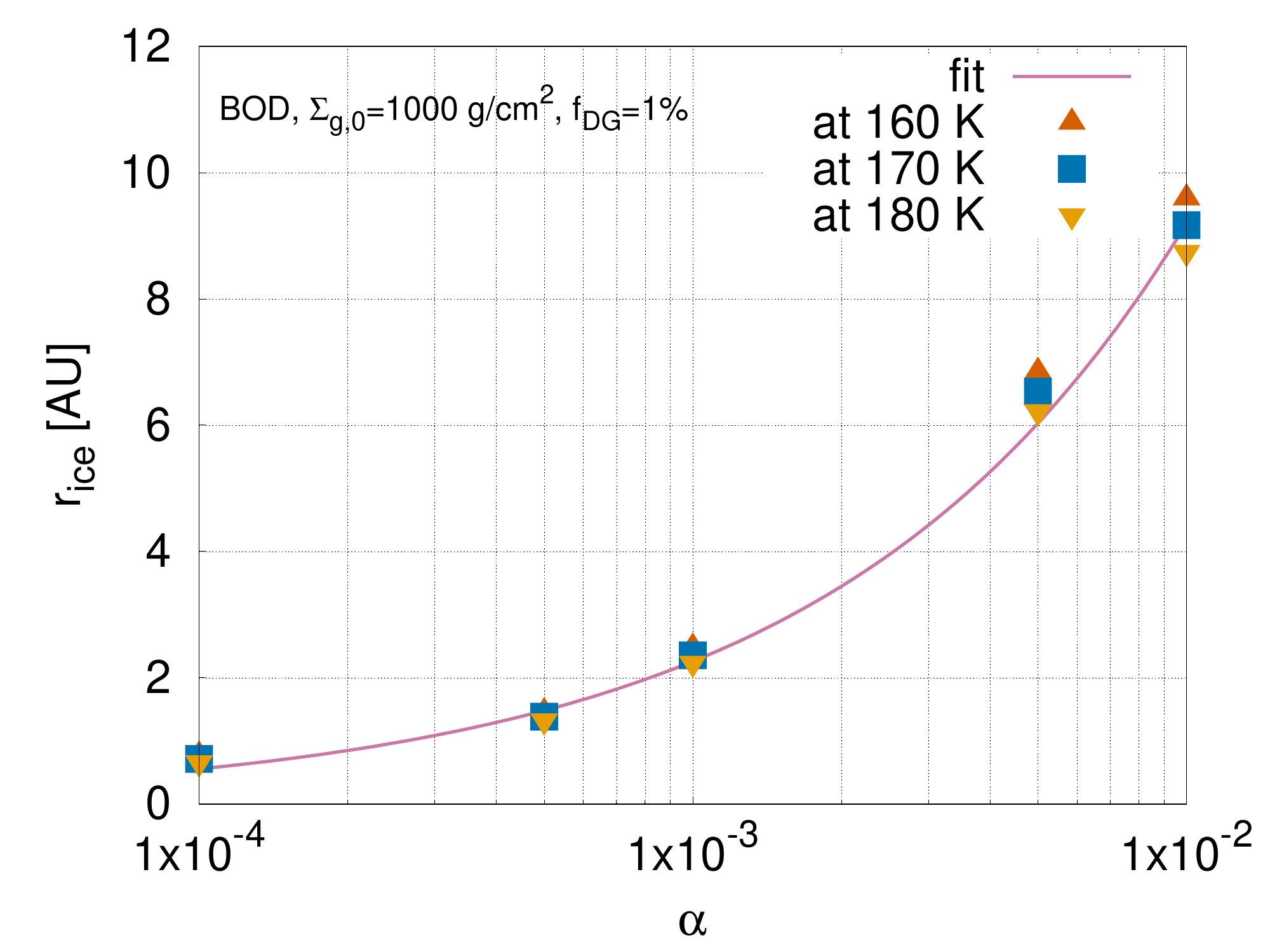}
\end{subfigure}
\begin{subfigure}{\columnwidth}
\includegraphics[width=1.05\textwidth]{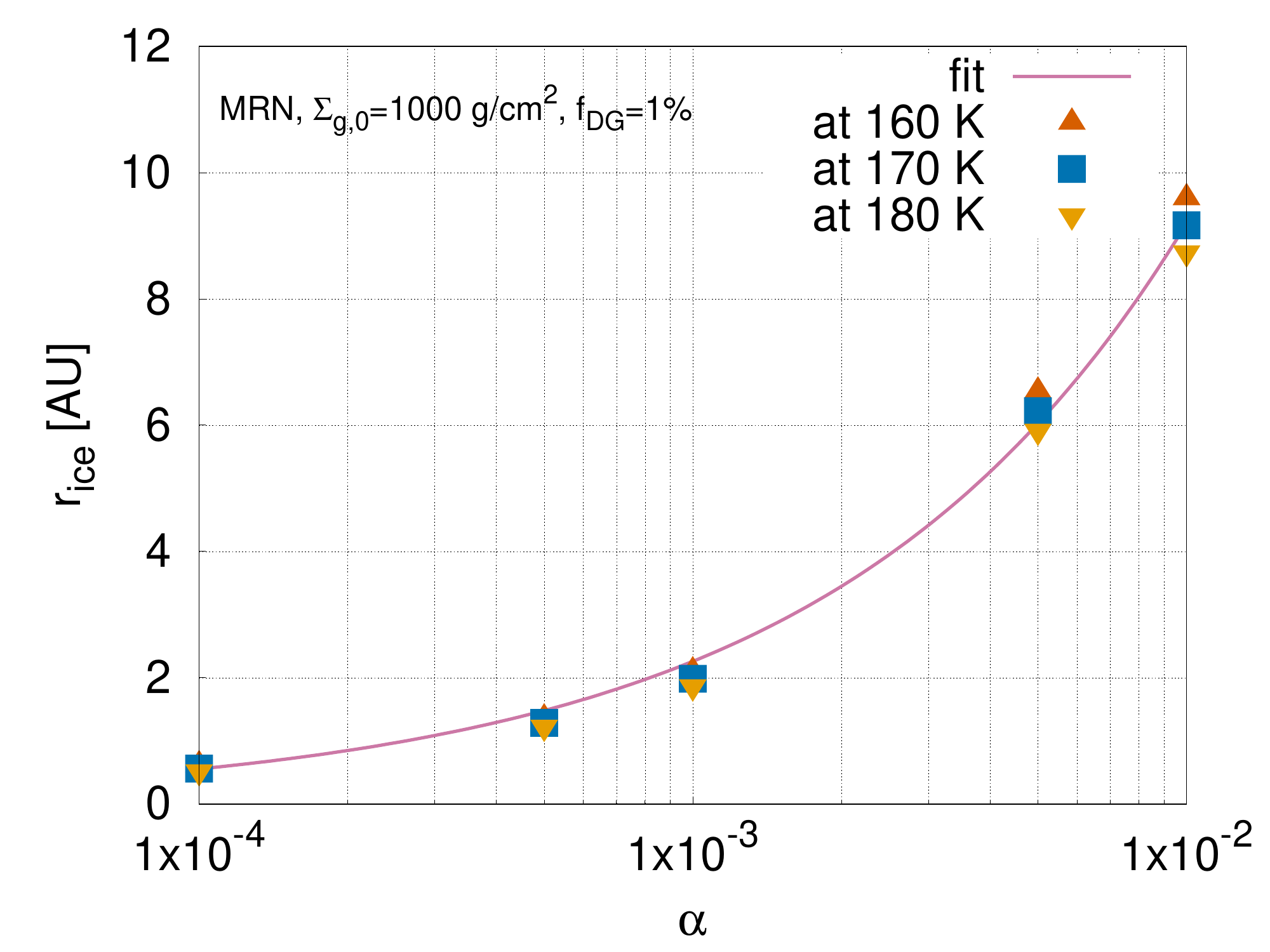}
\end{subfigure}
\par\bigskip

\begin{subfigure}{\columnwidth}
\includegraphics[width=\textwidth]{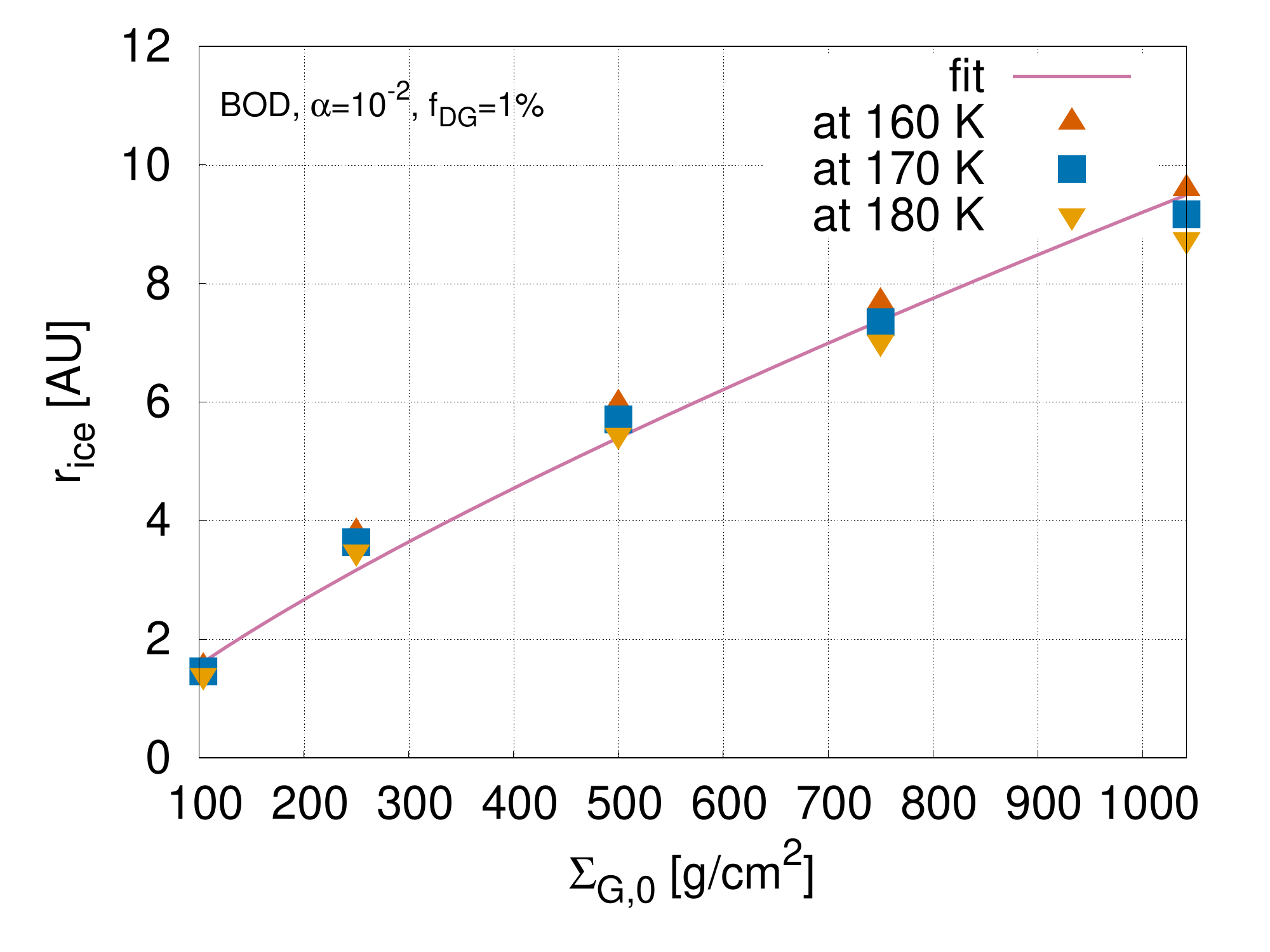}
\end{subfigure}
\begin{subfigure}{\columnwidth}
\includegraphics[width=\textwidth]{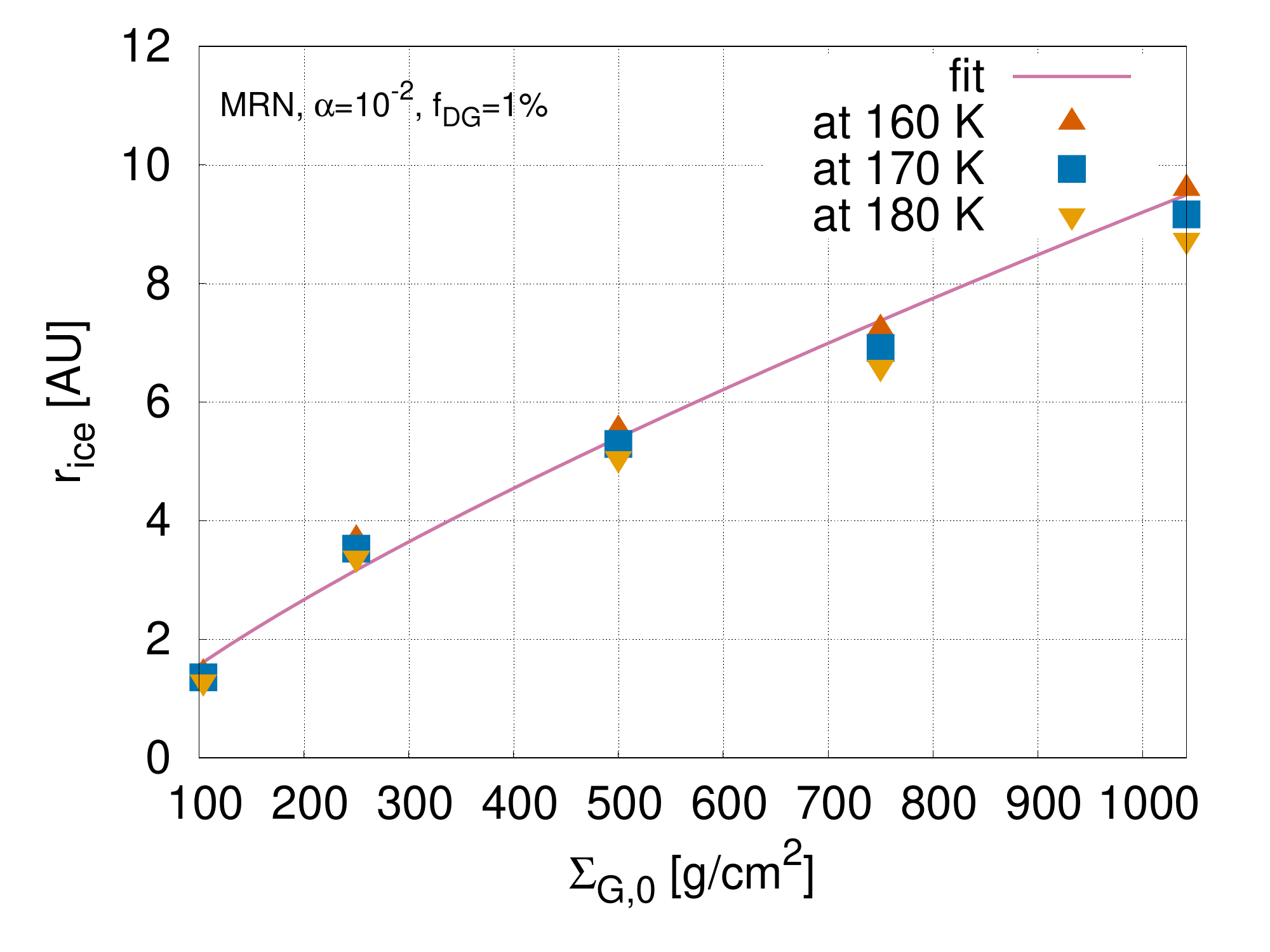}
\end{subfigure}
\par\bigskip

\begin{subfigure}{\columnwidth}
\includegraphics[width=1.03\textwidth]{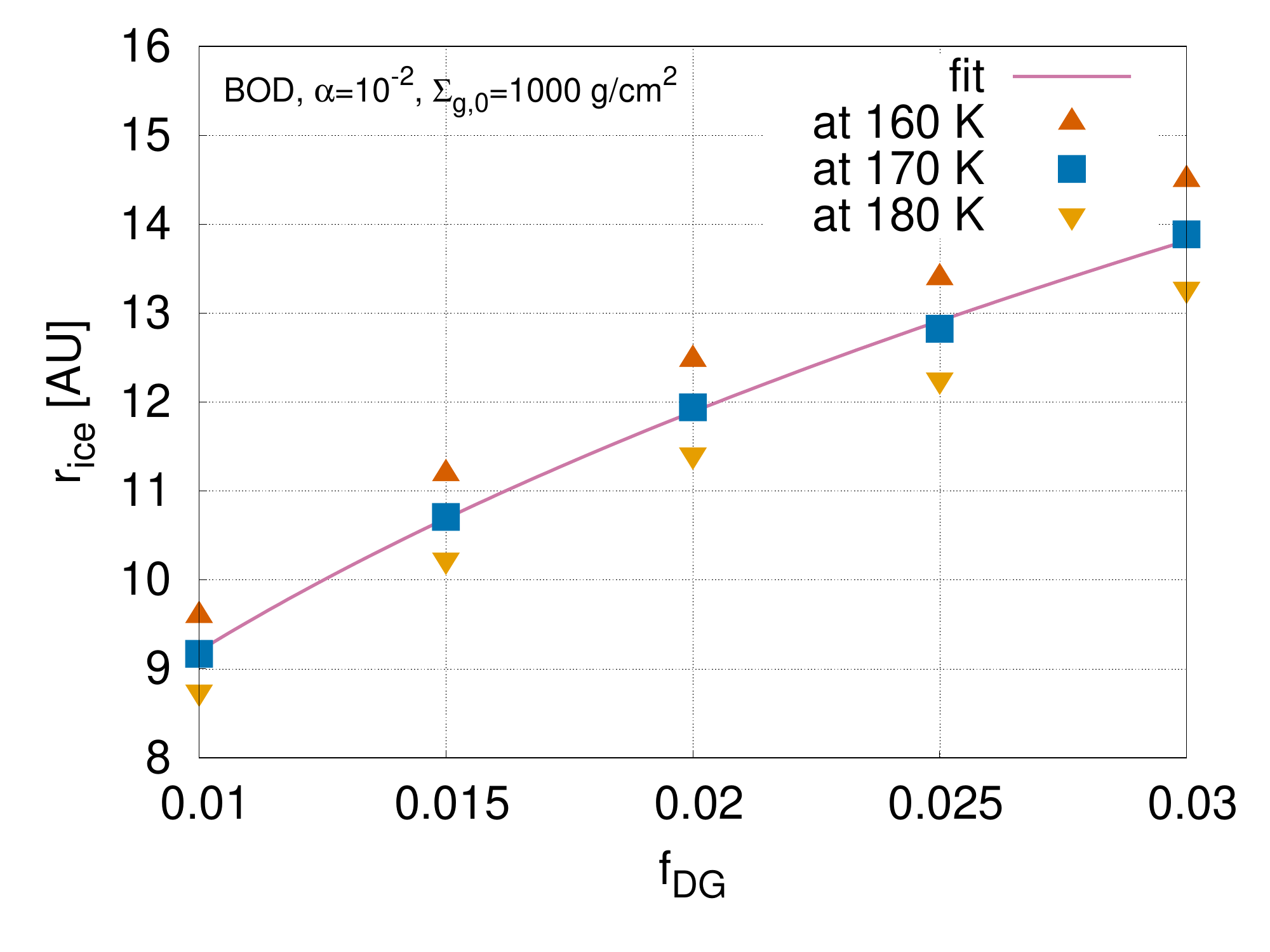}
\end{subfigure}
\begin{subfigure}{\columnwidth}
\includegraphics[width=1.03\textwidth]{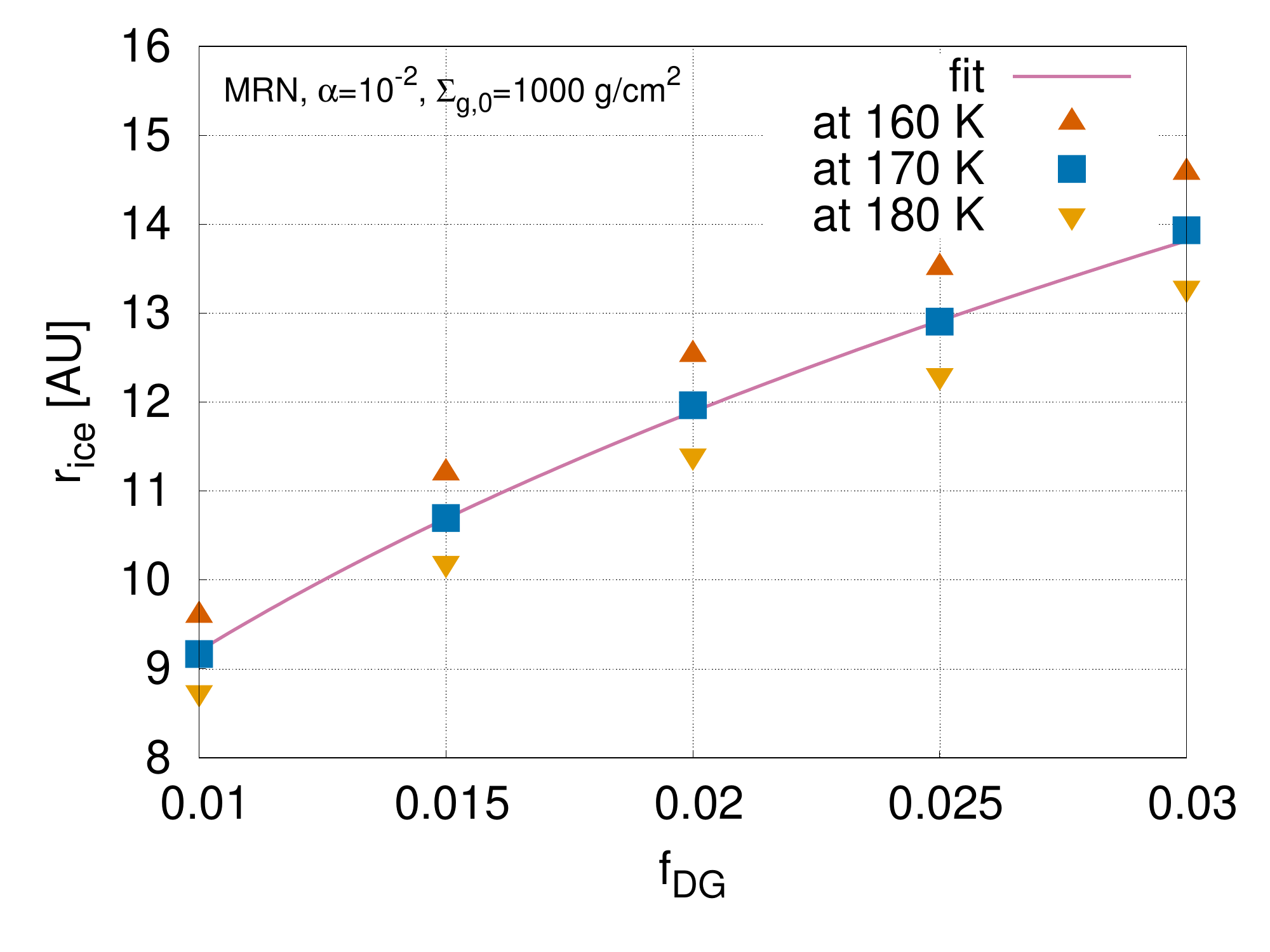}
\end{subfigure}
\par\bigskip

\caption{Iceline position as a function of $\alpha$-viscosity (top), initial gas surface density (middle) and dust-to-gas ratio (bottom) for the discs with the BOD distribution (left column) and the discs with the MRN distribution (right column). The iceline transition is defined as $T=(170 \pm 10) K$. The specific parameters used for the simulations presented in this plot  are shown in Table \ref{tab:Iceline_fitting_data}. The solid lines are the fits to each parameter and are the same for the discs with either one of the distributions. }
\label{Fig:IcelineFitting}
\end{figure*}

In Sec. \ref{subsec:Iceline} we present the position of the iceline as a function of the $\alpha$-viscosity parameter, the initial gas surface density, $\Sigma_{g,0}$ and the total dust-to-gas ratio $f_{DG}$.  In order to do this fitting, we used five simulations for each parameter (see Table \ref{tab:Iceline_fitting_data}). In Fig. \ref{Fig:IcelineFitting} we show the individual fitting over each one of the parameters. The fit to the three parameters writes

\be f = C \cdot \left(\frac{\alpha}{0.01}\right)^{p_1}\cdot\left(\frac{\Sigma_{g,0}}{1000~g/cm^2}\right)^{p_2}\left(\frac{f_{DG}}{0.01}\right)^{p_3}~,
\ee
with C=9.20$\pm$0.05 AU, $p_1$ = 0.61$\pm$0.03, $p_2$ = 0.77$\pm$0.03, $p_3$ = 0.37$\pm$0.01. The resulting fit is the same regardless of the grain size distribution used in the disc (solid line in Fig. \ref{Fig:IcelineFitting}).

\begin{table}[b]
\centering
\begin{tabular}{c|c|c}
\hline \hline 
{\boldmath$\alpha$} & {\boldmath$\Sigma_{g,0}$ [${g/cm^2}$]} &  {\boldmath$f_{DG}$} \\
 ($\Sigma_{g,0} = 1000~g/cm^2$, & ($\alpha=10^{-2}$ & ($\alpha=10^{-2}$, \\
 $f_{DG}=1\%$)  & $f_{DG}=1\%$) & $\Sigma_{g,0} = 1000~g/cm^2$) \\
\hline \hline 
$10^{-2}$	& 100	&  1\%\\
\hline
$5 \times 10^{-3}$ & 250	 &  1.5\%	\\
\hline
$10^{-3}$			& 500		&  2\%	\\
\hline
$5 \times10^{-4}$ 			& 750			&  2.5\%	\\
\hline
$10^{-4}$ 		& 1000			&  3\%	\\
\hline \hline 
\end{tabular}
\caption{Parameters used in the simulations performed for the fitting of the iceline position to $\alpha$-viscosity, initial gas surface density and total dust-to-gas ratio for the BOD and the MRN distribution.}
\label{tab:Iceline_fitting_data}
\end{table}

\section{The effect of settling}
\label{sec:AppC}

We implement in our work the effect of settling for the grains in the disc as described in Sect. \ref{subsec:vertical}, in order to vertically distribute the grains according to their sizes and the local disc parameters. This implies that the disc structure can be affected both by a change in the grain size due to the different opacities that each size provides (Fig. \ref{Fig:Loop}) and by a change in the settling efficiency of the given grain size. In order to test if both of these effects are significant factors that define the disc structure, we run one simulation where the disc only contains millimeter grains and compare with a disc which also contains only millimeter grains but does not take settling into account. Thus in this latter case, the millimeter grains are vertically distributed according to a constant dust-to-gas ratio throughout the whole disc. Additionally, we run a simulation where the opacities of the grains correspond to millimeter grains, but we assume that the grains are vertically distributed as micrometer grains (so we assume $s= 1\mu m$ in the equations describing settling, Eq. \ref{eq:vertical_distribution} - \ref{eq:Stokes}). We choose $\alpha=10^{-4}$ for which the settling of millimeter grains will be very effective. However, micrometer dust grains are not expected to be affected by settling even at this low $\alpha$-viscosity. The models also have an initial gas surface density of $\Sigma_{g,0} = 1000~g/cm^2$ and total dust-to-gas ratio $f_{DG} = 1\%$. 

\begin{figure} 
\includegraphics[width=\columnwidth]{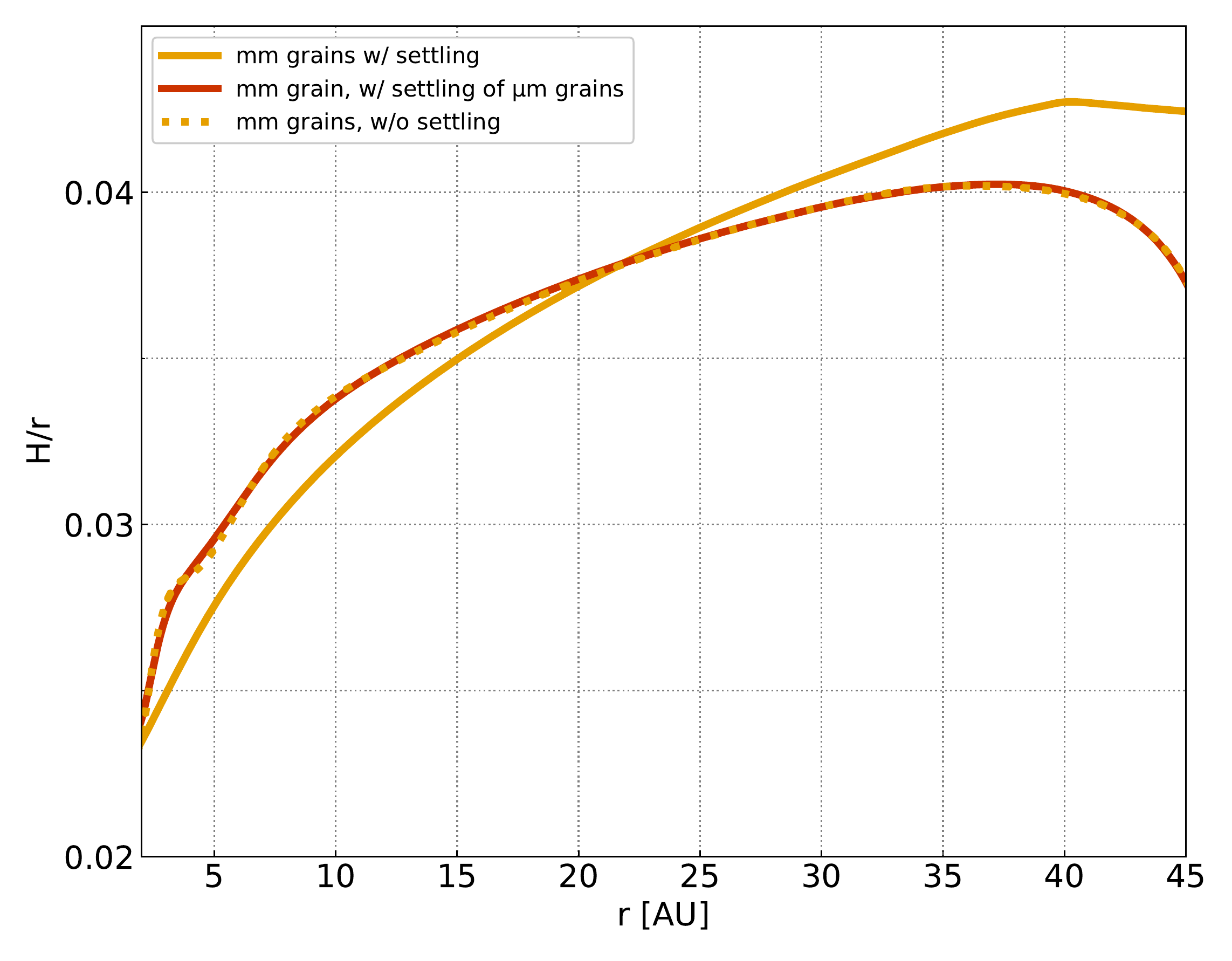}
\caption{Comparison of the aspect ratio as a function of orbital distance for discs with mm grains with settling, without settling and with the settling of $\mu m$ grains. The disc structure is almost the same when no settling is implemented and when the opacities correspond to mm grains, but the grains are distributed as micrometer sized dust (Eq. \ref{eq:vertical_distribution} - \ref{eq:Stokes}). When settling is included the millimeter grains are concentrated near the midplane, leaving the upper layers with diminished opacities and enhanced cooling rate. Without efficient settling the inner discs get hotter. However, this creates a shadow that prevents the efficient heating of the outermost regions from stellar irradiation.}
\label{Fig:settling_test}
\end{figure}

In Figure \ref{Fig:settling_test} we present the aspect ratio of these three disc models. The aspect ratio as a function of orbital distance for the disc where grains are vertically distributed as micrometer sized dust resembles the one of a disc where the dust-to-gas ratio is constant all over the disc. This is expected because micrometer sized dust is not significantly affected by settling even at the low $\alpha$-viscosity of $10^{-4}$ (see Fig. \ref{Fig:settling}). However, we find that the aspect ratio is lower in the inner regions of discs when the millimeter grains are allowed to settle with their corresponding properties. When settling is included, the millimeter grains are mainly concentrated near the midplane (see also Fig. \ref{Fig:settling}), while at higher altitudes the opacity diminishes. Without settling or with reduced efficiency of settling the opacity is similar at all altitudes, which leads to a reduced cooling rate and higher aspect ratio in the inner regions. 

Without settling of the millimeter grains according to their size properties, the outer regions are not sufficiently heated. Due to the increased aspect ratio in the inner disc, stellar irradiation to the outer disc is diminished, creating a shadow that cools down the outer region. At the same time the millimeter grains have very low opacity so they cannot absorb the stellar heating efficiently in the outer disc. The disc at the outermost radii might keep cooling down until it reaches the temperature of the surroundings \citep{2004A&A...417..159D}.
Hence including settling is very important to avoid such complications and inconsistencies in the disc structures. 

Different grain sizes lead to different disc structures even without any settling implemented. The distinctive structures of  discs with different grain sizes comes mainly by their individual opacities (Fig. \ref{Fig:Opacities}).  However, without settling the disc opacity above midplane is overestimated so the discs are hotter in the inner regions and thus do not allow the stellar irradiation to heat the outer regions.
In order to consistently take into account the influence of a grain size distribution to the resulting disc structures, it is important to include settling. 

\end{appendix}

\end{document}